\documentclass[12pt]{article}
\usepackage{amsmath,amssymb,amsfonts,amsthm,bbm}
\usepackage{epic,eepic,epsfig,longtable}
\usepackage{multirow,verbatim}
\usepackage{array}
\usepackage{graphicx}
\usepackage{floatrow}
\usepackage{subcaption}
\usepackage{caption}
\usepackage{paralist}
\usepackage{latexsym}
\usepackage{comment}
\usepackage{epsfig}
\usepackage{setspace}
\usepackage{CJK}
\usepackage{color}

\usepackage{geometry}
\usepackage{algorithm}
\usepackage{algpseudocode}
\usepackage{pdflscape}
\usepackage{rotating}
\usepackage{array, multirow}
\usepackage[authoryear,round]{natbib}
\usepackage{booktabs}
\usepackage{tabularx}
\def\spacingset#1{\renewcommand{\baselinestretch}%
{#1}\small\normalsize} \spacingset{1}

\makeatletter
\def\singlespace{\def\baselinestretch{1}\@normalsize}

\numberwithin{equation}{section}

\renewcommand{\hat}{\widehat}

\renewcommand{\hat}{\widehat}

\newcommand{\bfm}[1]{\ensuremath{\mathbf{#1}}}

\def\bb{\bfm b}   \def\bB{\bfm B}  
     
   \def\bD{\bfm D}  
     
\def\bff{\bfm f}  \def\bF{\bfm F}  \def\FF{\mathbb{F}}
   \def\bG{\bfm G}  
   \def\bH{\bfm H}  
   \def\bI{\bfm I}  
   \def\bJ{\bfm J}  
     
   \def\bL{\bfm L}

   \def\bR{\bfm R}

\def\bu{\bfm u}   \def\bU{\bfm U}  
\def\bv{\bfm v}   \def\bV{\bfm V}  
   \def\bW{\bfm W}  
\def\bx{\bfm x}   \def\bX{\bfm X}

\newcommand{\bfsym}[1]{\ensuremath{\boldsymbol{#1}}}

 \def\bbeta{\bfsym \beta}
 \def\bgamma{\bfsym \gamma}

 \def\bmu{\bfsym {\mu}}                 
 \def\bnu{\bfsym {\nu}}
 \def\btheta{\bfsym {\theta}}           
           \def\bepsilon{\bfsym \varepsilon}
              \def\bSigma{\bfsym \Sigma}
         \def\bLambda {\bfsym {\Lambda}}
           \def\bOmega {\bfsym {\Omega}}

\def\1{\bfsym{1}}	

\DeclareMathOperator{\argmin}{argmin}

\DeclareMathOperator{\E}{E}

\DeclareMathOperator{\supp}{supp}

\def\newpage{\vfill\eject}

\def\today{\ifcase\month\or
  January\or February\or March\or April\or May\or June\or
  July\or August\or September\or October\or November\or December\fi
  \space\number\day, \number\year}

\newdimen\biblioindent    \biblioindent=30pt

 at 8truept

\newcommand{\beq}{\begin{equation}}
  \newcommand{\eeq}{\end{equation}}
\newcommand{\beqn}{\begin{eqnarray}}
  \newcommand{\eeqn}{\end{eqnarray}}
\newcommand{\beqnn}{\begin{eqnarray*}}
  \newcommand{\eeqnn}{\end{eqnarray*}}

\def\lasso{{\rm LASSO}}

\allowdisplaybreaks
\usepackage{verbatim}
\def\tilde{\widetilde}

\def\FF{\mathcal{F}}
\def\[{\left [}  \def\]{\right ]} \def\({\left (}  \def\){\right )}
 \def\endpf{$\blacksquare$}
\def\hat{\widehat}

\newtheorem{assumption}{Assumption}
\newtheorem{theorem}{Theorem}

\newtheorem{proposition}{Proposition}
\theoremstyle{definition}

\newtheorem{remark}{Remark}

 \def \Diag {\mathrm{Diag}}
\def \det {\mathrm{det}}

\def \vec{\mathrm{vec}}

\title{Nonconvex High-Dimensional Time-Varying Coefficient Estimation for Noisy High-Frequency Observations with a Factor Structure}
\author{Minseok Shin$^a$ and Donggyu Kim$^b$\footnote{Corresponding author. E-mail address: donggyu.kim@ucr.edu.} \\
$^a$Pohang University of Science and Technology (POSTECH) \\
$^b$University of California, Riverside\\
}

\begin{document}
\maketitle
\begin{spacing}{1.85}

\begin{abstract}
In this paper, we propose a novel high-dimensional time-varying coefficient estimator for noisy high-frequency observations with a factor structure.
In high-frequency finance, we often observe that noises dominate the signal of underlying true processes and that covariates exhibit a factor structure due to their strong dependence.
Thus, we cannot apply usual regression procedures to analyze high-frequency observations. 
To handle the noises, we first employ a smoothing method for the observed dependent and covariate processes.
Then, to handle the strong dependence of the covariate processes, we apply Principal Component Analysis (PCA) and transform the highly correlated covariate structure into a weakly correlated structure.
However, the variables from PCA still contain non-negligible noises.
To manage these non-negligible noises and the high dimensionality, we propose a nonconvex penalized regression method for each local coefficient.
This method produces consistent but biased local coefficient estimators.
To estimate the integrated coefficients, we propose a debiasing scheme and obtain a debiased integrated coefficient estimator using debiased local coefficient estimators.
Then, to further account for the sparsity structure of the coefficients, we apply a thresholding scheme to the debiased integrated coefficient estimator.
We call this scheme the Factor Adjusted Thresholded dEbiased Nonconvex LASSO (FATEN-LASSO) estimator.
Furthermore, this paper establishes the concentration properties of the FATEN-LASSO estimator and discusses a nonconvex optimization algorithm. 
\end{abstract}

\noindent \textbf{Keywords:}  debias, diffusion process,  PCA,  LASSO,    smoothing, sparsity.

\section{Introduction} \label{SEC-1}
Regression models are widely used in statistical analysis.
With the wide availability of high-frequency data,  increasing attention has been paid to high-frequency regression.
The framework of high-frequency regression enables us to accommodate the time variation in the coefficient process, which is often observed in financial practice \citep{ferson1999conditioning, kalnina2023inference, reiss2015nonparametric}.
Thus, various statistical methods have been developed to analyze high-frequency regression.
For example, \citet{barndorff2004econometric, andersen2005framework} proposed a realized coefficient estimator, which was constructed using the ratio of realized covariance to realized variance.
\citet{mykland2009inference} estimated the integrated coefficient by aggregating the spot coefficients obtained from local blocks.
See also \citet{ait2020high, oh2026robust, reiss2015nonparametric}.
\citet{chen2018inference} suggested the statistical inference for volatility functionals of general It\^o semimartingales.
\citet{andersen2021recalcitrant} proposed the measure for market beta dispersion and studied the intra-day variation in market betas.
These models and estimation methods perform well under the assumption that the number of factors is finite.
Recently, \citet{chen2023realized} proposed the high-dimensional market beta estimation procedure with a large number of dependent variables, where the number of common factors diverges slowly as the number of dependent variables goes to infinity.

\begin{figure}[!ht]
\centering
\includegraphics[width = 1\textwidth]{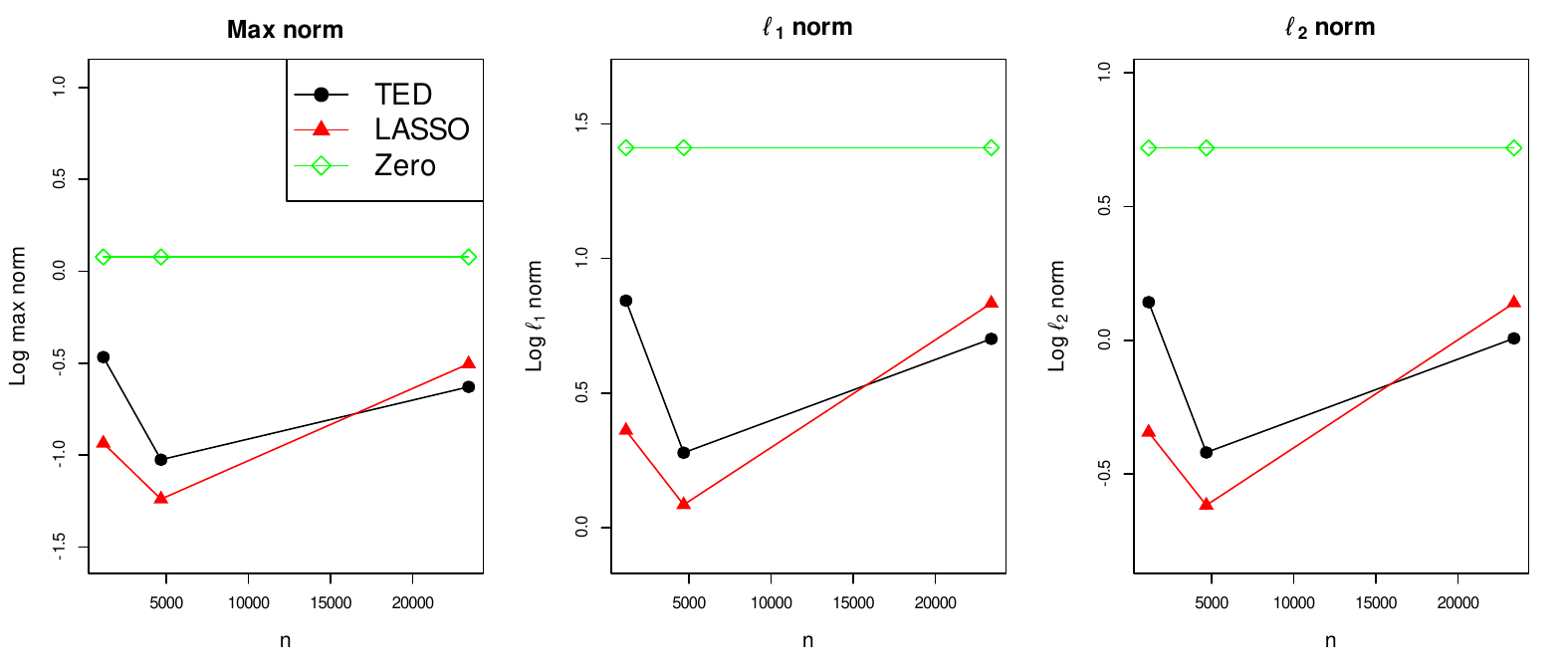}
\caption{The log max, $\ell_1$, and $\ell_2$ norm error plots of the TED (black dot), LASSO (red triangle), and Zero (green diamond) estimators for sample sizes $n=1170, 4680, 23400$.}\label{Beta_simulation_intro}
\end{figure}

However, in finance, we often encounter a large number of factor candidates \citep{bali2011maxing, cochrane2011presidential, harvey2016and, hou2020replicating, mclean2016does}.
This causes the curse of dimensionality, and the estimation methods designed for the finite dimension cannot consistently estimate the coefficients.
To overcome the curse of dimensionality,  LASSO \citep{tibshirani1996regression}, SCAD \citep{fan2001variable}, and the Dantzig selector \citep{candes2007dantzig} are often employed under the sparsity assumption on the model parameters.
\citet{loh2012high} introduced the nonconvex high-dimensional regression to handle noisy and missing variables.
However, these estimation methods cannot account for the time-varying property of coefficient processes.
Recently, to handle both the curse of dimensionality and the time-varying feature of the coefficient process, \citet{kim2026high} proposed a Thresholded dEbiased Dantzig (TED) estimator under the sparsity assumption on the coefficient process.
They first employed a time-localized Dantzig selector \citep{candes2007dantzig}  to estimate the instantaneous coefficient and then applied debiasing and truncation schemes to estimate the integrated coefficient.
However, the TED estimator cannot handle the microstructure noise of high-frequency data, since the noises and regression variables have an unbalanced order relationship.
For example, Figure \ref{Beta_simulation_intro} plots the log max, $\ell_1$, and $\ell_2$ norm errors of the TED, LASSO, and Zero estimators for estimating integrated coefficients with sample sizes $n=1170, 4680, 23400$, where the dependent and covariate processes are contaminated by microstructure noises.
The Zero estimator estimates the coefficients as zero.
The detailed simulation setting is described in Section \ref{SEC-4} and Appendix \ref{simulation-setup}, while the rank  of the covariate process is set to $0$ in this example.
{As seen in Figure \ref{Beta_simulation_intro}, the TED and LASSO estimators fail to estimate the integrated coefficients consistently.
For example, both estimators show the smallest errors at $n=4680$, while their errors increase when $n=23400$. 
This may be because, as the sample size $n$ increases, the effect of noise increases and can dominate the signal of the coefficients.}
Thus, handling  microstructure noise is essential when estimating  high-dimensional time-varying coefficient processes.
On the other hand, we often observe that the financial returns have strong comovements due to the factor structure.
When the covariates are highly correlated, the existing methods, such as LASSO, SCAD, and the Dantzig selector, fail to consistently estimate the coefficients \citep{barigozzi2024fnets, fan2020factor, kneip2011factor}.
That is, the direct application of the usual regression procedures cannot guarantee consistency.
These findings lead to the demand for developing an estimation method that can simultaneously handle the high dimensionality and time variation in the coefficient process, the microstructure noise of high-frequency data, and the factor structure in the covariate process.

In this paper, we develop a novel high-dimensional integrated coefficient estimator   with factor-based regression jump diffusion  processes contaminated by microstructure noises.
To handle the high dimensionality and time variation in the coefficient process, we impose a sparse structure on the coefficient process and assume that the coefficient process follows a diffusion process.
To accommodate the highly correlated structure of the covariate process, we impose an approximate factor structure \citep{bai2003inferential, fan2013large}, where the factor loading matrix process follows a diffusion process.
Due to the time-varying property of the coefficient process, we first estimate the instantaneous coefficients.
Specifically, since noises dominate the signals of the regression variables, we smooth the observed dependent and covariate processes using a kernel function.    
Then, we apply the Principal Component Analysis (PCA) \citep{ait2017using, bai2003inferential, dai2019knowing, fan2013large, fan2020factor} on the smoothed covariate process to separate the latent factor part from the idiosyncratic part. 
This procedure transforms the highly dependent covariates into weakly dependent ones.
Then, we perform a local regression procedure using the smoothed variables.
Due to the noises, the direct application of the LASSO procedure to the smoothed variables cannot guarantee that the derivative of the empirical loss function with the true parameter  goes to zero as the sample size goes to infinity, which is called the deviation condition. 
The deviation condition is essential to obtaining the consistency of the LASSO-type estimator. 
Thus, we adjust the bias in a loss function using the noise covariance matrix estimator and employ $\ell_1$-regularization to accommodate the sparsity of the coefficient process.
Due to the bias adjustment, it becomes a nonconvex optimization problem.
We demonstrate that the resulting instantaneous coefficient estimator achieves the sharp convergence rate.
However, the instantaneous coefficient estimator has another type of bias coming from the $\ell_1$-regularization.
To handle this bias, we employ a debiasing scheme and estimate the integrated coefficient using debiased instantaneous coefficient estimators.
However, the debiasing scheme causes non-sparsity of the integrated coefficient estimates.
To accommodate sparsity, the integrated coefficient estimator is further regularized. 
We call it the Factor Adjusted Thresholded dEbiased Nonconvex LASSO (FATEN-LASSO) estimator.
We show that the FATEN-LASSO estimator has a sharp convergence rate. 
Finally, to implement nonconvex optimization, we adopt the composite gradient descent method  \citep{agarwal2012fast} and investigate its properties.

The rest of the paper is organized as follows.
Section \ref{SEC-2}  introduces the high-dimensional factor-based regression jump diffusion process.
Section \ref{SEC-3} proposes the FATEN-LASSO estimator and establishes its concentration properties.
In Section \ref{SEC-4}, we conduct a simulation study to check the finite sample performance of the proposed FATEN-LASSO estimation procedure. 
In Section \ref{SEC-5}, we apply the proposed estimation procedure to high-frequency financial data.
The conclusion is presented in Section \ref{SEC-6}, and we provide technical proofs and miscellaneous materials in the Appendix.

\setcounter{equation}{0}
\section{The model setup} \label{SEC-2}
We first fix some notations.
For any  $p_1$ by $ p_2$ matrix $\bG = \left(G_{ij}\right)$, define
 \begin{equation*}
  	 \|\bG\|_1 = \max\limits_{1 \leq j \leq p_2}\sum\limits_{i = 1}^{p_1}|G_{ij}|, \quad \|\bG\|_\infty = \max\limits_{1 \leq i \leq p_1}\sum\limits_{j = 1}^{p_2}|G_{ij}|, \quad \text{ and } \quad \| \bG \| _{\max} = \max_{i,j} | G_{ij}|.
 \end{equation*}
We denote the Frobenius norm by $\|\bG\|_F = \sqrt{ \mathrm{tr}(\bG^{\top} \bG)}$, and the matrix spectral norm $\|\bG\|_2$ is denoted by the square root of the largest eigenvalue of $\bG\bG^\top$.
The vectorization of the matrix $\bG$, $\vec(\bG)$, is the column vector obtained by stacking the columns of $\bG$.
In addition, $\det(\bG)$ is   the determinant of $\bG$.
For any vector $\bx \in \mathbb{R}^{p_3}$, $\Diag\(\bx\)$ denotes a $p_3$ by $p_3$ diagonal matrix with the elements of $\bx$ on the main diagonal.
For any process $f(t)$ and $\Delta_n=1/n$, we define $\Delta_i ^n f = f(i \Delta_n) - f((i-1) \Delta_{n})$ for $1 \leq i \leq n$.
The sign function is defined as 
\begin{equation*}
\text{sign}(x) =
\begin{cases} 
-1 & \text{if } x < 0; \\
0 & \text{if } x = 0; \\
1 & \text{if } x > 0.
\end{cases}
\end{equation*}
We use the subscript 0 to represent the true parameters.
We  use $C$'s to denote generic positive constants whose values are free of $n$ and $p$ and may vary from appearance to appearance.

Let $Y(t)$ and $\bX(t)=\(X_1(t), \ldots, X_p(t)\)^{\top}$  be the true dependent process and the vector of the true $p$-dimensional  covariate process, respectively.
We consider the  regression diffusion model as follows:
\begin{eqnarray} \label{model-1}
&& dY(t)=   dY^c(t) + dY^J(t), \cr
&& dY^c(t)=\bbeta ^{\top}(t)  d\bX^c(t)+d Z(t), \quad \text{and} \quad dY^J(t) = J^Y(t) d \Lambda^Y(t),  
\end{eqnarray}
where $Y^c(t)$ and $Y^J(t)$ denote the continuous part and jump part of $Y(t)$, respectively, $J^Y(t)$ is a jump size process, $\Lambda^Y(t)$ is a Poisson process with a bounded intensity, $\bX^c(t)=\(X^c_1(t), \ldots, X^c_p(t)\)^{\top}$ is the continuous part of $\bX(t)$, $\bbeta(t)=\(\beta_1(t), \ldots, \beta_p(t)\)^{\top}$ is a coefficient process, and $Z(t)$ is a residual process.
The superscripts $c$ and $J$ represent the continuous and jump parts of the process, respectively.
The true covariate process $\bX(t)$ satisfies the following factor-based jump diffusion model:
\begin{eqnarray}\label{model-2}
	&&d\bX(t)=  d \bX^c(t) + d \bX^J(t),   \quad  d\bX^c(t)= \bmu(t) dt+ \bB(t)d\bff(t)+d\bu(t), \cr
	&&  d\bff(t) = \bnu_{f}(t) d\bW_f(t), \quad d\bu(t) = \bnu_{u}(t) d\bW_{u}(t), \quad \text{and} \quad d\bX^J(t) =  \bJ(t) d \bLambda(t), 
\end{eqnarray}
where  $\bX^J(t)$ denotes the jump part of $\bX(t)$, $\bJ(t)=\(J_1(t), \ldots, J_p(t)\)^{\top}$ is a jump process, and $\bLambda(t)=\(\Lambda_1(t), \ldots, \Lambda_p(t)\)^{\top}$ denotes a $p$-dimensional Poisson process with the bounded intensity processes. 
We note that $\bLambda(t)$ is a vector of $p$ individual Poisson processes and these  individual  processes are allowed to be dependent.
Additionally, $\bmu(t)$ is a $p$-dimensional drift process, $\bB(t)=\(B_{ij}(t)\)_{1\leq i \leq p, 1 \leq j \leq r}$ is a factor loading matrix process, $\bff(t)=\(f_1(t), \ldots, f_r(t)\)^{\top}$ is a latent factor process, $\bu(t)=\(u_1(t), \ldots, u_p(t)\)^{\top}$ is an  idiosyncratic process, $\bnu_{f}(t)$ and  $\bnu_{u}(t)$ are  $r$ by $q_1$ and $p$ by $q_2$ instantaneous volatility matrices, respectively, and $\bW_f(t)$ and $\bW_{u}(t)$ are $q_1$-dimensional and $q_2$-dimensional independent Brownian motions, respectively.
The residual process $Z(t)$ satisfies
\begin{equation}\label{model-3}
 dZ(t)= \nu_{z}(t) d W_{z}(t),
\end{equation}
where $\nu_{z}(t)$ is a one-dimensional instantaneous volatility process and $d W_{z}(t)$ is a one-dimensional independent Brownian motion.
The processes $\bmu(t)$, $\bnu_{f}(t)$, $\bnu_{u}(t)$, and $\nu_{z}(t)$ are predictable.
The coefficient process $\bbeta(t)$ and factor loading matrix process $\bB(t)$  satisfy the following diffusion models:
\begin{equation*}
	d \bbeta(t)= \bmu_{\beta}(t) dt + \bnu_{\beta}(t) d \bW_{\beta}(t) \quad \text{and} \quad d \vec\(\bB(t)\)= \bmu_{B}(t) dt + \bnu_{B}(t) d \bW_{B}(t), 
\end{equation*}
where $\bmu_{\beta}(t)$ and $\bmu_{B}(t)$ are $p$-dimensional and $pr$-dimensional drift processes, respectively, $\bnu_{\beta}(t)$ and $\bnu_{B}(t)$ are  $p$ by $q_3$ and $pr$ by $q_4$  instantaneous volatility matrix processes, respectively, $\bmu_{\beta}(t)$, $\bmu_{B}(t)$, $\bnu_{\beta}(t)$, and $\bnu_{B}(t)$ are predictable, and $\bW_{\beta}(t)$  and $\bW_{B}(t)$ are $q_3$-dimensional and $q_4$-dimensional independent Brownian motions, respectively.
In this paper, the parameter of interest is the following integrated coefficient:
\begin{equation*}	
	I \beta = (I \beta_j)_{j=1,\ldots, p}= \int_{0}^1 \bbeta(t) dt. 
\end{equation*}
{
The integrated coefficient can be viewed as the average of the spot coefficients over the period. 
Thus, $I\beta$ summarizes the overall effect of covariate movements on the dependent process over the given period.
}
In financial practices, there exist hundreds of potential factor candidates \citep{bali2011maxing, campbell2008search, cochrane2011presidential, harvey2016and, hou2020replicating, mclean2016does}. 
To accommodate the large set of factor candidates, we assume that the dimension of the covariate process, $p$, is large, which causes the curse of dimensionality.
To handle this issue, we impose the following {general} sparsity condition:
{
\begin{equation}\label{sparsity_beta}
	\sup_{0 \leq t \leq 1}\sum_{i=1}^p    |\beta_{i}(t) | ^{\delta}  \leq s_p  \quad \text{and} \quad 	\sum_{i=1}^p|I  \beta_{i} | ^{\delta}  \leq s_p \, \text{ a.s.},
\end{equation}
where $\delta \in [0,1)$, $s_p$ is allowed to diverge slowly in $p$, and $0^0$ is defined as 0. 
The boundary case $\delta=0$ corresponds to the exact sparsity condition, where only a small number of factors can affect the dependent process.
Since the beta process follows an It\^{o} diffusion process, the boundedness in \eqref{sparsity_beta} generally holds with high probability.
This implies that even without the almost sure sparsity condition, we can obtain the same results in this paper with high probability.
However, for simplicity, we assume the general sparsity condition \eqref{sparsity_beta}.
We note that, in finance applications, a large number of factors have often been found to explain asset returns \citep{campbell2008search, cochrane2011presidential, harvey2016and, hou2020replicating, jensen2023there, mclean2016does}. 
Under the general sparsity condition \eqref{sparsity_beta}, many coefficients are allowed to be nonzero, as long as their overall magnitude is sufficiently controlled, which is able to explain a large number of factors partially. 
}

Unfortunately, we cannot observe the true processes $\bX(t)$ and $Y(t)$, since the high-frequency data are contaminated by microstructure noises.
These noises result from market inefficiencies, such as the bid--ask spread, the rounding effect, and asymmetric information.
To account for this feature, we assume that the observed processes satisfy
\begin{equation}\label{noise}
 Y^{o}(t_{i}) = Y(t_{i}) + \epsilon^{Y}(t_{i}) \quad  \text{and}  \quad   \bX^{o}(t_{i}) = \bX(t_{i}) + \bepsilon^{X}(t_{i}) \quad  \text{ for }  i = 0, \ldots, n,
\end{equation}
where $t_{i} \in [0, 1]$ is the $i$th observation time point, $Y^{o}(t_{i})$ is the observed dependent process for time $t_i$, $\bX^o(t_{i})=\(X_1^o(t_{i}), \ldots, X_p^o(t_i)\)^{\top}$ is the observed covariate process for time $t_i$, and $\epsilon^{Y}(t_{i})$ and $\bepsilon^{X}(t_{i})=\(\epsilon^X_{1}(t_{i}), \ldots, \epsilon^X_{p}(t_{i})\)^{\top}$ are independent one-dimensional and $p$-dimensional  microstructure noises for $Y(t_i)$ and $\bX(t_i)$, respectively.
The noises are independent over time and have a mean of zero and variances of $\E\{\epsilon^{Y}(t_{i}) \} ^2=V^Y$ and  $\E\left\{  \bepsilon^{X}(t_{i})  \(\bepsilon^{X}(t_{i})\)^{\top}  \right\}=\bV^X$, where $\bV^X = \left( V^X_{jj'} \right)_{1 \leq j,j^{'} \leq p}$.
For simplicity, we assume that the observation time points are synchronized and equally spaced: $t_{i}-t_{i-1} = 1/n $ for $i=1, \ldots, n$.

\begin{remark}
It is  more realistic to consider  dependent microstructure noises and asynchronous observation time points. 
In fact, we can relax the conditions for the observation time points by employing the generalized sampling time \citep{ait2010high}, refresh time \citep{barndorff2011multivariate}, and previous tick \citep{zhang2011estimating} schemes. See also \citet{chen2023realized, fan2019structured}.
Then, the above condition can be extended to the non-synchronized and unequally spaced condition.
On the other hand, various studies, including \citet{chen2023realized, oh2026robust}, have developed  estimation methods that can handle both the time variation of the coefficient process and the dependence structure of microstructure noises.
For example, \citet{oh2026robust} addresses price-dependent and autocorrelated microstructure noise with a time-varying coefficient process.
In our setting, we can handle the dependent microstructure noise by employing the bias adjustment scheme proposed by \citet{oh2026robust} when estimating the covariance, $\bV^{X}$, of the microstructure noise.
That is, the dependent microstructure noise can be handled by introducing a robust  covariance estimator for the dependent structure of the microstructure noise. 
In this paper, to focus on developing an  integrated coefficient estimation method for dependent covariate processes, we assume the synchronized and equally spaced observation time and  independent microstructure noise conditions for simplicity.
\end{remark}

\section{Nonconvex high-dimensional high-frequency regression}\label{SEC-3}
\subsection{Integrated coefficient estimation procedure} \label{SEC-Estimation}
In this section, we propose a high-dimensional integrated coefficient estimation procedure in the presence of microstructure noises, factor structure in covariate processes, and jumps.
Recently, \citet{kim2026high} developed an integrated coefficient estimator that can handle the high dimensionality and time variation in the coefficient process without microstructure noises.
However, in practice, when employing higher-frequency observations, microstructure noises and strong dependence tend to be observed.  
To accommodate the noises,  we impose the noise structure as in \eqref{noise}.
To handle the strong dependence, we employ the approximate factor structure as in \eqref{model-2}.  
Based on the noisy and strongly correlated high-frequency observation structure, we propose an integrated coefficient estimation procedure as follows.
\subsubsection{{Step 1: Smoothed variables, jump truncation, and localization}}
Due to the time variation in the coefficient process, we first need to estimate the instantaneous coefficients.
To handle the high dimensionality of instantaneous coefficients, we usually employ a penalized regression method for the observed log-returns, $\Delta_i^n Y^o$ and $\Delta_i^n \bX^o$ \citep{kim2026high, shin2025robust}.
However, in the presence of noises, the noises dominate the signals of true log-returns.
This relationship ruins the regression structure in \eqref{model-1}.
To overcome this, we first construct smoothed variables for the observed processes.
Specifically, we consider the following:
\begin{equation*}
 \Delta_i^n\hat{Y} = \sum^{k_1-1}_{l=0}g\(\dfrac{l}{k_1}\)\Delta_{i+l+1} ^n Y^{o} \quad \text{and} \quad  \Delta_i^n \hat{X}_{j} = \sum^{k_1-1}_{l=0}g\(\dfrac{l}{k_1}\)\Delta_{i+l+1} ^n X^{o}_{j},
\end{equation*}
where the kernel function $g(x)$ is Lipschitz continuous and satisfies $g(0) = g(1) = 0$ and $\int_0^1\{g(t)\}^2dt$ \\  $> 0$, and $k_1$ is the bandwidth parameter for $g(x)$.
{
After smoothing, the continuous part is of order $\sqrt{k_1/n}$ caused by the local discretization, whereas the microstructure noise part is of order $1/\sqrt{k_1}$.
If $k_1$ is chosen too small, then the noise term dominates, so the smoothed variables remain too noisy and the resulting estimation error is driven by microstructure noise.
If $k_1$ is chosen too large, then the continuous term becomes too large and the overall error is driven by the continuous part.
Hence, in either case, the overall convergence rate is determined by the larger of the two terms and is therefore not optimal.
Choosing $k_1 = c_{k_1}n^{1/2}$ for some constant $c_{k_1}$ balances these two components, and this balance yields the optimal convergence rate \citep{christensen2010pre, fan2018robust, jacod2009microstructure, kim2016asymptotic}.
Then, we construct the localized variables as follows:}
\begin{equation*}
\mathcal{Y}_i = 
\begin{pmatrix}
\Delta_i^n\hat{Y}^{\text{trunc}} \\
\Delta_{i+1}^n\hat{Y}^{\text{trunc}} \\
 \vdots \\ 
 \Delta_{i+k_2-k_1}^n\hat{Y}^{\text{trunc}}
\end{pmatrix}
\quad \text{ and } \quad
\mathcal{X}_i = 
\begin{pmatrix}
\left(\Delta_i^n\hat{\bX}^{\text{trunc}}\right)^{\top} \\ 
\left(\Delta_{i+1}^n\hat{\bX}^{\text{trunc}}\right)^{\top}  \\  
 \vdots \\ 
\left(\Delta_{i+k_2-k_1}^n\hat{\bX}^{\text{trunc}}\right)^{\top}
\end{pmatrix},    
\end{equation*}
where $k_2$ is the number of observed log-returns used in each local window,
\begin{eqnarray*}
&& \Delta_i^n\hat{Y}^{\text{trunc}}=\Delta_i^n\hat{Y} \, \1 \(| \Delta_i^n\hat{Y}| \leq w_n \),
\quad
\Delta_i^n\hat{\bX}^{\text{trunc}}= \left( \Delta_i^n\hat{X}_j^{\text{trunc}}\right)_{j=1, \ldots, p}, \cr
&&  \Delta_i^n\hat{X}_j^{\text{trunc}} = \Delta_i^n\hat{X}_j \, \1 \(| \Delta_i^n\hat{X}_j| \leq v_{j,n} \),
\end{eqnarray*}
$\1\(\cdot\)$ is an indicator function, and $w_n$ and $v_{j,n}$, $j=1, \ldots, p$, are the truncation parameters to handle the jumps.
We choose $k_2 = c_{k_2}n^{3/4}$ for some constant $c_{k_2}$.
In addition, we choose 
\begin{equation}\label{jump_trunc1}
w_n = C_w s_p \sqrt{\log p} n^{-1/4} \quad \text{ and } \quad v_{j,n}=C_{j,v} \sqrt{\log p} n^{-1/4}
\end{equation}
for some large constants  $C_w$ and $C_{j,v}$, and $j=1, \ldots, p$.

{
\begin{remark}
We note that the choice of $k_2 = c_{k_2}n^{3/4}$ balances two types of errors in the local estimation step.
If $k_2$ is chosen too large, the local window covers a longer time span, and the local constant approximation of the underlying continuous time processes becomes less accurate.
In this case, the discretization error may dominate the convergence rate of the instantaneous coefficient estimator.
On the other hand, if $k_2$ is chosen too small, the local window contains a smaller number of non-overlapping smoothed variables.
That is, the effective sample size in the local estimation step becomes too small, and the local estimation error becomes large.
Thus, with $k_1=c_{k_1} n^{1/2}$, the choice of $k_2 = c_{k_2}n^{3/4}$ balances the discretization error and the local estimation error.
\end{remark}
}

{
\begin{remark}
The truncation parameters $w_n$ and $v_{j,n}$ are used to detect and truncate jumps after the smoothing step.
We do not truncate each observed log-return separately, because at the raw high-frequency level, the microstructure noise dominates the continuous return, which makes it difficult to recover the information in the continuous return.
After smoothing, however, the continuous signal becomes more stable, so jump detection becomes more effective.
For this reason, we apply the truncation step to the smoothed variables rather than to the raw returns.
The truncation levels in \eqref{jump_trunc1} are chosen to be large enough to detect jumps with high probability, while still being sharp enough to establish the restricted eigenvalue condition.
A more detailed discussion of these requirements is provided in Appendix \ref{jump-discuss}.
\end{remark}
}

\subsubsection{{Step 2: Factor-based decorrelation}}
Due to the highly correlated structure of the true covariate process $\bX(t)$, it is difficult to directly apply commonly used model selection methods, such as LASSO, SCAD, and the Dantzig selector \citep{barigozzi2024fnets, fan2020factor, fan2024latent, kneip2011factor}.
To address this, we employed a decorrelation step based on the factor  structure  \eqref{model-2}.
Specifically, using PCA \citep{ait2017using, dai2019knowing, fan2020factor, fan2013large}, we first estimate the factor loading matrix $\bB(i\Delta_n)$ and smoothed latent factor variable 
\begin{equation*}
\bF_i = 
\begin{pmatrix}
\sum^{k_1-1}_{l=0}g\(\dfrac{l}{k_1}\)\Delta_{i+l+1} ^n \bff^{\top}  \\ 
\sum^{k_1-1}_{l=0}g\(\dfrac{l}{k_1}\)\Delta_{i+l+2} ^n \bff^{\top}   \\  
 \vdots \\ 
\sum^{k_1-1}_{l=0}g\(\dfrac{l}{k_1}\)\Delta_{i+l+k_2-k_1+1} ^n \bff^{\top} 
\end{pmatrix}
\end{equation*}
as follows:
\begin{equation}\label{LSM_factor}
(\hat{\bB}_{i \Delta_n}, \hat{\bF}_i)=\arg \min_{\bB \in \mathbb{R}^{p \times r}, \bF \in \mathbb{R}^{(k_2 - k_1 +1) \times r}} \|\mathcal{X}_i-\bF \bB^{\top} \|_{F}^{2},
\end{equation}
subject to 
\begin{equation*}
p^{-1} \bB^{\top} \bB = \bI_r \quad \text{and} \quad \bF^{\top} \bF \text{ is an $r \times r$ diagonal matrix},
\end{equation*}
where $\bI_{r}$ is the $r$-dimensional identity matrix.
The above constraint is imposed to handle the identification problem for the latent factor and factor loading matrix.
The identifiability assumption is described in Assumption \ref{assumption1}(d).
Then, we estimate the smoothed idiosyncratic variable 
\begin{equation*}
\bU_{i} = 
\begin{pmatrix}
\sum^{k_1-1}_{l=0}g\(\dfrac{l}{k_1}\)\Delta_{i+l+1} ^n \bu^{\top}  \\ 
\sum^{k_1-1}_{l=0}g\(\dfrac{l}{k_1}\)\Delta_{i+l+2} ^n \bu^{\top}   \\  
 \vdots \\ 
\sum^{k_1-1}_{l=0}g\(\dfrac{l}{k_1}\)\Delta_{i+l+k_2-k_1+1} ^n \bu^{\top} 
\end{pmatrix}
\end{equation*}
by 
\begin{equation}\label{est_idio}
\hat{\bU}_i = \mathcal{X}_i-\hat{\bF}_i \hat{\bB}^{\top}_{i \Delta_n}
\end{equation}
and define 
\begin{equation*}
\hat{\bG}_i=\(\hat{\bU}_i, \hat{\bF}_i\).
\end{equation*}
We note that $\hat{\bB}_{i \Delta_n}$ and $\hat{\bF}_i$ estimate  $\bB\(i\Delta_n\)\bH_i$ and $\bF_i\(\bH_i^{\top}\)^{-1}$, respectively, where the nonsingular $r$ by $r$ matrix $\bH_i$ comes from handling the identification issue (see Proposition \ref{Prop3} in the Appendix).
Let $\bgamma(t)=\bB^{\top}(t)\bbeta(t)$.
From \eqref{model-1} and \eqref{model-2}, we have
\begin{equation*}
dY^c(t)=\bbeta ^{\top}(t)\bmu(t) dt+ \bbeta ^{\top}(t)d\bu(t)+\bgamma^{\top}(t)d\bff(t) + d Z(t).
\end{equation*}
Thus, based on  $\mathcal{Y}_i$ and $\hat{\bG}_i$, we can estimate $\bbeta(i\Delta_n)$ and $\bgamma(i\Delta_n)$ using the weakly correlated structure of $\hat{\bG}_i$.
Specifically, the local regression coefficient for  $\mathcal{Y}_{i}$ and $\hat{\bG}_i$ can be approximated by 
\begin{equation*}
\btheta\(i\Delta_n\)=\(\bbeta^{\top}(i\Delta_n), \bgamma^{\top}(i\Delta_n) \bH_i\)^{\top},
\end{equation*}
where $\btheta\(i\Delta_n\)$ is defined for $i=0, \ldots, n-k_2$.
This implies that we can estimate the instantaneous coefficient by choosing the first $p$ entries after estimating $\btheta\(i\Delta_n\)$.

\subsubsection{{Step 3: Nonconvex instantaneous coefficient estimation}}
We estimate  $\btheta\(i\Delta_n\)$ based on $\mathcal{Y}_{i}$ and  $\hat{\bG}_i$ as follows.
For each local regression, we need to handle the curse of dimensionality. 
To do this, we often utilize a penalized regression method, such as LASSO \citep{tibshirani1996regression} or Dantzig \citep{candes2007dantzig}, under the sparsity assumption.
However, they cannot consistently estimate instantaneous coefficients due to the bias from   microstructure noises in $\hat{\bU}_i$.
For example, the usual LASSO leads to the following instantaneous coefficient estimator at time $i \Delta_n$:
\begin{equation*}  
 \hat{\bbeta}_{i\Delta_n}^{\text{LASSO}}=\(\hat{\theta}_{i\Delta_n,j}^{\text{LASSO}}\)_{j=1,\ldots,p},
\end{equation*}
where
\begin{equation*}
 \hat{\btheta}_{i \Delta_n}^{\text{LASSO}} = \(\hat{\theta}_{i\Delta_n,j}^{\text{LASSO}}\)_{j=1,\ldots,p+r} =  \underset{\btheta \in \mathbb{R}^{p+r}}{\operatorname{arg \, min}} \, \dfrac{n}{k_1 k_2} \left\| \mathcal{Y}_{i}- \hat{\bG}_i \btheta \right\|^2 _{2} + \eta \left\| \btheta \right\| _{1} 
\end{equation*}
and $\eta>0$ is some regularization parameter.
To obtain the consistency of $\hat{\bbeta}_{i\Delta_n}^{\text{LASSO}}$, we need the deviation condition $\dfrac{2n}{k_1 k_2}\left\|\hat{\bG}_i^{\top} \hat{\bG}_i\btheta_{0, i \Delta_n} - \hat{\bG}_i^{\top}\mathcal{Y}_{i} \right\|_{\max} \overset{p}{\to} 0$. 
However, in the presence of noises, this condition cannot be satisfied, since  $\E \(\hat{\bU}_i^{\top}\hat{\bU}_i\)$ contains noise covariance terms for the covariate process $\bX(t)$.
Thus, we need to estimate the noise covariance matrix and adjust the bias.
The noise covariance matrix for the covariate process is estimated by
\begin{equation}\label{ins_variance}
 \hat{\bV}^X = \dfrac{1}{2n} \sum_{i=1}^{n} \Delta_i^n \bX^{\text{trunc}}   \(\Delta_i^n \bX^{\text{trunc}}\)^{\top},
\end{equation}
where 
\begin{equation*}
\Delta_i^n \bX^{\text{trunc}}=\left(\Delta_i^n X_j^{o} \, \1 \(| \Delta_i^n X_j^{o} | \leq v^{(2)}_{j,n} \)\right)_{j=1, \ldots, p}
\end{equation*}
and $v^{(2)}_{j,n}$, $j=1, \ldots, p$, are the truncation parameters to handle the jumps.
We utilize 
\begin{equation}\label{jump_trunc2}
v^{(2)}_{j,n}=C^{(2)}_{j,v} \sqrt{\log p}
\end{equation}
for some large constants $C^{(2)}_{j,v}$, $j=1, \ldots, p$.
{
\begin{remark}
As in \eqref{jump_trunc1}, the truncation parameter $v^{(2)}_{j,n}$ is used to detect and truncate jumps when estimating the noise covariance matrix.
Compared with \eqref{jump_trunc1}, the required order is different because we only need to estimate the noise covariance matrix from the observed log-returns.
The truncation level in \eqref{jump_trunc2} is chosen to be large enough to detect jumps with high probability, while remaining sharp enough to satisfy the restricted eigenvalue condition.
A more detailed discussion is provided in Appendix \ref{jump-discuss}.
\end{remark}
}

Then, the  instantaneous coefficient estimator at time $i\Delta_n$  is defined as follows:
  \begin{equation*}  
 \hat{\bbeta}_{i\Delta_n}=\(\hat{\theta}_{i\Delta_n,j}\)_{j=1,\ldots,p},
\end{equation*}
where
\begin{eqnarray}
&& \hat{\btheta}_{i\Delta_n}=\(\hat{\theta}_{i\Delta_n,j}\)_{j=1,\ldots,p+r} = \underset{\| \btheta \|_{1} \leq \rho}{\operatorname{arg \, min}} \, {\mathcal{L}}_{i}(\btheta) + \eta \left\| \btheta \right\| _{1}, \label{ins_beta} \\
&& {\mathcal{L}}_{i}(\btheta) = \dfrac{n}{2\phi k_1 k_2}\| \mathcal{Y}_{i} - \hat{\bG}_i \btheta \|_2^2 - \dfrac{n \zeta}{2\phi k_1^2} \btheta^{\top} \hat{\bV} \btheta, \quad
\hat{\bV} = 
\begin{pmatrix}
    \begin{array}{c|c}
        \hat{\bV}^X & \mathbf{0}_{p \times r} \\ \hline
        \mathbf{0}_{r \times p} & \mathbf{0}_{r \times r}
    \end{array}
\end{pmatrix}, \label{ins_beta2}
\end{eqnarray}
$\rho$ satisfies $\rho \geq \| \btheta_{0}(i\Delta_n) \|_{1}$, $\eta>0$ is the regularization parameter, ${\mathcal{L}}_{i}(\btheta)$ is the empirical loss function, $\phi = \frac{1}{k_{1}} \sum_{\ell = 0 }^{k_{1}-1} \left \{ g \( \frac{\ell}{k_{1}}\) \right \}^2$, and {$\zeta = k_1 \sum\limits_{l = 0}^{k_{1}-1}\left \{ g\left(\frac{l}{k_{1}}\right) - g\left(\frac{l+1}{k_{1}}\right)\right\} ^2 = O\(1\)$.}
The tuning parameters $\rho$ and $\eta$ will be specified in Theorem \ref{Thm1}.
For the empirical loss function ${\mathcal{L}}_{i}(\btheta)$ in \eqref{ins_beta2}, the deviation condition $\left\| \nabla {\mathcal{L}}_{i}(\btheta_{0}(i \Delta_n)) \right\|_{\infty} = \dfrac{n}{\phi k_1 k_2}\left\|\hat{\bG}_i^{\top} \hat{\bG}_i\btheta_{0}(i \Delta_n) - \hat{\bG}_i^{\top}\mathcal{Y}_{i} -\dfrac{k_2\zeta}{k_1}  \hat{\bV}\btheta_{0}(i \Delta_n) \right\|_{\max} \overset{p}{\to} 0$ is satisfied since the noise part in  $\hat{\bU}_i^{\top}\hat{\bU}_i$ is adjusted by the noise covariance estimator $\hat{\bV}^X$ (see Proposition \ref{Prop4} in the Appendix). 
Additionally, the Hessian matrix of the empirical loss function has the following structure:
\begin{equation*}
	\nabla^2 {\mathcal{L}}_{i}(\btheta) = \begin{pmatrix}
    \begin{array}{c|c}
        \dfrac{n}{\phi k_1 k_2} \hat{\bU}_i^{\top} \hat{\bU}_i -  \dfrac{n \zeta}{\phi k_1^2} \hat{\bV}^X  & \dfrac{n}{\phi k_1 k_2} \hat{\bU}_i^{\top} \hat{\bF}_i \\ \hline
       \dfrac{n}{\phi k_1 k_2} \hat{\bF}_i^{\top} \hat{\bU}_i & \dfrac{n}{\phi k_1 k_2} \hat{\bF}_i^{\top} \hat{\bF}_i
    \end{array}
\end{pmatrix}.
\end{equation*}
We note that the upper left term has the same form as the  pre-averaging realized volatility (PRV) \citep{christensen2010pre, jacod2009microstructure}, which can be one of the estimators for $\bSigma_u(t)=\bnu_{u}(t)\bnu_{u}^{\top}(t)$.
However, we cannot guarantee that $\nabla^2 {\mathcal{L}}_{i}(\btheta)$ is positive semidefinite due to the bias adjustment, which implies that  the objective function ${\mathcal{L}}_{i}(\btheta) + \eta \left\| \btheta \right\| _{1}$ can be unbounded from below.
To handle the unbounded problem, we impose a constraint on $\btheta$, such as $\| \btheta \|_{1} \leq \rho$,  for the nonconvex optimization problem \eqref{ins_beta}.
Theorem \ref{Thm1} shows that the instantaneous coefficient estimator $\hat{\bbeta}_{i \Delta_n}$ is consistent when we choose appropriate values for $\rho$ and $\eta$.
\subsubsection{{Step 4: Debiasing via CLIME}}
To estimate the integrated coefficient, we can consider the integration of $\hat{\bbeta}_{i \Delta_n}$. 
However, $\hat{\bbeta}_{i \Delta_n}$ is biased due to the regularization, so their integration fails to enjoy the law of large numbers property.
In other words, while the integration has the same convergence rate as  $\hat{\bbeta}_{i \Delta_n}$, the convergence rate is not fast enough.
To obtain a faster convergence rate, we apply the debiasing scheme to each value of $\hat{\bbeta}_{i \Delta_n}$ as follows.
We first estimate the inverse instantaneous idiosyncratic volatility matrix $\bOmega(i \Delta_n)=\bSigma_{u}^{-1}(i \Delta_n)$ based on the following constrained $\ell_1$-minimization for inverse matrix estimation (CLIME) \citep{cai2011constrained}:
\begin{equation} \label{CLIME}
	\hat{\bOmega}_{i \Delta_n} = \arg \min \| \bOmega\|_{1} \quad \text{s.t.} \quad  \left\|    \left(\dfrac{n}{\phi k_1 k_2} \hat{\bU}_i^{\top} \hat{\bU}_i - \dfrac{n \zeta}{\phi k_1^2}\hat{\bV}^X \right)\bOmega  - \bI  \right\|_{\max} \leq \tau, 
\end{equation}
where  $\tau$ is the tuning parameter that will be specified in Theorem \ref{Thm2}. 
Using the inverse instantaneous volatility matrix estimator $\hat{\bOmega}_{i \Delta_n}$, we adjust the instantaneous coefficient estimator $\hat{\bbeta}_{i \Delta_n}$ as follows:
\begin{equation}\label{debias}
	\tilde{ \bbeta}_{i \Delta_n} = \hat{\bbeta}_{i \Delta_n}  +  \dfrac{n}{\phi k_1 k_2}  \hat{\bOmega}_{i \Delta_n}^{\top}  \left\{  \hat{\bU}_i^{\top} \mathcal{Y}_{i} - \left(  \hat{\bU}_i^{\top} \mathcal{X}_i - \dfrac{k_2 \zeta}{k_1} \hat{\bV}^X \right) \hat{\bbeta}_{i \Delta_n} \right\}.
\end{equation}
Note that in \eqref{debias}, we used  $\dfrac{n}{\phi k_1 k_2} \hat{\bU}_i^{\top} \mathcal{X}_i- \dfrac{n \zeta}{\phi k_1^2}\hat{\bV}^X $  as the proxy for the instantaneous idiosyncratic volatility matrix  at time $i \Delta_n$, instead of $\dfrac{n}{\phi k_1 k_2} \hat{\bU}_i^{\top} \hat{\bU}_i- \dfrac{n \zeta}{\phi k_1^2}\hat{\bV}^X $.
This is because $\hat{\bU}_i^{\top}\(\mathcal{Y}_{i} - \mathcal{X}_i \hat{\bbeta}_{i \Delta_n}\)$ has fewer error terms than $\hat{\bU}_i^{\top}\(\mathcal{Y}_{i} - \hat{\bU}_i \hat{\bbeta}_{i \Delta_n}\)$.

\subsubsection{{Step 5: Integrated coefficient estimation and thresholding}}
We estimate the integrated coefficient as follows:
\begin{equation}\label{Inte}
	\hat{I \beta}  = \sum_{i=0}^{[1/(k_2 \Delta_n) ]-1}\tilde{\bbeta}_{i k_2 \Delta_n} k_2 \Delta_n.
\end{equation}
The debiased integrated coefficient estimator $\hat{I \beta}$ benefits from the law of large numbers property and has a faster convergence rate than the integration of the instantaneous coefficient estimators.
However, the bias adjustment leads to the non-sparse structure of the integrated coefficient estimator. 
To accommodate the sparse structure of the integrated coefficient, we employ the thresholding scheme as follows:
\begin{equation}\label{thres}
	\tilde{I\beta}_j= s (\hat{I\beta}_j)  \1 \(  |\hat{I\beta}_j  | \geq h_n  \) \quad \text{and} \quad  \tilde{I \beta} = \( \tilde{I\beta}_j \)_{j=1,\ldots,p},
\end{equation}
where $s(x)$ is the thresholding function  satisfying $|s (x)-x| \leq h_n$ and $h_n$ is a thresholding level that will be specified in Theorem \ref{Thm3}.
For the  thresholding function $s(x)$, we usually employ the soft thresholding function $s(x)= x- \text{sign}(x) h_n$ or hard thresholding function $s(x)=x$.
We utilize the  hard thresholding function in the empirical study.
We call this the Factor Adjusted Thresholded dEbiased Nonconvex LASSO (FATEN-LASSO) estimator. 
A summary of the FATEN-LASSO estimation procedure is presented in Appendix \ref{SEC-materials}.
We will discuss the choice of the tuning parameters in Appendix \ref{SEC-Tuning}.

\begin{remark}
In this paper, we allow the time variation of processes, such as the dependent, covariate, coefficient, and factor loading matrix.
The rank $r$ is assumed to be constant over time.
To allow the time variation of the rank $r$, the state heterogeneous structure \citep{chun2022state} can be considered for the covariate process.
That is, the rank $r$ is the same within the same state, but may change when the state changes.
However, extending this approach from the one-dimensional case, as in \citet{chun2022state}, to the high-dimensional case is not straightforward.
Thus, we leave this issue for a future study.
\end{remark}

 \subsection{Theoretical results}\label{Theory}
In this section, we show the  asymptotic properties of the proposed FATEN-LASSO estimator.
To investigate its asymptotic behaviors, the following assumptions are required.
  \begin{assumption}\label{assumption1}~
  \begin{itemize}

\item [(a)] $\bmu(t)$,  $\bSigma_f(t)=\bnu_{f}(t)\bnu_{f}^{\top}(t)$, $\bSigma_u(t)$, $\nu_z(t)$, $\bbeta(t)$, $\bmu_{\beta}(t)$,  $\bSigma_{\beta}(t) = \bnu_{\beta}(t)\bnu_{\beta}^{\top}(t)$, $\bB(t)$, and $\bSigma_{B}(t) = \bnu_{B}(t)\bnu_{B}^{\top}(t)$ are almost surely entry-wise bounded, and $ \| \bSigma_{u}^{-1}(t)\|_1 \leq C $  \text{ a.s.}

 \item [(b)] The noises $\epsilon_j^{X}(t_i)$, $j=1, \ldots, p$, and $\epsilon^{Y}(t_i)$ are sub-Gaussian with a bounded parameter.

{\item [(c)] For $\delta \in [0, 1)$, the drift process $\bmu_{\beta}(t) = \(\mu_{\beta,1}(t), \ldots, \mu_{\beta,p}(t) \)^{\top}$ and the volatility process $\bSigma_{\beta}(t)=\(\Sigma_{\beta, ij}(t)\)_{i,j=1, \ldots, p}$  satisfy the following sparsity condition: 
 \begin{equation*}
 \sup_{0 \leq t \leq 1}\sum_{i=1}^p    |\mu_{\beta, i}(t) |^{\delta}   \leq s_p   \quad \text{and} \quad  \sup_{0 \leq t \leq 1}\sum_{i=1}^p |\Sigma_{\beta, ii}(t)|^{\delta/2}    \leq s_p  \text{ a.s.}
 \end{equation*}
}

\item [(d)]  For $0 \leq i \leq [1/(k_2 \Delta_n) ]-1$, $\bSigma_f(i k_2 \Delta_n)$ has distinct eigenvalues bounded away from $0$ and $\|p^{-1}\bB^{\top}(i k_2 \Delta_n) \bB(i k_2 \Delta_n) - \bI_{r} \|_{2} \rightarrow 0$ as $p \rightarrow \infty$.
 

 \item[(e)]  
$\sup_{0 \leq t \leq 1} \left\|  \bSigma_{u}(t) \right\|_{1} \leq C$ and  $\left\| \bV^X \right\|_{1} \leq C$.

\item[(f)]  
For $\bb = \(1, \ldots, 1 \)^{\top} \in \mathbb{R}^p$, the random variable  $\( \bb^{\top}\int_{i\Delta_n}^{(i+k_2)\Delta_n} d\bLambda(t) dt \)$ is sub-exponential with a bounded parameter.

 \item [(g)] The volatility matrix processes, $\bSigma_f(t) = (\Sigma_{f,ij}(t) ) _{i,j=1,\ldots, r}$ and  $\bSigma_u(t) = (\Sigma_{u,ij}(t) ) _{i,j=1,\ldots, p}$,  satisfy the following H\"older condition:
  \begin{eqnarray*}
    && \max_{1 \leq i,j \leq r}| \Sigma_{f,ij}(t) -\Sigma_{f,ij}(s) | \leq  C  \sqrt{|t-s| \log p}     \,  \text{ a.s.,} \cr
    && \max_{1 \leq i,j \leq p}| \Sigma_{u,ij}(t) -\Sigma_{u,ij}(s) | \leq  C  \sqrt{|t-s| \log p}     \,  \text{ a.s.}
 \end{eqnarray*}

 \item[(h)] $Cn^{c_1}  \leq p \leq \exp ( n^{c_2})$ for some constants $c_1, c_2>0$, and $n^{-1/8} s_p \(\log p\)^{2} + p^{-1/2} s_p \sqrt{\log p}   \rightarrow 0$ as $n,p \rightarrow \infty$.

 \item [(i)] The inverse  idiosyncratic volatility matrix process $\bSigma_{u}^{-1}(t) = \bOmega(t) = (\omega_{ij}(t))_{i,j=1,\ldots,p}$ satisfies the following sparsity condition:
  \begin{equation*}
  	 \sup_{0 \leq t \leq 1} \max_{ 1 \leq i \leq p} \sum_{j=1}^p |\omega_{ij}(t) | ^{q} \leq s_{\omega, p } \,  \text{ a.s.},
 \end{equation*}
  where $q \in [0, 1)$ and $s_{\omega, p }$ is allowed to diverge slowly in $p$.

 
\end{itemize} 
  \end{assumption}

\begin{remark}\label{remark-5}
Assumption \ref{assumption1}(a) is the boundedness condition that implies the sub-Gaussian tails for the continuous processes, such as the latent factor process $\bff(t)$, the idiosyncratic process $\bu(t)$, and the coefficient process $\bbeta(t)$.
Sub-Gaussian assumptions are frequently used in high-dimensional statistics to obtain probability bounds.
Similarly, in  Assumption \ref{assumption1}(b), we impose sub-Gaussianity for the  noises.
The sub-Gaussianity assumption can be relaxed by incorporating robust estimation schemes, such as truncation or Huber loss–based M-estimators, which mitigate the influence of heavy-tailed observations \citep{fan2018robust, shin2025robust, shin2023adaptive}.
{
Assumption \ref{assumption1}(c) is the sparsity condition on the coefficient process, which is required to handle the discretization error of the instantaneous coefficient estimator $\hat{\bbeta}_{i \Delta_n}$. 
}
Assumption \ref{assumption1}(d) is the identifiability condition, which is often used to estimate the latent factor and factor loading matrix \citep{ait2017using, dai2019knowing}.
Assumption \ref{assumption1}(e) is the sparsity condition on the idiosyncratic volatility matrix and noise covariance matrix, which is required to harness the approximate factor model and to obtain the restricted eigenvalue condition for the LASSO-type estimator.
To obtain the restricted eigenvalue condition, we also need the technical condition Assumption \ref{assumption1}(f).
Assumption  \ref{assumption1}(g)  is the continuity condition, which is required to investigate the asymptotic properties of the estimators for the time-varying processes.
This condition can be obtained with high probability when the volatility processes $\bSigma_f(t)$ and $\bSigma_u(t)$ follow continuous It\^o diffusion models with bounded drift and volatility processes.
Finally, in Assumption  \ref{assumption1}(i), we impose the sparse structure on the inverse idiosyncratic volatility matrix process to investigate the asymptotic behaviors of the CLIME estimator.
\end{remark}

The following theorem establishes the asymptotic behaviors of the instantaneous coefficient estimator $ \hat{\bbeta}_{i \Delta_n}$.

\begin{theorem} \label{Thm1}
Under the models \eqref{model-1}--\eqref{noise} and Assumption \ref{assumption1}(a)--(g), let $k_1 = c_{k_1} n^{1/2}$ and $k_2 = c_{k_2} n^{3/4}$ for some constants $c_{k_1}$ and $c_{k_2}$. 
For any given positive constant $a$, choose $\rho = C_{\rho,a} s_p$ and $\eta=C_{\eta,a} \left\{ n^{-1/8} s_p \(\log p\)^{2} + p^{-1/2} s_p \sqrt{\log p} \right\}$  for some large constants $C_{\rho,a}$ and $C_{\eta,a}$.
Then,  for large $n$, we have
\begin{equation}\label{Thm1-result1}
	\max_{i}  \| \hat{\bbeta}_{i \Delta_n} - \bbeta_{0}(i \Delta_n) \|_{1} \leq  C s_p \eta^{{1-\delta}} \quad \text{and}  \quad \max_{i} \| \hat{\bbeta}_{i \Delta_n} - \bbeta_{0}(i \Delta_n) \|_{2} \leq C \sqrt{s_p}\eta^{{1-\delta/2}},  
\end{equation}
with the probability at least $1-p^{-a}$. 
\end{theorem}

\begin{remark}
{Theorem \ref{Thm1} provides $\ell_1$ and $\ell_2$ norm error bounds for $\hat{\bbeta}_{i \Delta_n}$.
Under the exact sparsity condition for the coefficient process (i.e., $\delta=0$), these bounds are of order $n^{-1/8}+p^{-1/2}$ with the sparsity level and log order terms.}
We note that for each local regression, the number of observed log-returns  is $Cn^{3/4}$, whereas the number of non-overlapping smoothed variables is $Cn^{1/4}$.
Due to the cost of managing the noises, we are able to use only  $Cn^{1/4}$ variables to estimate the instantaneous coefficient.
Thus, the optimal convergence rate is expected to be $n^{-1/8}$, and {the first term in $s_p \eta$, given by $n^{-1/8} s_p^2 \(\log p\)^{2}$, is near-optimal.
The second term, $p^{-1/2}s_p^2 \sqrt{\log p}$}, comes from estimating the latent factor process, and $p^{-1/2}$  is the same rate as in \citet{fan2020factor}.
When $n=O(p^4)$, the proposed instantaneous coefficient estimator can achieve the near-optimal convergence rate.
\end{remark}

As discussed in Section \ref{SEC-Estimation}, the instantaneous coefficient estimators are biased due to the regularization.
Thus, the integration of the instantaneous coefficient estimators cannot benefit from the law of large numbers property.
In other words, while the integration converges, the convergence rate is not fast enough.
To handle this issue, we utilize the debiasing scheme and obtain the debiased  integrated coefficient estimator, as  outlined in \eqref{debias} and \eqref{Inte}.
We establish the asymptotic property of the debiased  integrated coefficient estimator in the following theorem.

\begin{theorem} \label{Thm2}
Under the assumptions in Theorem \ref{Thm1} and Assumption \ref{assumption1}(i), for any given positive constant $a$, choose $\tau = C_{\tau,a}  \left\{ n^{-1/8}\(\log p\)^{2}  + p^{-1/2}\sqrt{\log p} \right\}$ for some constant  $C_{\tau,a}$.
Then, we have, with the probability at least $1-p^{-a}$,
\begin{equation}\label{Thm2-result1}
 \|  \hat{I\beta}  - I\beta_0 \| _{\max}  \leq  C b_n,
\end{equation}
where 
{$b_n= s_p^{2-\delta} \tau^{2-\delta} + s_p s_{\omega, p} \tau^{2-q}.$}
\end{theorem}

\begin{remark}
Theorem \ref{Thm2} provides the max norm error bound for the debiased integrated coefficient estimator.
When {both the coefficient process and the inverse volatility matrix process  satisfy the exact sparsity condition, that is, $\delta=q=0$,} the debiased integrated coefficient estimator has the convergence rate of $s_p \(s_p+s_{\omega, p}\) \left\{n^{-1/4}\(\log p\)^4 + p^{-1} \log p\right\}$.
In contrast, we have the convergence rate of $s_p^2 \left\{n^{-1/8} \(\log p\)^{2} + p^{-1/2} \sqrt{\log p}\right\}$ without a debiasing scheme.
In  high-dimensional statistics, the sparsity level is assumed to diverge relatively slowly, such as $\log p$. 
Thus, the debiased integrated coefficient estimator has the faster convergence rate than the integration of the instantaneous coefficient estimators.
 \end{remark}

Theorem \ref{Thm2} shows that the input-integrated coefficient estimator $\hat{I\beta}$ performs well because of the debiasing scheme.
Finally, to accommodate  the sparse structure, we utilize  the thresholding scheme and obtain the FATEN-LASSO estimator.
The following theorem establishes the $\ell_1$ convergence rate of the FATEN-LASSO estimator.

 \begin{theorem} \label{Thm3}
Under the assumptions in Theorem \ref{Thm2},  for any given positive constant $a$, choose $h_n= C_{h,a} b_n$ for some constant $C_{h,a}$, where $b_n$ is defined in Theorem \ref{Thm2}.
Then, we have, with   probability at least $1-p^{-a}$,
\begin{equation}\label{Thm3-result1}
 \|  \tilde{I\beta}  - I\beta_0 \| _{1}  \leq C   s_p b_n^{{1-\delta}}. 
\end{equation}
\end{theorem}


Theorem \ref{Thm3} shows the  $\ell_1$ norm error bound of the proposed FATEN-LASSO estimator.
When the exact sparsity condition is satisfied, that is, {$\delta=q=0$},  the FATEN-LASSO estimator has the convergence rate of $s_p^2 \(s_p+s_{\omega, p}\) \left\{n^{-1/4}\(\log p\)^4 + p^{-1} \log p\right\}$. 
We note that  in the presence of microstructure noises, $n^{-1/4}$ is  the optimal  convergence rate of the integrated coefficient estimator in the finite-dimensional setup.
Thus, when $n= O(p^4)$, the FATEN-LASSO estimator achieves the optimal convergence rate with up to $\log p$ and sparsity level orders.

\begin{remark}
When the true rank $r$ is 0, the FATEN-LASSO estimator has the optimal convergence rate up to the orders of $\log p$, $s_p$, and $s_{\omega,p}$.
We note that, since we do not need to estimate the latent factor for $r=0$, the $p^{-1}$ term does not appear.
{We also note that the convergence rate with respect to $n$ is mainly affected by the presence of microstructure noise.
For example, under exact sparsity and with microstructure noise, smoothing reduces the effective sample size and yields the $n^{-1/4}$ rate.
Without microstructure noise, the smoothing step is no longer needed, and the corresponding rate can improve to $n^{-1/2}$, up to logarithmic and sparsity-related terms.}
\end{remark}

\subsection{Implementation of the  FATEN-LASSO estimation procedure }

To implement the FATEN-LASSO estimation procedure, we need to solve the nonconvex optimization problem \eqref{ins_beta}.
However,  it is generally hard to obtain the global minimizer of a nonconvex optimization problem in polynomial time.
To handle this issue, we employ  the composite gradient descent method \citep{agarwal2012fast} as follows:
\begin{equation*}  
 \hat{\bbeta}_{i\Delta_n}^{t+1}=\(\hat{\theta}^{t+1}_{i\Delta_n,j}\)_{j=1,\ldots,p},
\end{equation*}
where
\begin{eqnarray}\label{gradient}
\hat{\btheta}_{i \Delta_n}^{t+1} &=&\(\hat{\theta}^{t+1}_{i\Delta_n,j}\)_{j=1,\ldots,p+r} \cr
&=& \underset{\left\|\btheta\right\|_1 \leq \rho}{\operatorname{arg \, min}} \left\{    \mathcal{L}_{i}\(\hat{\btheta}_{i \Delta_n}^{t}\) + \langle \nabla{\mathcal{L}}_{i}(\hat{\btheta}_{i \Delta_n}^{t}), \btheta - \hat{\btheta}_{i \Delta_n}^{t}\rangle +  \alpha_2 \left\|\btheta-\hat{\btheta}_{i \Delta_n}^{t} \right\|_2^2 + \eta \left\| \btheta \right\|_1       \right\},
\end{eqnarray}
$\hat{\btheta}_{i \Delta_n}^{0}$ is the initial parameter, and $\alpha_2 >0 $ is defined in Proposition \ref{Prop5} in the Appendix.
Then, we can obtain the following proposition.

\begin{proposition}\label{Prop1}
Under the assumptions in Theorem \ref{Thm1}, we have, with the probability at least $1-p^{-a}$,
\begin{equation}\label{Prop1-eq1}
 \max_{i} \|\hat{\bbeta}_{i \Delta_n}^{t} - \hat{\bbeta}_{i \Delta_n} \|_2 \leq C \sqrt{s_p} \eta^{{1-\delta/2}}
\end{equation}
{for all iterations $t$ satisfying
\begin{equation*}
t \geq C \left\{ \log \(1+\dfrac{\phi_i (\hat{\btheta}_{i \Delta_n}^{0} ) - \phi_i (\hat{\btheta}_{i \Delta_n})}{ \left\{n^{-1/4}\(\log p\)^4 + p^{-1} \right\} s_p^2 \eta^{2-2\delta} }\) + \log_2 \log_2 \( \dfrac{1}{ \left\{n^{-1/4}\(\log p\)^4 + p^{-1} \right\}s_p \eta^{1-2\delta} } \) \right\},
\end{equation*}}
where $\phi_i\( \btheta \) = {\mathcal{L}}_{i}(\btheta) + \eta \left\| \btheta \right\| _{1}$.
\end{proposition}
Proposition \ref{Prop1} shows that  the $\ell_2$ distance between the local minimizer $\hat{\bbeta}_{i \Delta_n}^{t}$ and global minimizer $\hat{\bbeta}_{i \Delta_n}$ has the same convergence rate as the statistical error of the global minimizer $\hat{\bbeta}_{i \Delta_n}$. 
That is, the local and global minimizers have the same convergence rate in terms of the $\ell_2$ norm.
Furthermore, the local minimizer $\hat{\bbeta}_{i \Delta_n}^{t}$  can be obtained in polynomial time.
Thus, the proposed FATEN-LASSO procedure is computationally feasible with theoretical guarantees.

\section{A simulation study} \label{SEC-4}
In this section, we conducted a simulation study to check the finite sample performance of the FATEN-LASSO estimator.
The data were generated with a frequency of $1/n^{all}$ {over the interval [0,1]} based on the factor-based   regression jump diffusion models in \eqref{model-1}--\eqref{model-3}.
The detailed simulation setup is presented in Appendix \ref{simulation-setup}.
Noise-contaminated high-frequency observations were generated as follows:
\begin{eqnarray*}
&& Y^{o}(t_{i}) = Y(t_{i}) + \epsilon^{Y}(t_{i}) \quad   \text{and} \quad  \bX^{o}(t_{i}) = \bX(t_{i}) + \bepsilon^{X}(t_{i}) \quad  \text{ for }  i = 0, \ldots, n,
\end{eqnarray*}
where $\epsilon^{Y}(t_{i})$ and $\epsilon^X_{j}(t_{i})$ were obtained from an independent normal distribution with a mean of zero and a standard deviation of $0.005 \sqrt{\int_{0}^{1} \[\bbeta ^{\top}(t)\bB(t) \bSigma_f(t) \bB^{\top}(t)\bbeta(t) + \bbeta ^{\top}(t) \bSigma_u(t) \bbeta(t)+  \nu_z^{2}(t)\]dt}$ and  $0.005 \sqrt{\(\int_{0}^{1} \[\bB(t)\bSigma_f(t) \bB^{\top}(t) + \bSigma_u(t)\] dt\)_{jj}}$, respectively.
We set $p=100$, $s_p=[\log p]$, $r=3$, $n^{all}=23400$, and we varied $n$ from $1170$  to $23400$.
To obtain the FATEN-LASSO estimator, we employed the hard thresholding function $s\(x\)=x$ and implemented the tuning parameter choice procedure described in Appendix \ref{SEC-Tuning}. 
{To tune the parameters selected by the out-of-sample criterion, we additionally generated the processes over the subsequent period $[1,2]$. 
This additional sample was used only for tuning-parameter selection. 
For each candidate tuning parameter set, we computed the out-of-sample $R^2$ based on 10-min returns. 
In the simulation study, each 10-min return was constructed from 600 consecutive returns.
To reduce the computational cost, the tuning step was conducted using 100 Monte Carlo iterations, whereas the estimation errors were evaluated using 500 iterations.}

{For comparison, we consider ten benchmark estimators: FATEC-LASSO, FATN-LASSO, FAEN-LASSO, TEN-LASSO, TV-RIDGE, TV-PCR, TV-ENET, TED, SV-LASSO, and LASSO.
Specifically, FATEC-LASSO removes the noise-covariance bias adjustment in the local estimation step, FATN-LASSO omits the debiasing step, FAEN-LASSO omits the final thresholding step, and TEN-LASSO removes the factor-adjustment step.
TV-RIDGE, TV-PCR, and TV-ENET are local-window-based estimators that use ridge regression, principal component regression, and elastic net, respectively, instead of the local nonconvex LASSO step.
TED directly uses the observed log-returns and therefore does not account for microstructure noise or the factor structure.
SV-LASSO applies LASSO to the smoothed variables, while the standard LASSO serves as a simple benchmark based on the observed log-returns.
The formal definitions of these benchmark estimators and their implementation details are presented in Appendix \ref{SEC-Benchmark}.
All tuning parameters were selected separately for each estimator and each sample size $n$. 
The average estimation errors under the max norm, $\ell_1$ norm, and $\ell_2$ norm were calculated through 500 iterations.
}

\begin{figure}[!ht]
\centering
\includegraphics[width = 1\textwidth]{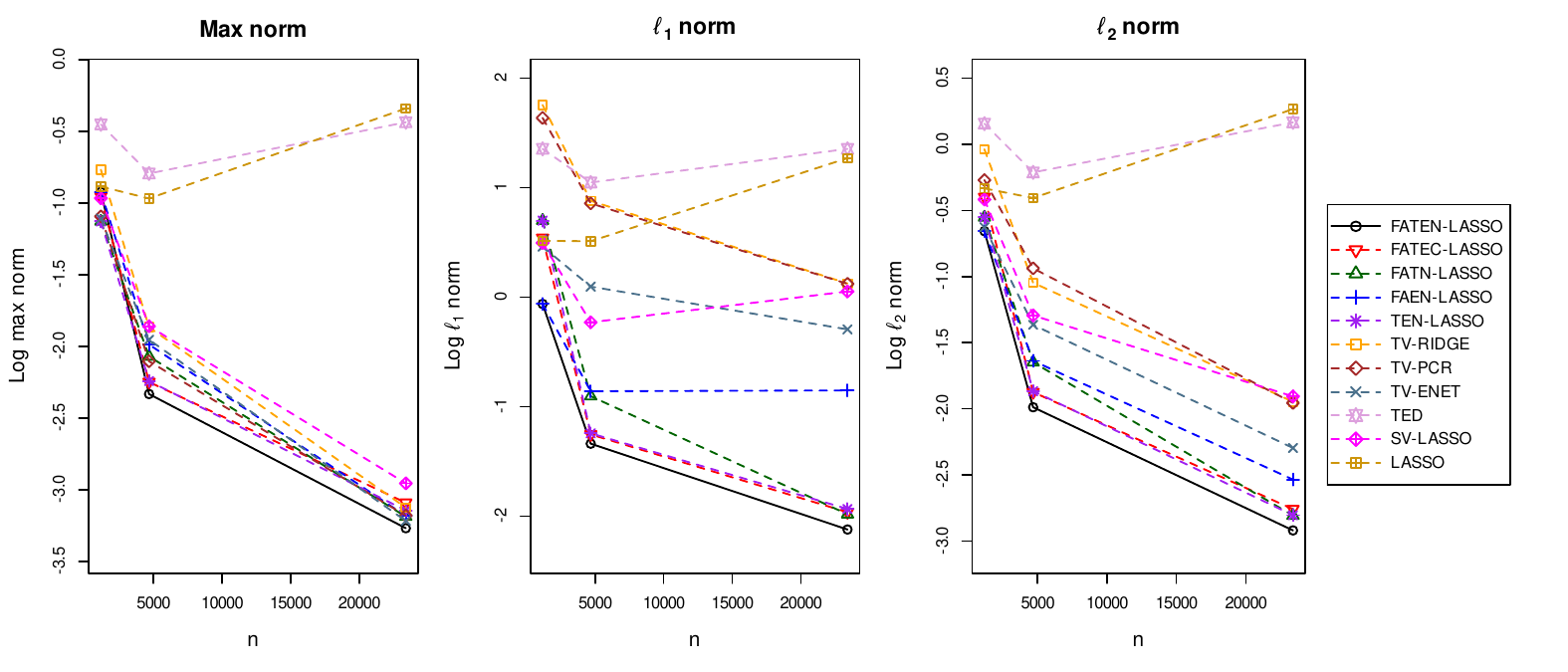}
\caption{The log max, $\ell_1$, and $\ell_2$ norm error plots of the FATEN-LASSO and benchmarks for $p=100$ and $n=1170, 4680, 23400$.}\label{Beta_simulation}
\end{figure}

Figure \ref{Beta_simulation} plots the log max, $\ell_1$, and $\ell_2$ norm errors of FATEN-LASSO and the benchmark estimators with $n=1170, 4680, 23400$.
As seen in Figure \ref{Beta_simulation}, the estimation errors of FATEN-LASSO decrease as the sample size $n$ increases.
As expected, FATEN-LASSO performs best overall among the eleven estimators across the three error norms.
This may be because only the FATEN-LASSO estimator can fully handle the factor structure in the covariate process, the microstructure noise of high-frequency data, and the time variation in the coefficient process.
In contrast, TED and LASSO estimators are not consistent.
One possible explanation for this is that the proportion of the noise in log-returns increases as the sample size $n$ increases.
These results suggest that the proposed FATEN-LASSO estimator can effectively handle the strongly correlated structure of the covariates, microstructure noise, and the time-varying coefficient process when estimating high-dimensional integrated coefficients.
{In Appendix \ref{SEC-sensitivity}, we conduct a sensitivity analysis for the tuning parameters used in the FATEN-LASSO estimator. 
The results show that the estimator is reasonably stable in most cases, although some tuning parameters require more careful selection than others.
Nevertheless, both the in-sample criterion and the out-of-sample $R^2$ criterion lead to relatively small estimation errors overall.}

\section{An empirical study} \label{SEC-5}
We applied the proposed FATEN-LASSO estimator to real high-frequency trading data, collected from January 2013 to December 2019.
We obtained stock price data from the End of Day website  (https://eoddata.com/), futures price data from the FirstRate Data website (https://firstratedata \\
.com/), and firm fundamentals from the Center for Research in Security Prices (CRSP)/Compustat Merged Database.
We collected 1-min log-price data using the previous tick scheme \citep{wang2010vast, zhang2011estimating} and excluded half trading days.
For the dependent process, we considered  five assets: Apple Inc. (AAPL), Berkshire Hathaway Inc. (BRK.B), Amazon.com, Inc. (AMZN), Alphabet Inc. (GOOG), and Exxon Mobil Corporation (XOM). 
These assets have the largest market values in the following Global Industry Classification Standard (GICS) sectors: information technology, financials, consumer discretionary, communication services, and energy. 
{For the covariate process, we employed the high-frequency factor zoo dataset of \citet{aleti2022high}, which is available from the author's website. 
Following \citet{aleti2022high}, we considered three types of portfolios: characteristic-sorted portfolios \citep{y2022open, jensen2023there}, industry portfolios \citep{fama1997industry}, and six Fama--French portfolios \citep{carhart1997persistence, fama2015five}. 
From this dataset, we included all 90 characteristic-sorted portfolios in the value, size, profitability, investment, and momentum groups, together with all 48 industry portfolios and the six Fama--French portfolios. 
The groups, symbols, and descriptions of the characteristic-sorted and industry portfolios are reported in Table \ref{Table4} in Appendix \ref{SEC-materials}.
The six Fama--French portfolios are MKT, HML, SMB, RMW, CMA, and MOM, which correspond to the market, value, size, profitability, investment, and momentum factors, respectively.
In summary, we used five assets for the dependent process and 144 portfolios for the covariate process.}

{
When implementing the FATEN-LASSO estimator, we use the tuning parameter selection procedure described in Appendix \ref{SEC-Tuning}. 
To select the tuning parameters based on the out-of-sample $R^2$ criterion, we use only the 2013 subsample. 
Specifically, for each candidate tuning parameter set, we estimate the monthly integrated coefficients for each of the five assets using the data from month $m$ in 2013, and then compute the out-of-sample $R^2$ based on the 10-min log-returns in month $m+1$, for $m=1,\ldots,11$. 
We then average the resulting out-of-sample $R^2$ over the 11 month pairs and the five assets, and choose the tuning parameter set that maximizes this average. 
The selected values are $c_{k_1}= 1/8$, $c_J=5$, $c_J^{(2)}=1$, $\rho=10$, and $c_{\alpha_2}=1$.
Note that in the empirical study, we set $k_2 = 390$ so that the local window for estimating the instantaneous coefficients corresponds to one trading day under 1-min sampling. 
This choice is intended to mitigate the effect of potential discontinuities in the coefficient process across consecutive trading days.
For comparison, we consider eleven estimators: FATEN-LASSO and the ten benchmark estimators described in Appendix \ref{SEC-Benchmark}. 
Specifically, for each of the five assets, we obtain the monthly integrated coefficients using the FATEN-LASSO, FATEC-LASSO, FATN-LASSO, FAEN-LASSO, TEN-LASSO, TV-RIDGE, TV-PCR, TV-ENET, TED, SV-LASSO, and LASSO estimators. 
For all benchmark estimators, the tuning parameters based on the out-of-sample $R^2$ criterion are selected using the 10-min log-returns, following the same procedure as for FATEN-LASSO. 
For all estimators, the coefficients over non-trading periods are set to zero.
}

\begin{table}[!ht]
\caption{{Annual average in-sample $R^2$ and out-of-sample $R^2$ over monthly and one-day-ahead horizons for the FATEN-LASSO and benchmark estimators, averaged across the five assets.}}
\label{Table1}
\centering
\resizebox{\textwidth}{!}{%
{
\begin{tabular}{c c c c c c c c c c c c}
\hline
& \multicolumn{11}{c}{In-sample $R^2$ (monthly window)} \\ \cline{2-12}
& FATEN-LASSO & FATEC-LASSO & FATN-LASSO & FAEN-LASSO & TEN-LASSO & TV-RIDGE & TV-PCR & TV-ENET & TED & SV-LASSO & LASSO \\ \hline
whole period & 0.543 & 0.542 & 0.543 & 0.543 & 0.543 & 0.540 & 0.495 & 0.542 & 0.003 & 0.527 & 0.007 \\
2014         & 0.502 & 0.509 & 0.502 & 0.502 & 0.501 & 0.505 & 0.464 & 0.509 & 0.000 & 0.503 & 0.004 \\
2015         & 0.498 & 0.499 & 0.498 & 0.498 & 0.500 & 0.494 & 0.443 & 0.497 & 0.005 & 0.476 & 0.010 \\
2016         & 0.555 & 0.554 & 0.555 & 0.556 & 0.557 & 0.547 & 0.500 & 0.554 & 0.001 & 0.555 & 0.008 \\
2017         & 0.534 & 0.528 & 0.534 & 0.535 & 0.535 & 0.530 & 0.494 & 0.530 & 0.003 & 0.515 & 0.006 \\
2018         & 0.595 & 0.593 & 0.595 & 0.595 & 0.596 & 0.592 & 0.540 & 0.593 & 0.002 & 0.571 & 0.005 \\
2019         & 0.571 & 0.566 & 0.571 & 0.571 & 0.571 & 0.571 & 0.528 & 0.566 & 0.005 & 0.541 & 0.009 \\ \hline
& \multicolumn{11}{c}{Out-of-sample $R^2$ (monthly horizon)} \\ \cline{2-12}
& FATEN-LASSO & FATEC-LASSO & FATN-LASSO & FAEN-LASSO & TEN-LASSO & TV-RIDGE & TV-PCR & TV-ENET & TED & SV-LASSO & LASSO \\ \hline
whole period & 0.484 & 0.469 & 0.484 & 0.484 & 0.484 & 0.473 & 0.433 & 0.469 & 0.002 & 0.460 & 0.007 \\
2014         & 0.423 & 0.412 & 0.423 & 0.422 & 0.422 & 0.416 & 0.380 & 0.413 & 0.000 & 0.411 & 0.003 \\
2015         & 0.414 & 0.400 & 0.414 & 0.414 & 0.412 & 0.412 & 0.360 & 0.398 & 0.003 & 0.411 & 0.008 \\
2016         & 0.504 & 0.477 & 0.504 & 0.506 & 0.505 & 0.485 & 0.424 & 0.475 & 0.001 & 0.472 & 0.008 \\
2017         & 0.480 & 0.467 & 0.480 & 0.481 & 0.482 & 0.471 & 0.440 & 0.468 & 0.004 & 0.460 & 0.007 \\
2018         & 0.562 & 0.546 & 0.562 & 0.561 & 0.563 & 0.550 & 0.515 & 0.545 & 0.000 & 0.528 & 0.004 \\
2019         & 0.520 & 0.513 & 0.520 & 0.520 & 0.519 & 0.506 & 0.481 & 0.515 & 0.002 & 0.474 & 0.009 \\ \hline
& \multicolumn{11}{c}{Out-of-sample $R^2$ (one-day-ahead horizon)} \\ \cline{2-12}
& FATEN-LASSO & FATEC-LASSO & FATN-LASSO & FAEN-LASSO & TEN-LASSO & TV-RIDGE & TV-PCR & TV-ENET & TED & SV-LASSO & LASSO \\ \hline
whole period & 0.425 & 0.410 & 0.425 & 0.425 & 0.425 & 0.419 & 0.375 & 0.409 & 0.002 & 0.414 & 0.006 \\
2014         & 0.360 & 0.339 & 0.360 & 0.359 & 0.353 & 0.355 & 0.315 & 0.336 & 0.001 & 0.347 & 0.002 \\
2015         & 0.375 & 0.344 & 0.375 & 0.374 & 0.377 & 0.374 & 0.300 & 0.342 & 0.002 & 0.385 & 0.008 \\
2016         & 0.489 & 0.479 & 0.489 & 0.490 & 0.492 & 0.487 & 0.441 & 0.477 & 0.002 & 0.475 & 0.009 \\
2017         & 0.329 & 0.336 & 0.329 & 0.327 & 0.329 & 0.334 & 0.299 & 0.338 & 0.003 & 0.332 & 0.008 \\
2018         & 0.505 & 0.483 & 0.505 & 0.505 & 0.509 & 0.486 & 0.445 & 0.480 & 0.000 & 0.489 & 0.003 \\
2019         & 0.493 & 0.481 & 0.493 & 0.493 & 0.489 & 0.476 & 0.449 & 0.480 & 0.003 & 0.458 & 0.009 \\ \hline
\end{tabular}%
}}
\end{table}

{
We compare the performance of the FATEN-LASSO estimator and the benchmark estimators using an in-sample $R^2$ measure and two out-of-sample $R^2$ measures. 
The in-sample $R^2$ is computed based on the 10-min log-returns within each month using the monthly integrated coefficients estimated from the same month. 
For the first out-of-sample measure, we use the integrated coefficients estimated from month $m$ to fit the 10-min log-returns over month $m+1$, which yields a monthly-horizon out-of-sample $R^2$. 
For the second out-of-sample measure, we use the same coefficients to fit the 10-min log-returns on the first trading day of month $m+1$, which yields a one-day-ahead out-of-sample $R^2$. 
Since the tuning parameters are selected using the 2013 subsample, we exclude 2013 from the performance comparison. 
We then compute the annual average $R^2$ across the five assets and also report the average over the whole period.
Table \ref{Table1} reports the annual average in-sample $R^2$ and the annual average out-of-sample $R^2$ over the monthly and one-day-ahead horizons for the FATEN-LASSO estimator and the benchmark estimators.
From Table \ref{Table1}, we can see that FATEN-LASSO, FATN-LASSO, FAEN-LASSO, and TEN-LASSO perform similarly and achieve the best performance overall.
By contrast, TED and LASSO perform substantially worse, possibly because they do not account for microstructure noise in the 1-min high-frequency data.
Furthermore, the relatively weaker performance of FATEC-LASSO compared to FATEN-LASSO suggests that the noise-covariance bias adjustment helps improve the performance of the estimator.
These results suggest that handling microstructure noise is particularly important in terms of the $R^2$ measures.
On the other hand, in this empirical study, the gains in $R^2$ from the debiasing, thresholding, and factor-adjustment steps are relatively limited.
One possible explanation for the similar performance of FATEN-LASSO and FATN-LASSO is that the debiasing term is numerically small in this empirical application.
Nevertheless, this step is required to ensure the law of large numbers for the estimation of the integrated coefficient.
In particular, the estimated inverse instantaneous volatility matrices tend to be highly sparse, making the resulting adjustment limited.
This may be due to the heavy-tailedness of the high-frequency data, which can make the input estimators used to construct the inverse instantaneous volatility matrices unstable.
For the thresholding step, one possible reason for the similar performance of FATEN-LASSO and FAEN-LASSO is the presence of weak factors that have relatively small effects on individual assets.
Finally, one possible explanation for the similar performance of FATEN-LASSO and TEN-LASSO is that the gain from incorporating the common factor is limited relative to the estimation errors involved in selecting the rank $r$ and recovering the factor and idiosyncratic components.
Specifically, the selected rank $r$ is zero in 72.6\% of all months for the FATEN-LASSO estimator.
Consistent with these interpretations, the debiasing, thresholding, and factor adjustment steps do not substantially change the $\ell_1$ norms of the estimated integrated coefficients, as shown in Table \ref{Table2}.
While debiasing and thresholding are theoretically required to achieve improved convergence rates, their practical impact appears limited.
The factor adjustment step can be viewed as a generalization of the baseline model that accounts for latent factor structures; however, its impact appears limited in this dataset.
}

\begin{table}[!ht] 
\caption{{Average $\ell_1$ norms of the monthly integrated coefficients obtained from FATEN-LASSO and the benchmark estimators for each of the five assets, based on 144 factors over 84 months.}}
\label{Table2}
\centering
\resizebox{\textwidth}{!}{%
{
\begin{tabular}{c c c c c c c c c c c c}
\hline
& \multicolumn{11}{c}{$\ell_1$ norm} \\ \cline{2-12}
& FATEN-LASSO & FATEC-LASSO & FATN-LASSO & FAEN-LASSO & TEN-LASSO & TV-RIDGE & TV-PCR & TV-ENET & TED & SV-LASSO & LASSO \\ \hline
AAPL  & 4.776 & 10.139 & 4.776 & 4.848 & 4.361 & 9.319  & 10.915 & 10.301 & 0.013 & 5.604 & 0.006 \\
BRK.B & 4.647 & 10.761 & 4.647 & 4.681 & 4.398 & 7.935  & 8.692  & 10.708 & 0.015 & 6.116 & 0.018 \\
AMZN  & 8.952 & 15.546 & 8.952 & 8.988 & 8.667 & 12.197 & 14.027 & 15.605 & 0.088 & 8.954 & 0.028 \\
GOOG  & 5.654 & 11.564 & 5.654 & 5.860 & 5.406 & 9.375  & 11.532 & 11.946 & 0.112 & 5.088 & 0.061 \\
XOM   & 5.845 & 12.836 & 5.845 & 5.896 & 5.691 & 9.553  & 10.753 & 12.761 & 0.070 & 7.204 & 0.029 \\ \hline
\end{tabular}%
}}
\end{table}

{
Table \ref{Table2} reports the average $\ell_1$ norms of the estimated monthly integrated coefficients for FATEN-LASSO and the benchmark estimators, computed over 144 factors and 84 months for each of the five assets.
From Table \ref{Table2}, we find that although the TED and LASSO estimators have the smallest $\ell_1$ norms, this is likely because they do not account for microstructure noise and therefore estimate many coefficients as zero.
Among the remaining estimators, the nonconvex local-window-based estimators, namely FATEN-LASSO, FATN-LASSO, FAEN-LASSO, and TEN-LASSO, yield similarly small $\ell_1$ norms.
Combining the results in Tables \ref{Table1} and \ref{Table2}, we can conjecture that FATEN-LASSO, FATN-LASSO, FAEN-LASSO, and TEN-LASSO are better able to capture market dynamics than the other estimators while maintaining relatively parsimonious models.
In Appendix \ref{empirical-FATEN}, we investigate the integrated coefficient estimates obtained from the FATEN-LASSO procedure.
The results show the time-varying property of the coefficient process and its sparsity in terms of $\ell_1$ norms.
The findings are also consistent with the multi-factor models \citep{asness2013value, carhart1997persistence, eugene1992cross, fama2015five} and factor zoo analysis \citep{jensen2023there}.
}

\section{Conclusion}\label{SEC-6}
In this paper, we developed a novel FATEN-LASSO estimation procedure that can accommodate the microstructure noise of high-frequency data, factor structure of the covariate process, and time variation in the high-dimensional coefficient process. 
To handle the noise and factor structure, we first smoothed the observed variables and applied PCA to the smoothed covariate process.
Then, to estimate the instantaneous coefficient, we employed the nonconvex optimization with the smoothed variables.
We showed that the proposed instantaneous coefficient estimator can handle the noises, factor structure, and the high-dimensional time-varying coefficient process with the sharp convergence rate.
To handle the bias from the $\ell_1$-regularization, we utilized the debiasing scheme and obtained the debiased integrated coefficient estimator using debiased instantaneous coefficient estimators.
Then, we further regularized the debiased integrated coefficient estimator to account for the sparse structure of the coefficient process.
We showed that the proposed FATEN-LASSO estimator achieves the sharp convergence rate.
This property still holds even if the factor structure in the covariate processes does not exist.

{In the empirical study, FATEN-LASSO, FATN-LASSO, FAEN-LASSO, and TEN-LASSO exhibit similar performance and perform best overall in terms of the $R^2$ measures, while maintaining relatively parsimonious models.
This finding suggests that, when estimating high-dimensional integrated coefficients based on high-frequency data, it is important to handle both the time variation in the coefficient process and microstructure noise.
However, the additional gains from the debiasing, thresholding, and factor-adjustment steps appear limited. 
One possible explanation is that the debiasing term is numerically small, possibly because the heavy-tailedness of high-frequency data makes the estimation of the inverse instantaneous volatility matrices unstable.
The limited gain from thresholding may be due to the presence of weak factors that have relatively minor impacts on each asset.
Furthermore, the limited gain from factor adjustment may be due to the errors involved in selecting the rank $r$ and recovering the common factor and idiosyncratic components.
These observations suggest several directions for future research.
First, it would be important to develop a robust debiasing procedure for heavy-tailed high-frequency data.
Second, it would be interesting to investigate prediction methods based on separate estimation of the factor and idiosyncratic components.
However, such an approach would require additional handling of the future factor component and the associated rotation issue.
It would be important but challenging to develop a theoretically justified prediction method that estimates the rank, common factors, and idiosyncratic components and performs well in practice.}

\bibliography{myReferences}

@article{wang2010vast,
  title={Vast volatility matrix estimation for high-frequency financial data},
  author={Wang, Yazhen and Zou, Jian},
  journal={The Annals of Statistics},
  volume={38},
  number={2},
  pages={943--978},
  year={2010},
  publisher={JSTOR}
}

@article{fan2013large,
  title={Large covariance estimation by thresholding principal orthogonal complements},
  author={Fan, Jianqing and Liao, Yuan and Mincheva, Martina},
  journal={Journal of the Royal Statistical Society: Series B (Statistical Methodology)},
  volume={75},
  number={4},
  pages={603--680},
  year={2013},
  publisher={Wiley Online Library}
}

@article{kim2016asymptotic,
  title={Asymptotic theory for large volatility matrix estimation based on high-frequency financial data},
  author={Kim, Donggyu and Wang, Yazhen and Zou, Jian},
  journal={Stochastic Processes and their Applications},
  volume={126},
  pages={3527–-3577},
  year={2016},
  publisher={Elsevier}
}

@article{fan2018robust,
  title={Robust high-dimensional volatility matrix estimation for high-frequency factor model},
  author={Fan, Jianqing and Kim, Donggyu},
  journal={Journal of the American Statistical Association},
  volume={113},
  number={523},
  pages={1268--1283},
  year={2018},
  publisher={Taylor \& Francis}
}

@article{fan2019structured,
  title={Structured volatility matrix estimation for non-synchronized high-frequency financial data},
  author={Fan, Jianqing and Kim, Donggyu},
  journal={Journal of Econometrics},
  volume={209},
  number={1},
  pages={61--78},
  year={2019},
  publisher={Elsevier}
}

@article{ait2017using,
  title={Using principal component analysis to estimate a high dimensional factor model with high-frequency data},
  author={A{\"\i}t-Sahalia, Yacine and Xiu, Dacheng},
  journal={Journal of Econometrics},
  volume={201},
  number={2},
  pages={384--399},
  year={2017},
  publisher={Elsevier}
}

@article{jacod2009microstructure,
  title={Microstructure noise in the continuous case: the pre-averaging approach},
  author={Jacod, Jean and Li, Yingying and Mykland, Per A and Podolskij, Mark and Vetter, Mathias},
  journal={Stochastic Processes and their Applications},
  volume={119},
  number={7},
  pages={2249--2276},
  year={2009},
  publisher={Elsevier}
}

@article{barndorff2011multivariate,
  title={Multivariate realised kernels: consistent positive semi-definite estimators of the covariation of equity prices with noise and non-synchronous trading},
  author={Barndorff-Nielsen, Ole E and Hansen, Peter Reinhard and Lunde, Asger and Shephard, Neil},
  journal={Journal of Econometrics},
  volume={162},
  number={2},
  pages={149--169},
  year={2011},
  publisher={Elsevier}
}

@article{christensen2010pre,
  title={Pre-averaging estimators of the ex-post covariance matrix in noisy diffusion models with non-synchronous data},
  author={Christensen, Kim and Kinnebrock, Silja and Podolskij, Mark},
  journal={Journal of Econometrics},
  volume={159},
  number={1},
  pages={116--133},
  year={2010},
  publisher={Elsevier}
}

@article{ait2010high,
  title={High-frequency covariance estimates with noisy and asynchronous financial data},
  author={A{\"\i}t-Sahalia, Yacine and Fan, Jianqing and Xiu, Dacheng},
  journal={Journal of the American Statistical Association},
  volume={105},
  number={492},
  pages={1504--1517},
  year={2010},
  publisher={Taylor \& Francis}
}

@article{zhang2011estimating,
  title={Estimating covariation: Epps effect, microstructure noise},
  author={Zhang, Lan},
  journal={Journal of Econometrics},
  volume={160},
  number={1},
  pages={33--47},
  year={2011},
  publisher={Elsevier}
}

@article{ait2020high,
  title={High-frequency factor models and regressions},
  author={A{\"\i}t-Sahalia, Yacine and Kalnina, Ilze and Xiu, Dacheng},
  journal={Journal of Econometrics},
  volume={216},
  number={1},
  pages={86--105},
  year={2020},
  publisher={Elsevier}
}

@article{ait2019principal,
  title={Principal component analysis of high-frequency data},
  author={A{\"\i}t-Sahalia, Yacine and Xiu, Dacheng},
  journal={Journal of the American Statistical Association},
  volume={114},
  number={525},
  pages={287--303},
  year={2019},
  publisher={Taylor \& Francis}
}

@article{cochrane2011presidential,
  title={Presidential address: Discount rates},
  author={Cochrane, John H},
  journal={The Journal of Finance},
  volume={66},
  number={4},
  pages={1047--1108},
  year={2011},
  publisher={Wiley Online Library}
}

@article{candes2007dantzig,
  title={The Dantzig selector: Statistical estimation when p is much larger than n},
  author={Candes, Emmanuel and Tao, Terence},
  journal={The Annals of Statistics},
  volume={35},
  number={6},
  pages={2313--2351},
  year={2007},
  publisher={Institute of Mathematical Statistics}
}

@article{cai2011constrained,
  title={A constrained $\ell_1$ minimization approach to sparse precision matrix estimation},
  author={Cai, Tony and Liu, Weidong and Luo, Xi},
  journal={Journal of the American Statistical Association},
  volume={106},
  number={494},
  pages={594--607},
  year={2011},
  publisher={Taylor \& Francis}
}

@article{barndorff2004econometric,
  title={Econometric analysis of realized covariation: High frequency based covariance, regression, and correlation in financial economics},
  author={Barndorff-Nielsen, Ole E and Shephard, Neil},
  journal={Econometrica},
  volume={72},
  number={3},
  pages={885--925},
  year={2004},
  publisher={Wiley Online Library}
}

@article{andersen2005framework,
  title={A framework for exploring the macroeconomic determinants of systematic risk},
  author={Andersen, Torben G and Bollerslev, Tim and Diebold, Francis X and Wu, Jin},
  journal={American Economic Review},
  volume={95},
  number={2},
  pages={398--404},
  year={2005},
  publisher={American Economic Association}
}

@article{reiss2015nonparametric,
  title={Nonparametric test for a constant beta between It{\^o} semi-martingales based on high-frequency data},
  author={Rei{\ss}, Markus and Todorov, Viktor and Tauchen, George},
  journal={Stochastic Processes and their Applications},
  volume={125},
  number={8},
  pages={2955--2988},
  year={2015},
  publisher={Elsevier}
}

@article{mykland2009inference,
  title={Inference for continuous semimartingales observed at high frequency},
  author={Mykland, Per A and Zhang, Lan},
  journal={Econometrica},
  volume={77},
  number={5},
  pages={1403--1445},
  year={2009},
  publisher={Wiley Online Library}
}

@article{andersen2021recalcitrant,
  title={Recalcitrant betas: Intraday variation in the cross-sectional dispersion of systematic risk},
  author={Andersen, Torben G and Thyrsgaard, Martin and Todorov, Viktor},
  journal={Quantitative Economics},
  volume={12},
  number={2},
  pages={647--682},
  year={2021},
  publisher={Wiley Online Library}
}

@article{harvey2016and,
  title={… and the cross-section of expected returns},
  author={Harvey, Campbell R and Liu, Yan and Zhu, Heqing},
  journal={The Review of Financial Studies},
  volume={29},
  number={1},
  pages={5--68},
  year={2016},
  publisher={Oxford University Press}
}

@article{mclean2016does,
  title={Does academic research destroy stock return predictability?},
  author={McLean, R David and Pontiff, Jeffrey},
  journal={The Journal of Finance},
  volume={71},
  number={1},
  pages={5--32},
  year={2016},
  publisher={Wiley Online Library}
}

@article{hou2020replicating,
  title={Replicating anomalies},
  author={Hou, Kewei and Xue, Chen and Zhang, Lu},
  journal={The Review of Financial Studies},
  volume={33},
  number={5},
  pages={2019--2133},
  year={2020},
  publisher={Oxford University Press}
}

@article{tibshirani1996regression,
  title={Regression shrinkage and selection via the lasso},
  author={Tibshirani, Robert},
  journal={Journal of the Royal Statistical Society: Series B (Methodological)},
  volume={58},
  number={1},
  pages={267--288},
  year={1996},
  publisher={Wiley Online Library}
}

@article{fama2015five,
  title={A five-factor asset pricing model},
  author={Fama, Eugene F and French, Kenneth R},
  journal={Journal of Financial Economics},
  volume={116},
  number={1},
  pages={1--22},
  year={2015},
  publisher={Elsevier}
}

@article{carhart1997persistence,
  title={On persistence in mutual fund performance},
  author={Carhart, Mark M},
  journal={The Journal of Finance},
  volume={52},
  number={1},
  pages={57--82},
  year={1997},
  publisher={Wiley Online Library}
}

@article{asness2013value,
  title={Value and momentum everywhere},
  author={Asness, Clifford S and Moskowitz, Tobias J and Pedersen, Lasse Heje},
  journal={The Journal of Finance},
  volume={68},
  number={3},
  pages={929--985},
  year={2013},
  publisher={Wiley Online Library}
}

@article{fan2001variable,
  title={Variable selection via nonconcave penalized likelihood and its oracle properties},
  author={Fan, Jianqing and Li, Runze},
  journal={Journal of the American statistical Association},
  volume={96},
  number={456},
  pages={1348--1360},
  year={2001},
  publisher={Taylor \& Francis}
}

@article{bali2011maxing,
  title={Maxing out: Stocks as lotteries and the cross-section of expected returns},
  author={Bali, Turan G and Cakici, Nusret and Whitelaw, Robert F},
  journal={Journal of Financial Economics},
  volume={99},
  number={2},
  pages={427--446},
  year={2011},
  publisher={Elsevier}
}

@article{campbell2008search,
  title={In search of distress risk},
  author={Campbell, John Y and Hilscher, Jens and Szilagyi, Jan},
  journal={The Journal of Finance},
  volume={63},
  number={6},
  pages={2899--2939},
  year={2008},
  publisher={Wiley Online Library}
}

@article{chen2018inference,
  title={Inference for Volatility Functionals of Multivariate It\^{o} Semimartingales Observed with Jump and Noise},
  author={Chen, Richard Y},
  journal={arXiv preprint arXiv:1810.04725},
  year={2018}
}

@article{kim2026high,
  title={High-Dimensional Time-Varying Coefficient Estimation in Diffusion Models},
  author={Kim, Donggyu and Oh, Minseog and Shin, Minseok},
  journal={Econometric Reviews},
  pages={1--21},
  year={2026},
  publisher={Taylor \& Francis}
}

@article{loh2012high,
  title={High-Dimensional Regression With Noisy and Missing Data: Provable Guarantees With Nonconvexity},
  author={Loh, Po-Ling and Wainwright, Martin J},
  journal={The Annals of Statistics},
  volume={40},
  number={3},
  pages={1637--1664},
  year={2012},
  publisher = {Institute of Mathematical Statistics}
}

@article{agarwal2012fast,
  title={Fast global convergence of gradient methods for high-dimensional statistical recovery},
  author={Agarwal, Alekh and Negahban, Sahand and Wainwright, Martin J},
  journal={The Annals of Statistics},
  volume={40},
  number={5},
  pages={2452--2482},
  year={2012},
  publisher={Institute of Mathematical Statistics}
}

@article{shin2025robust,
  title={Robust High-Dimensional Time-Varying Coefficient Estimation},
  author={Shin, Minseok and Kim, Donggyu},
  journal={Econometric Theory},
  pages={1--45},
  year={2025},
  publisher={Cambridge University Press}
}

@article{shin2023adaptive,
  title={Adaptive robust large volatility matrix estimation based on high-frequency financial data},
  author={Shin, Minseok and Kim, Donggyu and Fan, Jianqing},
  journal={Journal of Econometrics},
  volume={237},
  number={1},
  pages={105514},
  year={2023},
  publisher={Elsevier}
}

@article{kalnina2023inference,
  title={Inference for nonparametric high-frequency estimators with an application to time variation in betas},
  author={Kalnina, Ilze},
  journal={Journal of Business \& Economic Statistics},
  volume={41},
  number={2},
  pages={538--549},
  year={2023},
  publisher={Taylor \& Francis}
}

@article{ferson1999conditioning,
  title={Conditioning variables and the cross section of stock returns},
  author={Ferson, Wayne E and Harvey, Campbell R},
  journal={The Journal of Finance},
  volume={54},
  number={4},
  pages={1325--1360},
  year={1999},
  publisher={Wiley Online Library}
}

@article{chen2023realized,
  title={Realized regression with asynchronous and noisy high frequency and high dimensional data},
  author={Chen, Dachuan and Mykland, Per A and Zhang, Lan},
  journal={Journal of Econometrics},
  volume={239},
  number={2},
  pages={105446},
  year={2024},
  publisher={Elsevier}
}

@article{fan2020factor,
  title={Factor-adjusted regularized model selection},
  author={Fan, Jianqing and Ke, Yuan and Wang, Kaizheng},
  journal={Journal of Econometrics},
  volume={216},
  number={1},
  pages={71--85},
  year={2020},
  publisher={Elsevier}
}

@article{kneip2011factor,
  title={Factor models and variable selection in high-dimensional regression analysis},
  author={Kneip, Alois and Sarda, Pascal},
  journal={The Annals of Statistics},
  volume={39},
  number={5},
  pages={2410–2447},
  year={2011},
  publisher = {Institute of Mathematical Statistics}
}

@article{barigozzi2024fnets,
  title={FNETS: Factor-adjusted network estimation and forecasting for high-dimensional time series},
  author={Barigozzi, Matteo and Cho, Haeran and Owens, Dom},
  journal={Journal of Business \& Economic Statistics},
  volume={42},
  number={3},
  pages={890--902},
  year={2024},
  publisher={Taylor \& Francis}
}

@article{chun2022state,
  title={State heterogeneity analysis of financial volatility using high-frequency financial data},
  author={Chun, Dohyun and Kim, Donggyu},
  journal={Journal of Time Series Analysis},
  volume={43},
  number={1},
  pages={105--124},
  year={2022},
  publisher={Wiley Online Library}
}

@article{dai2019knowing,
  title={Knowing factors or factor loadings, or neither? Evaluating estimators of large covariance matrices with noisy and asynchronous data},
  author={Dai, Chaoxing and Lu, Kun and Xiu, Dacheng},
  journal={Journal of Econometrics},
  volume={208},
  number={1},
  pages={43--79},
  year={2019},
  publisher={Elsevier}
}

@article{eugene1992cross,
  title={The cross-section of expected stock returns},
  author={Fama, Eugene and French, Kenneth},
  journal={The Journal of Finance},
  volume={47},
  number={2},
  pages={427--465},
  year={1992}
}

@article{jensen2023there,
  title={Is there a replication crisis in finance?},
  author={Jensen, Theis Ingerslev and Kelly, Bryan and Pedersen, Lasse Heje},
  journal={The Journal of Finance},
  volume={78},
  number={5},
  pages={2465--2518},
  year={2023},
  publisher={Wiley Online Library}
}

@article{fan2024latent,
  title={Are latent factor regression and sparse regression adequate?},
  author={Fan, Jianqing and Lou, Zhipeng and Yu, Mengxin},
  journal={Journal of the American Statistical Association},
  volume={119},
  number={546},
  pages={1076--1088},
  year={2024},
  publisher={Taylor \& Francis}
}

@article{bai2003inferential,
  title={Inferential theory for factor models of large dimensions},
  author={Bai, Jushan},
  journal={Econometrica},
  volume={71},
  number={1},
  pages={135--171},
  year={2003},
  publisher={Wiley Online Library}
}

@article{oh2026robust,
  title={Robust realized integrated beta estimator with application to dynamic analysis of integrated beta},
  author={Oh, Minseog and Kim, Donggyu and Wang, Yazhen},
  journal={Journal of Econometrics},
  pages={105810},
  year={2026},
  publisher={Elsevier}
}

@article{aleti2022high,
  title={The high-frequency factor zoo},
  author={Aleti, Saketh},
  journal={Available at SSRN 4021620},
  year={2022}
}

@article{fama1997industry,
  title={Industry costs of equity},
  author={Fama, Eugene F and French, Kenneth R},
  journal={Journal of Financial Economics},
  volume={43},
  number={2},
  pages={153--193},
  year={1997},
  publisher={Elsevier}
}

@article{y2022open,
  title={Open source cross-sectional asset pricing},
  author={Y. Chen, Andrew and Zimmermann, Tom},
  journal={Critical Finance Review},
  volume={11},
  number={02},
  pages={207--264},
  year={2022},
  publisher={Emerald Publishing Limited}
}
\end{spacing}

\newpage
\appendix

\vspace*{1cm}
\begin{center}
{\LARGE Supplement to ``Nonconvex High-Dimensional Time-Varying Coefficient Estimation for Noisy High-Frequency Observations with a Factor Structure'' }
\end{center}
\vspace{1em}

\begin{spacing}{1.9}

\section{Discussion on tuning parameter selection}\label{SEC-Tuning}
In this section, we discuss the process of selecting the tuning parameters for the FATEN-LASSO estimator.
To construct the smoothed variables, we set
{\begin{equation*}
g(x) = x  \wedge (1 - x), \quad  k_1= \left\lfloor c_{k_1} n^{1/2} \right\rfloor, \quad  k_2 = \left\lfloor c_{k_2} n^{3/4} \right\rfloor,
\end{equation*}
where $c_{k_1}$ and $c_{k_2}$ are tuning parameters, and $\lfloor x \rfloor$ denotes the floor function, which maps a real number $x$ to the greatest integer less than or equal to $x$.
For the jump truncation parameters in \eqref{jump_trunc1} and \eqref{jump_trunc2}, we use
 \begin{equation}\label{jump_trunc3}
w_n = c_{J} \, \text{sd}\left(\Delta_i^n\hat{Y}\right), \quad  v_{j,n} = c_{J} \, \text{sd}\left(\Delta_i^n\hat{X}_j\right), \quad  v_{j,n}^{(2)} = c_{J}^{(2)} \, \text{sd}\left(\Delta_i^n X_j^o \right),
\end{equation}
where $c_{J}$ and $c_{J}^{(2)}$ are tuning parameters,} and  $\text{sd}$ denotes the sample standard deviation.
After obtaining $\hat{\bG}_i$, each column of $\mathcal{Y}_i$ and $\hat{\bG}_i$ is standardized to have a mean of zero and a variance of 1. 
For $j=1,\ldots,p$, the $j$-th entry of $\Delta_i^n \hat{\bX}^{\mathrm{trunc}}$ is divided by the standard deviation of the $j$-th column of $\hat{\bG}_i$. 
After obtaining the FATEN-LASSO estimator, we rescale the coefficient estimates accordingly.
For the nonconvex optimization, we implement the composite gradient descent algorithm in \eqref{gradient} with $10^3$ iterations.
{We treat $\rho$ as a tuning parameter and set
\begin{equation*}
\alpha_2 = c_{\alpha_2} \lambda_{\max}\!\left( \frac{n}{\phi k_1 k_2}\hat{\bG}_i^\top \hat{\bG}_i \right),
\end{equation*}
where $c_{\alpha_2}>0$ is a tuning parameter.}

{
In the simulation study, we select
$
(c_{k_1}, c_{k_2}, c_J, c_J^{(2)}, \rho, c_{\alpha_2})
$
by maximizing the out-of-sample $R^2$. 
The candidate grids are
\begin{eqnarray*}
c_{k_1} \in \left\{\frac{1}{16}, \frac{1}{8}, \frac{1}{4}, \frac{1}{2}, 1\right\}, \quad c_{k_2} \in \{1,2,3,4,5\}, \quad c_J \in \{1,2,3,4,5\}, \cr
c_J^{(2)} \in \{1,2,3,4,5\}, \quad \rho \in \{5,10,15,20,25\}, \quad c_{\alpha_2} \in \{0.5, 1, 1.5, 2, 2.5\}.
\end{eqnarray*}
We initialize the search at the middle points of the corresponding grids, namely,
\begin{equation*}
(c_{k_1}, c_{k_2}, c_J, c_J^{(2)}, \rho, c_{\alpha_2}) = \left(\frac{1}{4}, 3, 3, 3, 15, 1.5\right).
\end{equation*}
Starting from these initial values, we update one tuning parameter at a time while holding the others fixed, and choose the candidate value that yields the largest out-of-sample $R^2$.
Here, the out-of-sample $R^2$ is computed using 10-min returns.
We repeat full sweeps over the six parameters until the selected values no longer change.
In the empirical study, we follow the same procedure, except that $k_2$ is fixed at $390$, corresponding to one trading day under 1-min sampling. 
Accordingly, $c_{k_2}$ is not tuned in the empirical analysis.
The remaining tuning parameters, namely $c_{\eta}$, $c_{\tau}$, $c_h$, and $r$, are selected by in-sample data-driven criteria in both the simulation and empirical studies. 
Specifically, we set
\begin{equation}\label{tuning}
\eta = c_{\eta} n^{-1/8} (\log p)^2,
\quad
\tau = c_{\tau} n^{-1/8} (\log p)^2,
\quad
h_n = c_h n^{-1/4},
\end{equation}
where $c_{\eta}$, $c_{\tau}$, and $c_h$ are tuning parameters.
We choose $c_{\eta} \in [10^{-6},10^6]$ by minimizing the corresponding Bayesian information criterion (BIC). 
For each time  $i\Delta_n$, we initialize the composite gradient descent algorithm at
\begin{equation*}
\hat{\btheta}_{i\Delta_n}^{(0)} = (0,\ldots,0)^\top \in \mathbb{R}^{p+r}
\end{equation*}
for the largest candidate value of $\eta$. 
For each subsequent value of $\eta$, we use the solution obtained at the previous value as the initial value; that is, we employ a warm-start scheme along the solution path.
We select $c_{\tau} \in [10^{-2},10^2]$, which minimizes the following loss function:
\begin{equation*}
\operatorname{tr}\!\left[
\left(
\left\{
\frac{n}{\phi k_1 k_2}\hat{\bU}_i^{\top}\hat{\bU}_i
-
\frac{n\zeta}{\phi k_1^2}\hat{\bV}^{X}
\right\}
\hat{\bOmega}_{i\Delta_n}
-
\bI_p
\right)^2
\right],
\end{equation*}
where $\bI_p$ is the $p$-dimensional identity matrix.
To select $c_h$, we employ a leave-one-block-out cross-validation scheme. 
For each block $j$, we define the corresponding leave-one-block-out integrated coefficient as
\begin{equation*}
\widehat{I\beta}_{(-j)}
=
\sum_{i=0,\; i\neq j}^{[1/(k_2 \Delta_n)]-1}
\tilde{\bbeta}_{i k_2 \Delta_n}
\frac{k_2 \Delta_n}{1-k_2 \Delta_n}.
\end{equation*}
We then define
\begin{equation*}
\widetilde{I\beta}_{(-j)}(c_h)
=
s (\widehat{I\beta}_{(-j)})
\1 \(
|\widehat{I\beta}_{(-j)} | \ge h_n
\).
\end{equation*}
We choose $c_h \in [0.01,10]$ by minimizing
\begin{equation*}
k_2 \Delta_n \sum_{j=0}^{[1/(k_2\Delta_n)]-1} \left( \mathcal{Y}_j - \mathcal{X}_j \widetilde{I\beta}_{(-j)}(c_h)
\right)^\top \left( \mathcal{Y}_j - \mathcal{X}_j \widetilde{I\beta}_{(-j)}(c_h) \right).
\end{equation*}
Finally, to determine the rank $r$, we compute the FATEN-LASSO estimator $\widetilde{I\beta}^{\,r}$ for $r=0,1,2,3,4,5$ and choose $r$ by minimizing
\begin{equation}\label{rank}
\frac{1}{n-k_1+1} \sum_{i=0}^{n-k_1} \left( \Delta_i^n \hat{Y}^{\mathrm{trunc}} - \widetilde{I\beta}^{\,r\top}\Delta_i^n \hat{\bX}^{\mathrm{trunc}} \right)^2.
\end{equation}
}

\section{Benchmark estimators and implementation details}\label{SEC-Benchmark}
{In this section, we describe the benchmark estimators considered in the numerical study and explain how they are implemented.
We begin with four variants of the FATEN-LASSO estimator: FATEC-LASSO, FATN-LASSO, FAEN-LASSO, and TEN-LASSO.
These estimators are constructed by removing or replacing one component of the FATEN-LASSO procedure at a time.}
We first consider the Factor Adjusted Thresholded dEbiased Convex LASSO (FATEC-LASSO) estimator.
It uses the same estimation procedure as the FATEN-LASSO estimator, {except that it does not adjust the bias induced by the noise covariance terms when estimating the instantaneous coefficient.}
Specifically, $\hat{\btheta}^{\text{FATEC}}_{i\Delta_n}$ is defined by
\begin{equation}\label{TEC} 
 \hat{\btheta}^{\text{FATEC}}_{i \Delta_n} = {\operatorname{arg \, min}} \, \dfrac{n}{2\phi k_1 k_2}\left\| \mathcal{Y}_{i} - \hat{\bG}_i \btheta \right\|_2^2  + \eta^{\text{FATEC}} \left\| \btheta \right\| _{1},
\end{equation}
where the regularization parameter $\eta^{\text{FATEC}}=c_{\eta}^{\text{FATEC}} n^{-1/8} \(\log p \)^{2}$ and $c_{\eta}^{\text{FATEC}} \in [10^{-6}, 10^6]$ was selected by minimizing the corresponding BIC.
{Although FATEC-LASSO can partially mitigate the effect of microstructure noise through smoothed variables and can accommodate the strong dependence in the covariate process through factor adjustment, it does not satisfy the deviation condition due to the bias induced by the noise covariance terms.}
Consequently, the instantaneous coefficient estimator is inconsistent.
{We next consider the Factor Adjusted Thresholded Nonconvex LASSO (FATN-LASSO) estimator, which removes the CLIME-based debiasing step from FATEN-LASSO.
Specifically, after obtaining the instantaneous coefficient estimates, FATN-LASSO directly integrates them and then applies the same thresholding step as FATEN-LASSO.
We then consider the Factor Adjusted dEbiased Nonconvex LASSO (FAEN-LASSO) estimator, which removes the final thresholding step from FATEN-LASSO.
Specifically, after obtaining the debiased integrated coefficient estimator, FAEN-LASSO uses it directly without applying the thresholding function.
Finally, we consider the Thresholded dEbiased Nonconvex LASSO (TEN-LASSO) estimator.
TEN-LASSO is obtained as the special case of FATEN-LASSO with $r=0$.
Therefore, TEN-LASSO can be viewed as the version of FATEN-LASSO without factor adjustment.}

{As a second class of benchmark estimators, we consider three local-window-based estimators: Time-Varying Ridge (TV-RIDGE), Time-Varying Principal Component Regression (TV-PCR), and Time-Varying Elastic Net (TV-ENET).
These estimators use the same local-window structure as FATEN-LASSO, but replace the local nonconvex LASSO step with ridge regression, principal component regression (PCR), and Elastic Net, respectively.
For TV-RIDGE, the instantaneous coefficient estimator is defined by
\begin{equation}\label{TV-RIDGE}
\hat{\bbeta}^{\text{TV-RIDGE}}_{i \Delta_n}
=
\operatorname*{arg\,min}_{\bbeta}
\left\{
\dfrac{n}{2\phi k_1 k_2}\left\| \mathcal{Y}_{i} - \mathcal{X}_i \bbeta \right\|_2^2
+
\dfrac{\eta^{\text{RIDGE}}}{2}\|\bbeta\|_2^2
\right\},
\end{equation}
where $\eta^{\text{RIDGE}} \in [10^{-6},10^6]$ was selected by minimizing the corresponding BIC.
For TV-PCR, let $\mathcal{X}_i = \hat{\bL}_i \hat{\bD}_i \hat{\bR}_i^\top$ be the singular value decomposition of $\mathcal{X}_i$, and let $\hat{\bL}_{i,q}$, $\hat{\bD}_{i,q}$, and $\hat{\bR}_{i,q}$ denote the matrices formed by the first $q$ principal components.
We define
\begin{equation}\label{TV-PCR}
\hat{\bbeta}^{\text{TV-PCR}}_{i \Delta_n}
=
\hat{\bL}_{i,q}\hat{\bD}_{i,q}^{-1}\hat{\bR}_{i,q}^{\top}\mathcal{Y}_i,
\end{equation}
with the convention that $\hat{\bbeta}^{\text{TV-PCR}}_{i\Delta_n}=\mathbf{0}$ when $q=0$.
The number of retained principal components was selected from
$
q \in \{0,1,\ldots,\min(k_2-k_1,p)\}
$
by minimizing the corresponding BIC.
For TV-ENET, the instantaneous coefficient estimator is defined by
\begin{equation}\label{TV-ENET}
\hat{\bbeta}^{\text{TV-ENET}}_{i \Delta_n}
=
\operatorname*{arg\,min}_{\bbeta}
\left\{
\dfrac{n}{2\phi k_1 k_2}\left\| \mathcal{Y}_{i} - \mathcal{X}_i \bbeta \right\|_2^2
+
\eta^{\text{ENET}}
\left(
\alpha^{\text{ENET}} \|\bbeta\|_1 + \dfrac{1-\alpha^{\text{ENET}}}{2}\|\bbeta\|_2^2
\right)
\right\},
\end{equation}
where $\eta^{\text{ENET}} = c_{\eta}^{\text{ENET}} n^{-1/8}(\log p)^2$, $c_{\eta}^{\text{ENET}} \in [10^{-6},10^6]$, and  $\alpha^{\text{ENET}} \in \{0.1, 0.3, 0.5, 0.7, 0.9\}$ is the mixing parameter.
The parameters $c_{\eta}^{\text{ENET}}$ and $\alpha^{\text{ENET}}$ were selected by minimizing the corresponding BIC.
When implementing TV-RIDGE, TV-PCR, and TV-ENET based on BIC, we use the following degrees-of-freedom approximations. 
For TV-RIDGE, we use the effective degrees of freedom
\begin{equation*}
\mathrm{df}(\eta^{\text{RIDGE}}) = \sum_{\ell=1}^{\operatorname{rank}\left\{\mathcal{X}_{i}\right\}} \frac{d_{i,\ell}^2} {d_{i,\ell}^2 + (k_2-k_1+1)\eta^{\text{RIDGE}}}, 
\end{equation*}
where $d_{i,\ell}$ are the nonzero singular values of $\mathcal{X}_i$.
For TV-PCR, the degrees of freedom are taken as the number of retained principal components.
For TV-ENET, we use an active-set-based ridge-trace approximation for the degrees of freedom. 
Specifically, for each $(\eta^{\text{ENET}},\alpha^{\text{ENET}})$, let
\begin{equation*}
A_i(\eta^{\text{ENET}},\alpha^{\text{ENET}})
=
\left\{
j:\hat{\bbeta}^{\text{TV-ENET}}_{i\Delta_n,j}(\eta^{\text{ENET}},\alpha^{\text{ENET}})\neq 0
\right\}
\end{equation*}
be the active set, and let $\tilde d_{i,\ell}$ denote the nonzero singular values of $\mathcal{X}_{i,A_i(\eta^{\text{ENET}},\alpha^{\text{ENET}})}$.
We set
\begin{equation*}
\mathrm{df}(\eta^{\text{ENET}},\alpha^{\text{ENET}})
=
\sum_{\ell=1}^{\operatorname{rank}\left\{\mathcal{X}_{i,A_i(\eta^{\text{ENET}},\alpha^{\text{ENET}})}\right\}}
\frac{\tilde d_{i,\ell}^{2}}
{\tilde d_{i,\ell}^{2} + (k_2-k_1+1)\eta^{\text{ENET}}(1-\alpha^{\text{ENET}})}.
\end{equation*}
For each of TV-RIDGE, TV-PCR, and TV-ENET, the integrated coefficient estimator was constructed by averaging the corresponding instantaneous coefficient estimators over time.
These estimators accommodate time variation in the coefficient process through the local-window construction and address high dimensionality through shrinkage or dimension reduction. 
Moreover, TV-PCR and TV-ENET can partially accommodate strong dependence among the covariates, and TV-RIDGE can stabilize estimation under such dependence. 
However, unlike FATEN-LASSO, they do not explicitly exploit the latent factor structure, nor do they incorporate the debiasing and thresholding steps.
Consequently, under the current numerical design, they may not recover the integrated coefficients as accurately as FATEN-LASSO.}

{As a final group of benchmark estimators, we consider three additional estimators: TED, SV-LASSO, and the standard LASSO.
We first consider the TED estimator \citep{kim2026high}, which can handle the time variation in the coefficient process and the curse of dimensionality.
Specifically, using the observed log-returns, we utilize the Dantzig selector \citep{candes2007dantzig} to obtain the instantaneous coefficient estimator.
Then, we employ debiasing and truncation schemes to obtain the integrated coefficient estimator.
The detailed estimation procedure is presented in Algorithm 1 of \citet{kim2026high}.
Since TED directly uses the observed log-returns, it does not account for microstructure noise or the factor structure.
We next consider the Smoothed-Variable LASSO (SV-LASSO) estimator, which applies LASSO to the smoothed regression variables:
\begin{equation}\label{SV-LASSO}
\widetilde{I \beta}^{\text{SV-LASSO}}
=
\operatorname*{arg\,min}_{\bbeta}
\left\{
\dfrac{n}{2k_1(n-k_1+1)}\sum_{i=0}^{n-k_1}
\left(
\Delta_i^n\hat{Y}^{\text{trunc}} - (\Delta_i^n\hat{\bX}^{\text{trunc}})^{\top}\bbeta \right)^2 + \eta^{\text{SV-LASSO}} \|\bbeta\|_1
\right\},
\end{equation}
where $\eta^{\text{SV-LASSO}} = c_{\eta}^{\text{SV-LASSO}} n^{-1/4}\sqrt{\log p}$ and $c_{\eta}^{\text{SV-LASSO}} \in [10^{-6},10^6]$ was selected by minimizing the corresponding BIC.
The SV-LASSO estimator can partially mitigate the effect of microstructure noise through the use of smoothed variables, but it does not account for time variation in the coefficient process or the factor structure.
We also include the standard LASSO estimator \citep{tibshirani1996regression} as follows:}
\begin{equation}\label{LASSO}
	\tilde{I \beta}^{\lasso}= \argmin _{\bbeta} \left \{ \dfrac{1}{2n} \sum^{n-1}_{i=0} \(\Delta_{i+1}^n Y^{\text{trunc2}}- \( \Delta_{i+1} ^n \bX^{\text{trunc2}}\)^{\top}  \bbeta \)^2 + \eta^{\lasso} \| \bbeta \|_1 \right \},
\end{equation}  
where $\Delta_{i}^n Y^{\text{trunc2}} = \Delta_{i} ^n Y^o \, \1 \(|\Delta_{i} ^n Y^o | \leq w_n^{(3)} \)$, $\Delta_{i} ^n \bX^{\text{trunc2}} = \( \Delta_{i} ^n X_j^{o}\, \1 \(| \Delta_{i} ^n X_j^{o}| \leq v_{j,n}^{(3)}\)   \)_{j=1,\ldots, p}$, and the regularization parameter $\eta^{\lasso} \in [10^{-6}, 10^6]$ was selected by minimizing the corresponding BIC.
{We choose
\begin{equation*}
 w_n^{(3)} = c_J^{(3)} n^{-0.47} \sqrt{BV^Y}  \quad \text{and} \quad  v_{j,n}^{(3)} = c_J^{(3)} n^{-0.47} \sqrt{BV_j},
\end{equation*}
where $c_J^{(3)}$ is a tuning parameter}, and $BV^Y$ and $BV_j$ are the bipower variations defined by
\begin{equation*}
BV^Y = \dfrac{\pi}{2}\sum_{i=2}^{n} | \Delta_{i-1} ^n Y^{o}| \cdot | \Delta_{i} ^n Y^{o}| \quad \text{and} \quad BV_j = \dfrac{\pi}{2}\sum_{i=2}^{n} | \Delta_{i-1} ^n X_j^{o}| \cdot | \Delta_{i} ^n X_j^{o}|. 
\end{equation*}
This choice of truncation parameters is commonly used in the literature \citep{ait2020high, ait2019principal}.
The LASSO estimator can handle high dimensionality; however, it cannot account for microstructure noise, the factor structure, or time variation in the coefficient process.
{Table \ref{Table3} summarizes the tuning parameters used in FATEN-LASSO and the benchmark estimators. 
For each estimator, we distinguish between the tuning parameters selected by in-sample data-driven criteria and those selected by the out-of-sample $R^2$ criterion.}

\begin{table}[!ht]
{\centering
\caption{{Tuning parameters for FATEN-LASSO and the benchmark estimators. In the empirical study, we set $k_2=390$ (one day).}}
\label{Table3}
\footnotesize
\renewcommand{\arraystretch}{1.4}
\begin{tabular}{lll}
\hline
Estimator & Selected by in-sample criteria & Selected by out-of-sample $R^2$ criterion \\
\hline
FATEN-LASSO 
& $c_{\eta}, c_{\tau}, c_h, r$ 
& $c_{k_1}, c_{k_2}, c_J, c_J^{(2)}, \rho, c_{\alpha_2}$ \\
FATEC-LASSO 
& $c_{\eta}, c_{\tau}, c_h, r$ 
& $c_{k_1}, c_{k_2}, c_J, c_J^{(2)}$ \\
FATN-LASSO 
& $c_{\eta}, c_h, r$ 
& $c_{k_1}, c_{k_2}, c_J, c_J^{(2)}, \rho, c_{\alpha_2}$ \\
FAEN-LASSO 
& $c_{\eta}, c_{\tau}, r$ 
& $c_{k_1}, c_{k_2}, c_J, c_J^{(2)}, \rho, c_{\alpha_2}$ \\
TEN-LASSO 
& $c_{\eta}, c_{\tau}, c_h$ 
& $c_{k_1}, c_{k_2}, c_J, c_J^{(2)}, \rho, c_{\alpha_2}$ \\
TV-RIDGE 
& $\eta^{\text{RIDGE}}$ 
& $c_{k_1}, c_{k_2}, c_J$ \\
TV-PCR 
& $q$ 
& $c_{k_1}, c_{k_2}, c_J$ \\
TV-ENET 
& $c_{\eta}, \alpha$ 
& $c_{k_1}, c_{k_2}, c_J$ \\
TED 
& $c_{\lambda}, c_{\tau}, c_h$ 
& $c_{k}, c_J^{(3)}$ \\
SV-LASSO 
& $c_{\eta}$ 
& $c_{k_1}, c_J$ \\
LASSO 
& $\eta^{\text{LASSO}}$ 
& $c_J^{(3)}$ \\
\hline
\end{tabular}
}
\end{table}

\section{Simulation setup}\label{simulation-setup}
We considered the following factor-based regression jump diffusion models in \eqref{model-1}--\eqref{model-3}.
\begin{eqnarray*}
&& dY(t)= \bbeta ^{\top}(t)  d\bX^c(t)+d Z(t)+J^Y(t) d \Lambda^Y(t), \quad d\bX(t)=  d \bX^c(t) + d \bX^J(t), \cr
&&  d\bX^c(t)= \bB(t)d\bff(t)+d\bu(t), \quad  d \bX^J(t) =  \bJ(t) d \bLambda(t),  \quad  d\bff(t) = \bnu_{f}(t) d\bW_f(t), \cr
&& d\bu(t) = \bnu_{u}(t) d\bW_{u}(t), \quad \text{and} \quad    dZ(t)= \nu_z(t) d W_{z}(t),
\end{eqnarray*}
where the jump sizes $J^Y(t)$ and $J_j(t)$ were generated from the independent normal distribution with a mean of zero and a standard deviation of $0.01$, and $\Lambda^Y(t)$ and $\Lambda_{j}(t)$ were generated using Poisson processes with intensities  of  $20$  and  $\(15, \ldots, 15\)^{\top}$, respectively.
The initial values $Y(0)$ and $X_j(0)$ were set as zero, and $\nu_z(t)$ was generated from the following Ornstein--Uhlenbeck process:
\begin{equation*}
	d \nu_z(t)= 4\(0.15-\nu_z(t)\)dt + 0.03 d W_{z}(t),
\end{equation*}
where $\nu_z(0)=0.15$ and $W_{z}(t)$ is an independent Brownian motion.
For the volatility processes $\bnu_{f}(t)$ and $\bnu_{u}(t)$, we first generated the following Ornstein--Uhlenbeck processes:
\begin{eqnarray*}
	&& d \xi_f(t)= 3\(0.25-\xi_f(t)\)dt + 0.05 d W_{f}^{\xi}(t), \cr
	&& d \xi_u(t)= 5\(0.5-\xi_u(t)\)dt + 0.1 d W_{u}^{\xi}(t),
\end{eqnarray*}
where $\xi_f(0)=0.25$, $\xi_u(0)=0.45$, and $W_{f}^{\xi}(t)$ and $W_{u}^{\xi}(t)$ are $r$-dimensional and $p$-dimensional independent Brownian motions, respectively.
Then, $\bnu_{f}(t)$ and $\bnu_{u}(t)$ were taken to be the Cholesky decompositions  of $\bSigma_f(t)=(\Sigma_{f,ij}(t))_{1 \leq i,j \leq r}$ and $\bSigma_u(t)=(\Sigma_{u,ij}(t))_{1 \leq i,j \leq p}$, respectively, where $\Sigma_{f,ij}(t)=\xi_f(t) 0.5^{|i-j|}$ and $\Sigma_{u,ij}(t)=0.25 \xi_u(t) \bI_p$.
For the coefficient process $\bbeta(t)$, we considered the following model:
\begin{equation*}
d \bbeta(t)= \bmu_{\beta}(t) dt + \bnu_{\beta}(t) d \bW_{\beta}(t),
\end{equation*}
where $\bmu_{\beta}(t)=\(\mu_{\beta,1}(t), \ldots, \mu_{\beta,p}(t)\)^{\top}$  is a drift process,  $\bnu_{\beta}(t)=\(\nu_{\beta,ij}(t)\)_{1 \leq i,j \leq p}$ is an instantaneous volatility matrix process, and $\bW_{\beta}(t)$ is a $p$-dimensional independent Brownian motion.
To generate $\bnu_{\beta}(t)$, we first generated the following Ornstein--Uhlenbeck process:
\begin{equation*}
	d \varphi_{\beta}(t)= 3\(0.1-\varphi_{\beta}(t)\)dt + 0.02 d W_{\beta}^{\varphi}(t),
\end{equation*}
where $\varphi_{\beta}(0)=0.1$ and $W_{\beta}^{\varphi}(t)$ is an independent Brownian motion.
Then, we set $\(\nu_{\beta,ij}(t)\)_{1 \leq i,j \leq  s_p }$ as $\varphi_{\beta}(t) \bI_{ s_p }$,  where $\bI_{ s_p }$ is the $ s_p$-dimensional identity matrix.
For  $j=1,\ldots, s_p$, we took $\beta_{j}(0) =1$ and $\mu_{j,\beta}(t)= 0.05$ for $0 \leq t \leq 1$, whereas we set $\beta_{j}(t)=0$ for $j= s_p +1, \ldots, p$.
For the factor loading matrix process $\bB(t)$, we employed the following model:
\begin{equation*}
d \vec\(\bB(t)\)= \bmu_{B}(t) dt + \bnu_{B}(t) d \bW_{B}(t),
\end{equation*}
where $\bmu_{B}(t)=\(\mu_{B,1}(t), \ldots, \mu_{B,pr}(t)\)^{\top}$  is a drift process,  $\bnu_{B}(t)=\(\nu_{B,ij}(t)\)_{1 \leq i,j \leq pr}$ is an instantaneous volatility matrix process, and $\bW_{B}(t)$ is a $pr$-dimensional independent Brownian motion.
To obtain $\bnu_{B}(t)$, we first employed the following Ornstein--Uhlenbeck process:
\begin{equation*}
	d \varphi_{B}(t)= 3\(0.07-\varphi_{B}(t)\)dt + 0.02 d W_{B}^{\varphi}(t),
\end{equation*}
where $\varphi_{B}(0)=0.07$ and $W_{B}^{\varphi}(t)$ is an independent Brownian motion.
Then, we set $\(\nu_{B,ij}(t)\)_{1 \leq i,j \leq  pr }$ as $\varphi_{B}(t) \bI_{ pr }$,  where $\bI_{pr}$ is the $pr$-dimensional identity matrix.
The initial value $B_{ij}(0)$ is obtained from an independent normal distribution with a mean of zero and a standard deviation of $0.3$, and we set $\mu_{B,ij}(t)= 0.01$ for $1\leq i \leq p$, $1 \leq j \leq r$, and $0 \leq t \leq 1$.

{
\section{Sensitivity analysis for tuning parameters}\label{SEC-sensitivity}
\label{sensitivity}
In this section, we investigate the sensitivity of the FATEN-LASSO estimator with respect to the choice of tuning parameters with $n=23400$.
We vary one tuning parameter at a time while fixing the remaining tuning parameters at the values obtained by the data-driven criterion discussed in Appendix \ref{SEC-Tuning}.
We first consider the tuning parameters selected by the in-sample criterion.
Figure \ref{fig:tuning_is} plots the max, $\ell_1$, and $\ell_2$ norm errors with respect to the choice of $c_\eta$, $c_\tau$, $c_h$, and $r$.
The horizontal dashed line represents the estimation error obtained from the tuning parameter selection procedure discussed in Appendix \ref{SEC-Tuning}.
We note that, because the data-driven tuning parameter is determined by the data, the resulting estimation error does not necessarily coincide with the errors examined in the sensitivity analysis, where one tuning parameter is varied independently of the data-driven selection.
We find that the estimator is relatively stable with respect to $c_\tau$ and $r$.
For the rank $r$, choosing $r=0$, which ignores the factor structure, produces the largest errors, while choosing larger values yields relatively robust results.
The errors are more sensitive to the choices of $c_\eta$ and $c_h$, possibly because these parameters directly affect the regularization and thresholding steps.
The effect of $c_\eta$ is not monotone, possibly because the optimal value of $c_\eta$ can vary with $r$, and the selection of $c_\eta$ can affect the selection of $r$.
Nevertheless, the selected values lead to relatively small errors overall.
We next consider the tuning parameters selected by the out-of-sample $R^2$ criterion.
Figures \ref{fig:tuning_oos_1} and \ref{fig:tuning_oos_2} report the max, $\ell_1$, and $\ell_2$ norm errors with respect to the choice of $c_{k_1}$, $c_{k_2}$, $c_J$, $c_J^{(2)}$, $\rho$, and $c_{\alpha_2}$. 
The red circles indicate the selected tuning values.
We find that, in most cases, the estimation errors are reasonably stable around the selected values.
The performance is more sensitive when $c_{k_1}$ and $c_J$ are chosen too small.
This may be because a small $c_{k_1}$ does not sufficiently reduce the effect of microstructure noise, and a small $c_J$ can remove too much signal from the continuous component.
Despite this sensitivity, the tuning parameters selected by the out-of-sample $R^2$ criterion lead to relatively small estimation errors.
Taken together, the results show that the FATEN-LASSO estimator is reasonably stable over a wide range of values for most tuning parameters.
The results also indicate that $c_\eta$, $c_h$, $c_{k_1}$, and $c_J$ require relatively careful selection.
\begin{figure}[!ht]
\centering
\includegraphics[width=0.84\textwidth]{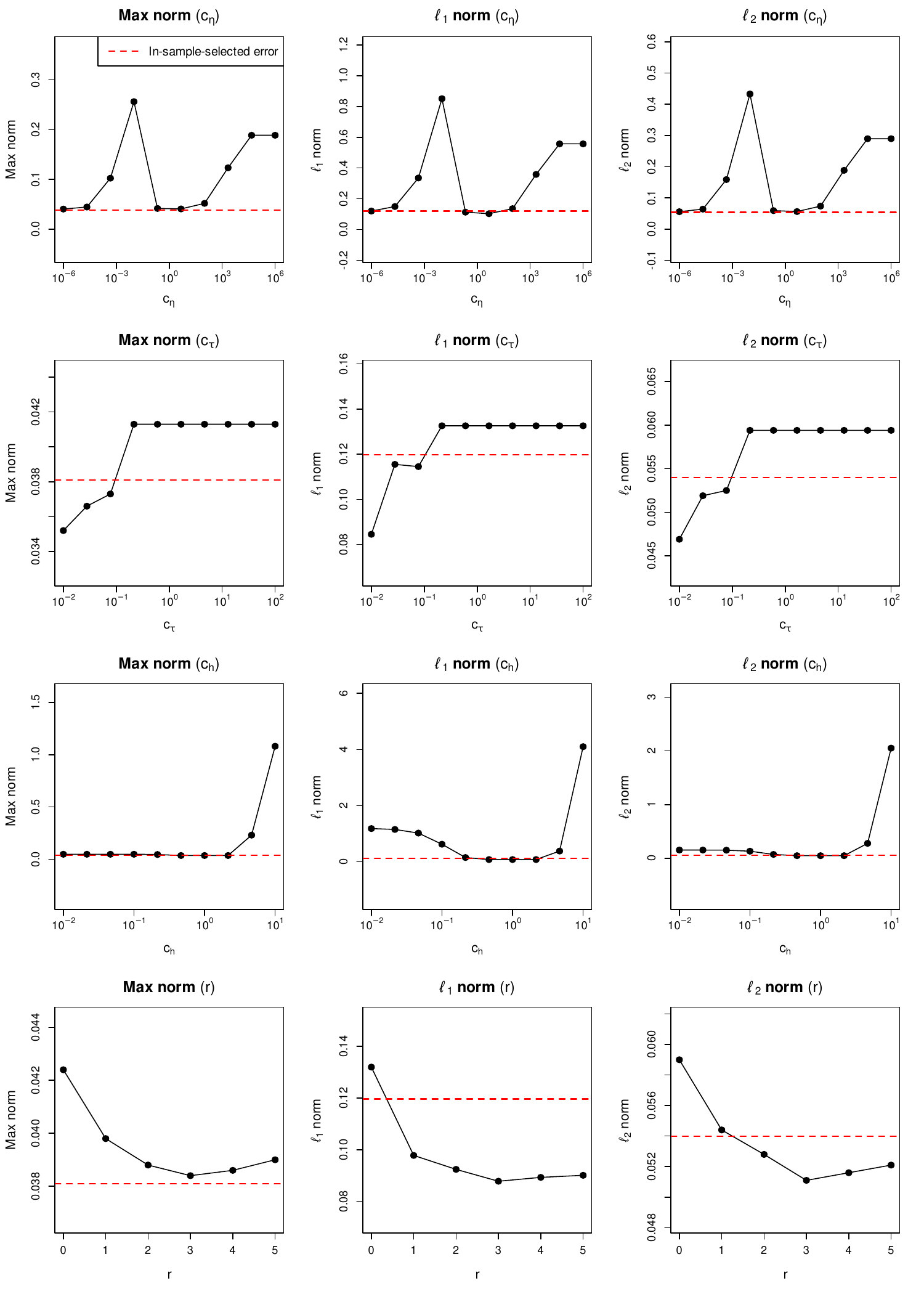}
\caption{
{Sensitivity analysis for $c_\eta$, $c_\tau$, $c_h$, and $r$ selected by the in-sample criterion with $n=23400$.
Each solid line varies one tuning parameter while fixing the others at their selected values.
The horizontal dashed lines indicate the errors from the data-driven tuning parameter selection.
The x-axis is shown on a logarithmic scale for $c_\eta$, $c_\tau$, and $c_h$.}
}
\label{fig:tuning_is}
\end{figure}
\begin{figure}[!ht]
\centering
\includegraphics[width=0.84\textwidth]{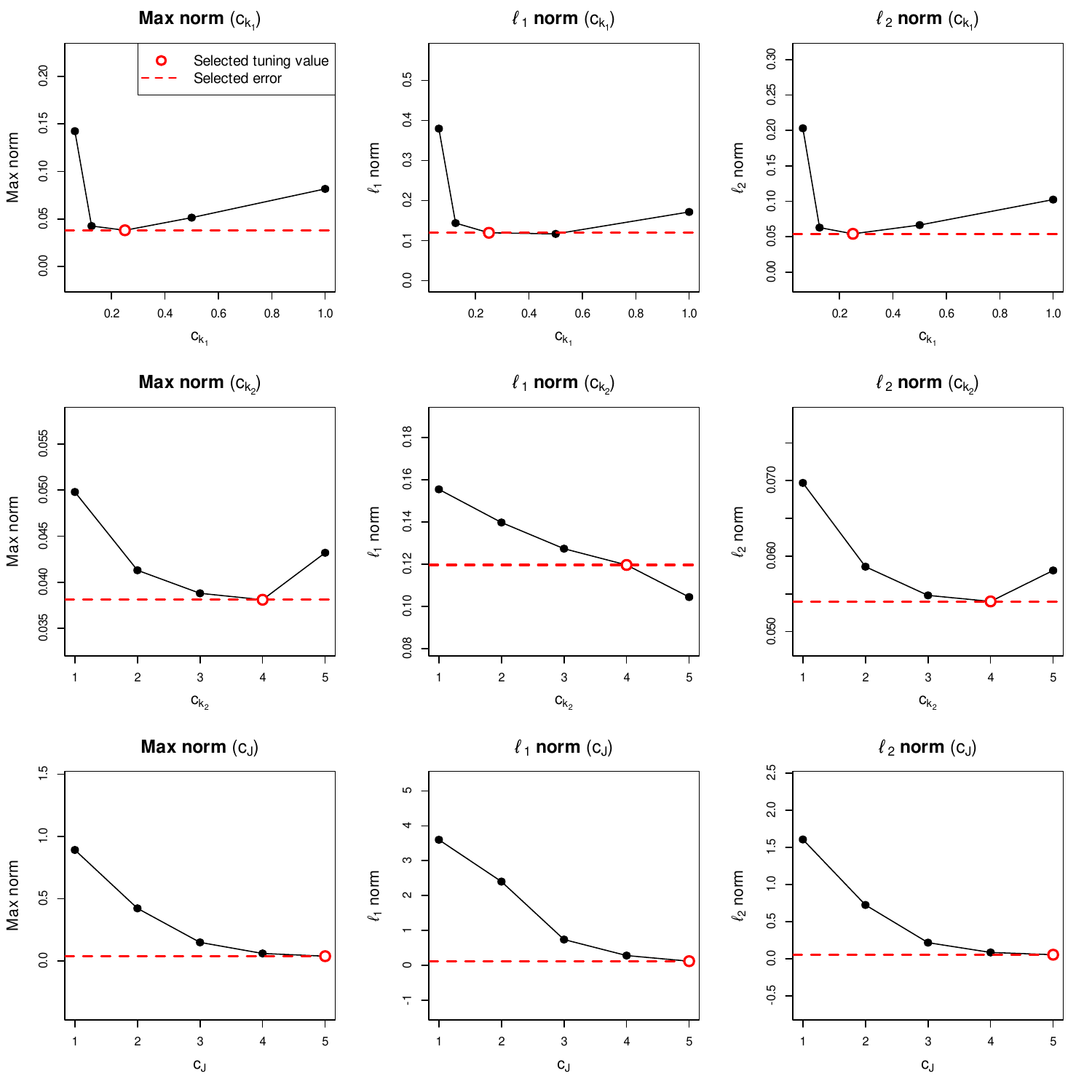}
\caption{
{Sensitivity analysis for $c_{k_1}$, $c_{k_2}$, and $c_J$ selected by the out-of-sample $R^2$ criterion with $n=23400$.
Each solid line varies one tuning parameter while fixing the others at their selected values.
The red circles indicate the selected tuning values, and the horizontal dashed lines indicate the corresponding errors.
}
}
\label{fig:tuning_oos_1}
\end{figure}
\begin{figure}[!ht]
\centering
\includegraphics[width=0.84\textwidth]{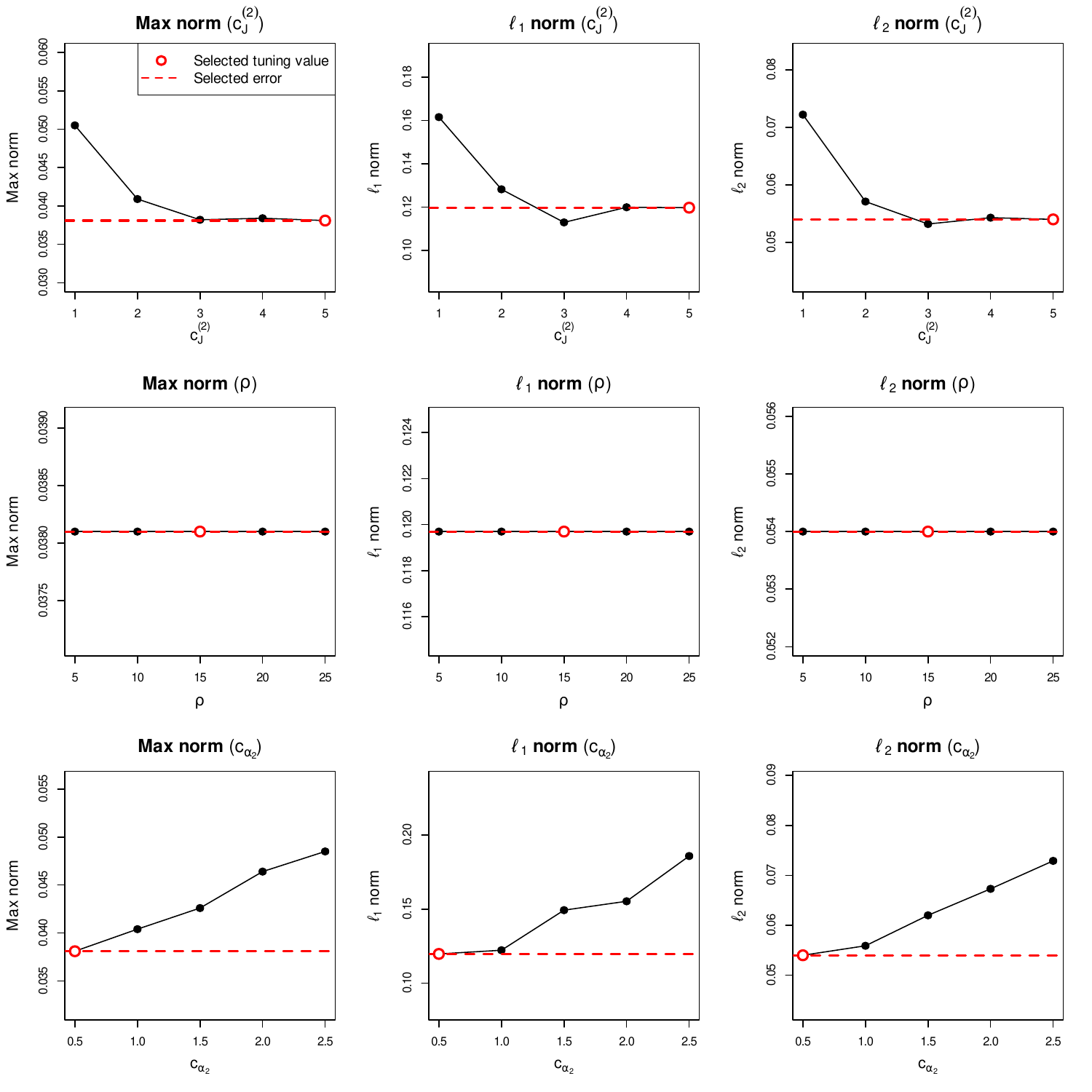}
\caption{
{Sensitivity analysis for $c_J^{(2)}$, $\rho$, and $c_{\alpha_2}$ selected by the out-of-sample $R^2$ criterion with $n=23400$.
Each solid line varies one tuning parameter while fixing the others at their selected values.
The red circles indicate the selected tuning values, and the horizontal dashed lines indicate the corresponding errors.
}
}
\label{fig:tuning_oos_2}
\end{figure}
}

\section{Empirical study for the FATEN-LASSO estimates}\label{empirical-FATEN}
\begin{figure}[ht!]
\centering
\includegraphics[height=1.17 \textwidth, width = 1\textwidth]{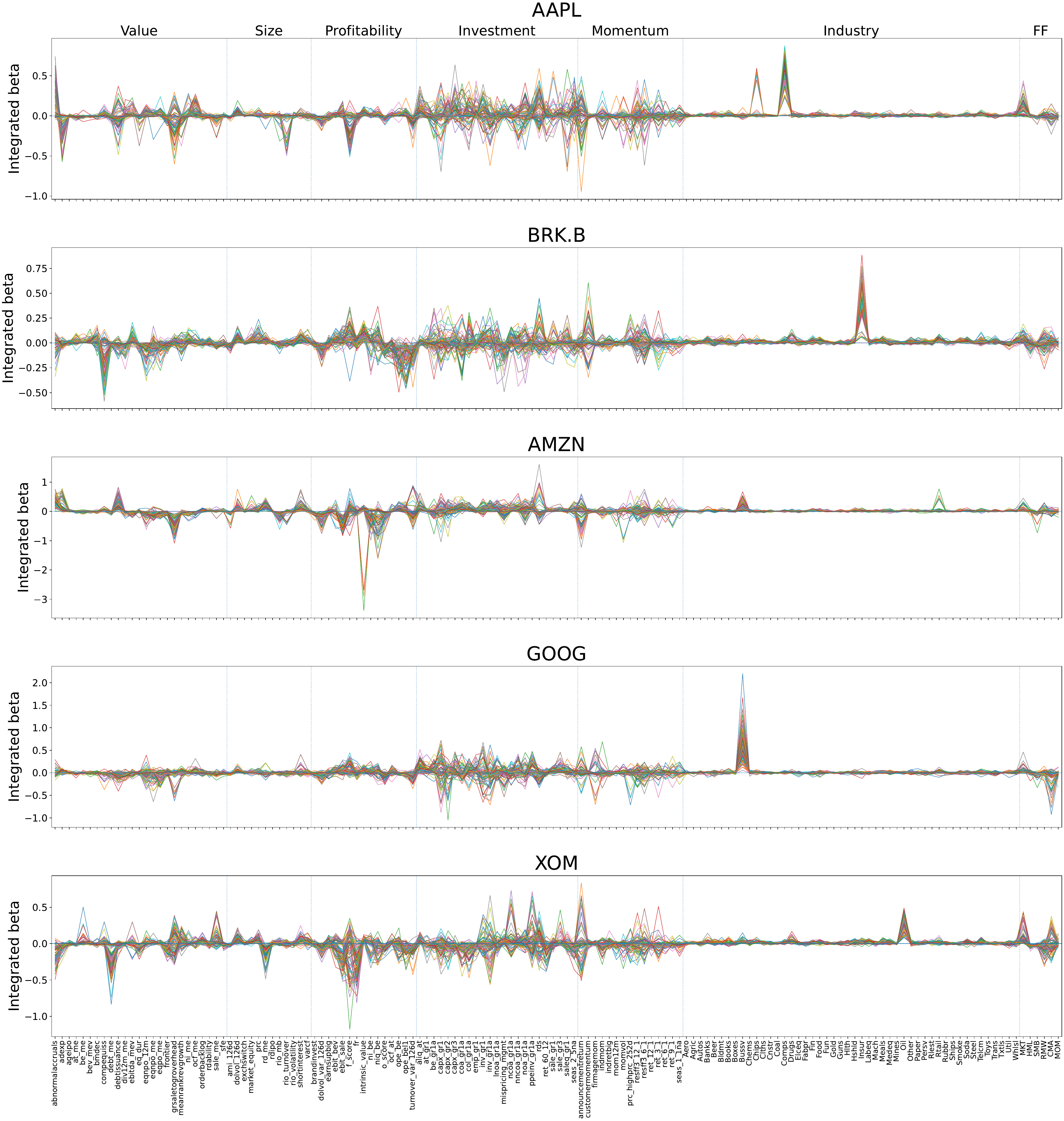}
\caption{{Monthly integrated coefficient estimates from FATEN-LASSO for the five assets over 144 factors. 
In each panel, each line corresponds to one month and traces the 144 integrated coefficient estimates across the factors.
Vertical dashed lines separate the value, size, profitability, investment, momentum, industry, and Fama--French groups.}}
\label{fig:allvalue}
\end{figure}

\begin{figure}[ht!]
\centering
\includegraphics[height=1.17 \textwidth, width = .95\textwidth]{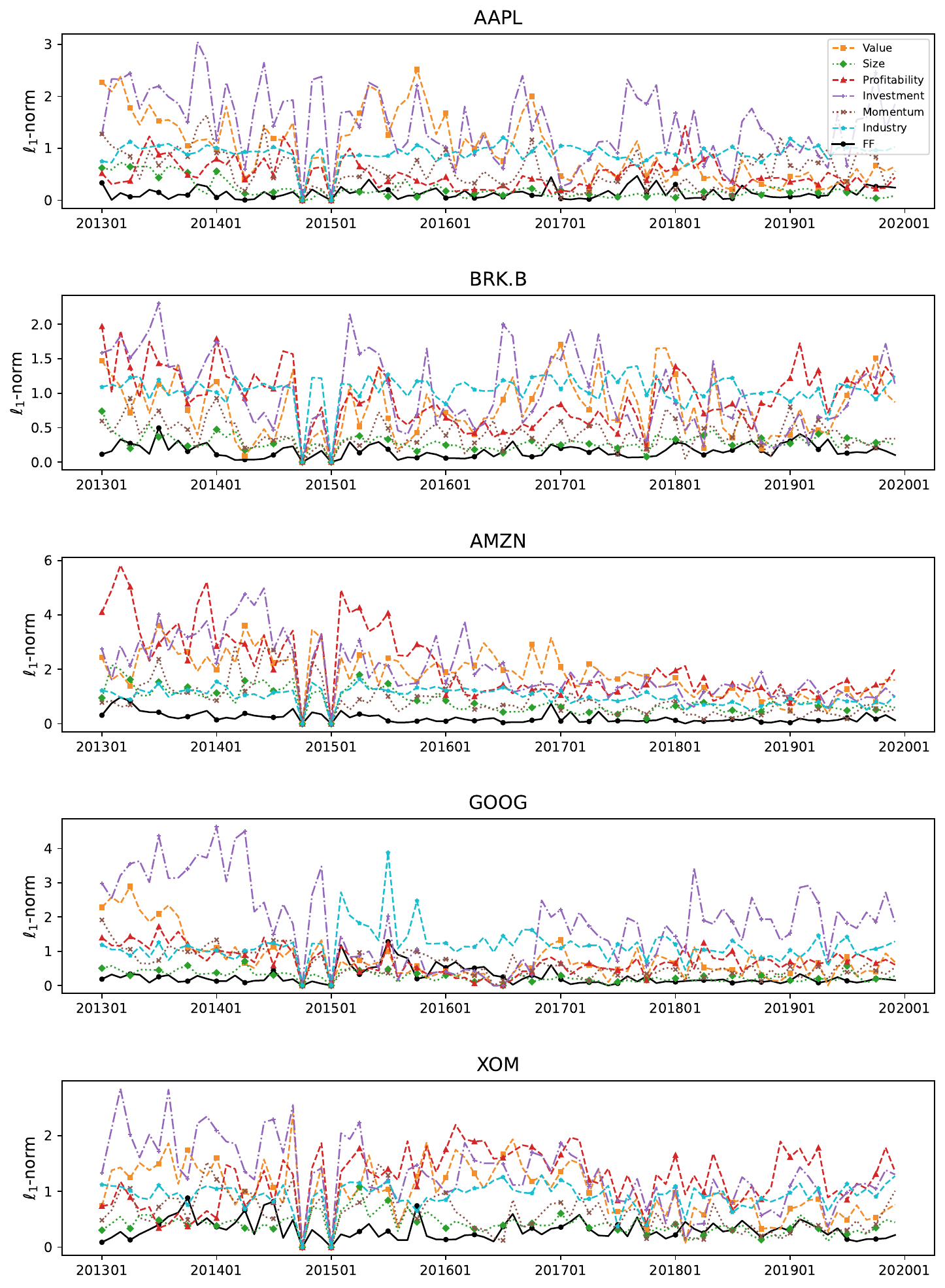}
\caption{{Groupwise $\ell_1$ norms of the monthly integrated coefficient estimates from the FATEN-LASSO procedure  for five assets across seven factor groups: value, size, profitability, investment, momentum, industry, and Fama--French.}}
\label{fig:emp_sparsity}
\end{figure}

\begin{figure}[t]
    \centering
    \includegraphics[width= 1 \textwidth]{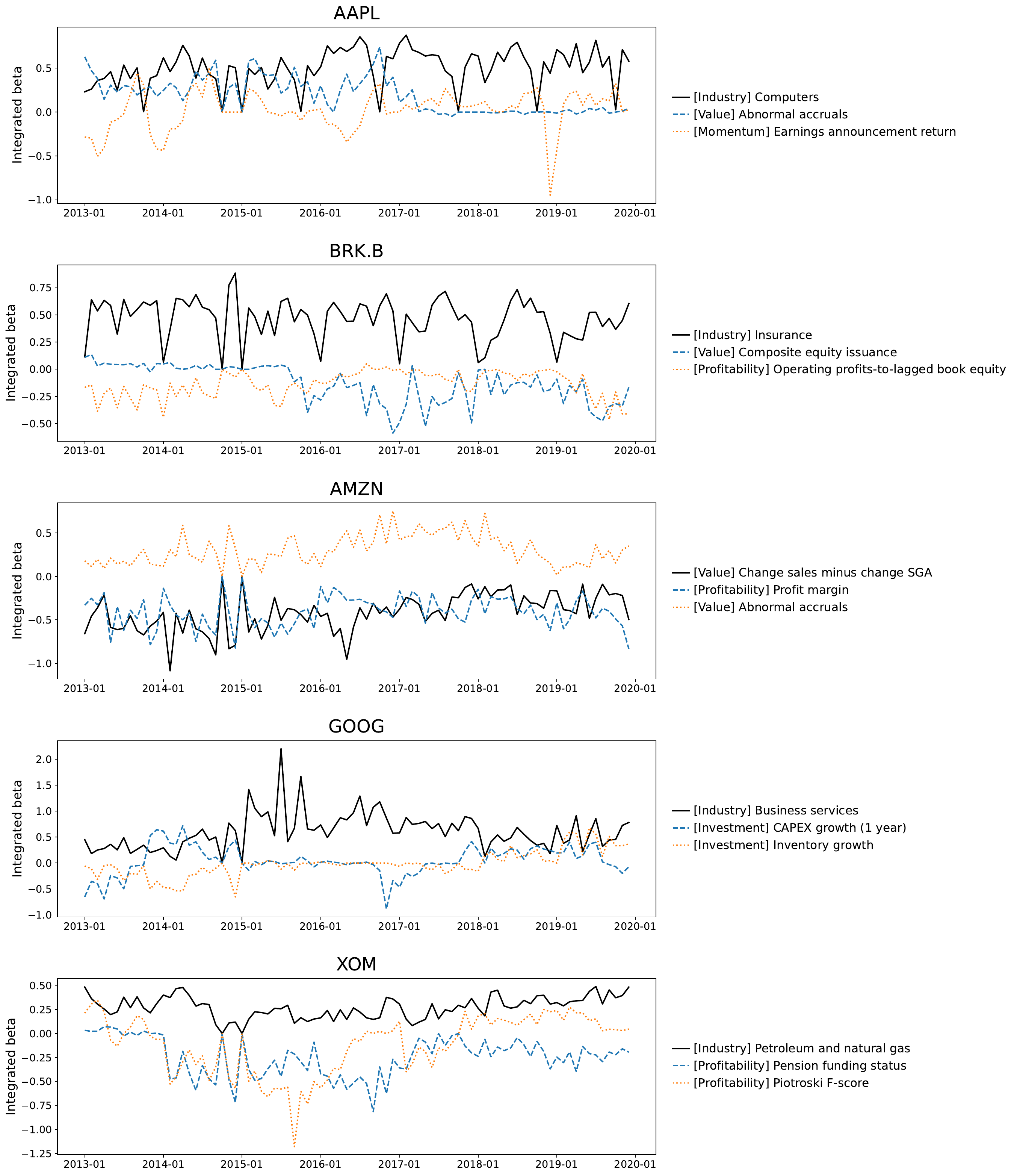}
    \caption{{Monthly integrated coefficient estimates from the FATEN-LASSO procedure for the three factors with the largest absolute sums over time, selected from the 144 factors for each of the five assets.}}
    \label{fig:mainfactors}
\end{figure}

In this section, we explore the integrated coefficient estimates from the FATEN-LASSO procedure.
{Figure \ref{fig:allvalue} plots the monthly integrated coefficients from the FATEN-LASSO estimator for the five assets across the 144 factors. 
Figure \ref{fig:emp_sparsity} reports the monthly groupwise $\ell_1$ norms of the FATEN-LASSO estimates for the seven factor groups: value, size, profitability, investment, momentum, industry, and Fama--French.
Figures \ref{fig:allvalue} and \ref{fig:emp_sparsity} illustrate the time variation of the coefficient process and its sparsity in terms of the groupwise $\ell_1$ norms.
We find that all seven groups exhibit positive $\ell_1$ norms over most periods.
This finding is consistent with the results from multi-factor models \citep{asness2013value, carhart1997persistence, eugene1992cross, fama2015five} and factor zoo analysis \citep{jensen2023there}. 
We note that the Fama--French group has the smallest $\ell_1$ norms overall and also contains the smallest number of factors (six).
To investigate the coefficient dynamics of the most influential factors, Figure \ref{fig:mainfactors} plots, for each asset, the monthly integrated coefficient estimates of the three factors with the largest absolute sums over time.
For AAPL, the three factors are Computers, Abnormal accruals, and Earnings announcement return; for BRK.B, they are Insurance, Composite equity issuance, and Operating profits-to-lagged book equity; for AMZN, they are Change sales minus change SGA, Profit margin, and Abnormal accruals; for GOOG, they are Business services, CAPEX growth (1 year), and Inventory growth; and for XOM, they are Petroleum and natural gas, Pension funding status, and Piotroski F-score. 
Overall, the dominant factors are economically interpretable and vary substantially across assets. 
For example, industry portfolios such as Computers, Insurance, Business services, and Petroleum and natural gas appear as the top factors for AAPL, BRK.B, GOOG, and XOM, respectively, whereas AMZN is characterized more by accounting- and profitability-related factors. 
This may be because sector-specific factors are more important for these four firms, whereas AMZN is more influenced by firm-specific operating and accounting factors.
}



\section{Proofs}\label{SEC-proof}
\subsection{Proof of Theorem \ref{Thm1}}\label{Proof-Thm1}
Without loss of generality, it suffices to show the statement for fixed $i$.
For simplicity, we denote the true $\bbeta\(i\Delta_n\)$, $\bB\(i\Delta_n\)$, $\bgamma\(i\Delta_n\)$, and $\btheta(i\Delta_n)$ by $\bbeta_{0}= (\beta_{0,j})_{j=1, \ldots, p}$, $\bB_{0}= (B_{0,ij})_{1\leq i \leq p, 1 \leq j \leq r}$, $\bgamma_{0}= (\gamma_{0,j})_{j=1, \ldots, r}$, and $\btheta_{0}= (\theta_{0,j})_{j=1, \ldots, p+r}$, respectively.
Let
\begin{eqnarray*}
&& \mathcal{Y}^c_i = 
\begin{pmatrix}
\sum^{k_1-1}_{l=0}g\(\dfrac{l}{k_1}\)\Delta_{i+l+1} ^n Y^c  \\ 
\sum^{k_1-1}_{l=0}g\(\dfrac{l}{k_1}\)\Delta_{i+l+2} ^n Y^c   \\  
 \vdots \\ 
\sum^{k_1-1}_{l=0}g\(\dfrac{l}{k_1}\)\Delta_{i+l+k_2-k_1+1} ^n Y^c  
\end{pmatrix}, \quad
\mathcal{X}^c_i = 
\begin{pmatrix}
\sum^{k_1-1}_{l=0}g\(\dfrac{l}{k_1}\)\Delta_{i+l+1} ^n \bX^{c\top}  \\ 
\sum^{k_1-1}_{l=0}g\(\dfrac{l}{k_1}\)\Delta_{i+l+2} ^n \bX^{c\top}   \\  
 \vdots \\ 
\sum^{k_1-1}_{l=0}g\(\dfrac{l}{k_1}\)\Delta_{i+l+k_2-k_1+1} ^n \bX^{c\top} 
\end{pmatrix},
\cr
&& \tilde{\mathcal{X}}_i = 
\begin{pmatrix}
\sum^{k_1-1}_{l=0}g\(\dfrac{l}{k_1}\)\int_{(i+l) \Delta_n } ^{ (i+l +1)  \Delta_n}   (\bbeta(t)- \bbeta_0)  ^{\top} d\bX^c(t)  \\ 
\sum^{k_1-1}_{l=0}g\(\dfrac{l}{k_1}\)\int_{(i+l+1) \Delta_n } ^{(i+l+2)  \Delta_n}   (\bbeta(t)- \bbeta_0)  ^{\top}  d\bX^c(t) \\  
 \vdots \\ 
\sum^{k_1-1}_{l=0}g\(\dfrac{l}{k_1}\)\int_{(i+l+k_2-k_1) \Delta_n } ^{(i+l+k_2-k_1+1)  \Delta_n}   (\bbeta(t)- \bbeta_0) ^{\top}  d\bX^c(t)
\end{pmatrix}
, \cr
&& \mathcal{Z}_i = 
\begin{pmatrix}
\sum^{k_1-1}_{l=0}g\(\dfrac{l}{k_1}\)\Delta_{i+l+1} ^n Z  \\ 
\sum^{k_1-1}_{l=0}g\(\dfrac{l}{k_1}\)\Delta_{i+l+2} ^n Z   \\  
 \vdots \\ 
\sum^{k_1-1}_{l=0}g\(\dfrac{l}{k_1}\)\Delta_{i+l+k_2-k_1+1} ^n Z 
\end{pmatrix},
\cr
&& \tilde{\bF}_i = 
\begin{pmatrix}
\sum^{k_1-1}_{l=0}g\(\dfrac{l}{k_1}\)\left[\int_{(i+l) \Delta_n } ^{ (i+l +1)  \Delta_n}   (\bB(t)- \bB_0) d\bff(t) \right]^{\top} \\ 
\sum^{k_1-1}_{l=0}g\(\dfrac{l}{k_1}\)\left[\int_{(i+l+1) \Delta_n } ^{(i+l+2)  \Delta_n} (\bB(t)- \bB_0) d\bff(t) \right]^{\top}  \\  
 \vdots \\ 
\sum^{k_1-1}_{l=0}g\(\dfrac{l}{k_1}\)\left[\int_{(i+l+k_2-k_1) \Delta_n } ^{(i+l+k_2-k_1+1)  \Delta_n}   (\bB(t)- \bB_0) d\bff(t) \right]^{\top}
\end{pmatrix}.
\end{eqnarray*}
Then, we have
  \begin{equation*}
 	\mathcal{Y}^c_i  = \mathcal{X}^c_i \bbeta_0 + \mathcal{Z}_i + \mathcal{\tilde{X}}_i \quad \text{and} \quad \mathcal{X}^c_i = \bF_i \bB_{0}^{\top} + \bU_{i} + \tilde{\bF}_i.
 \end{equation*}
Also, let 
\begin{eqnarray*}
&& \mathcal{E}_i^{Y} = 
\begin{pmatrix}
\sum^{k_1-1}_{l=0}g\(\dfrac{l}{k_1}\)\Delta_{i+l+1}^n \epsilon^{Y}   \\ 
\sum^{k_1-1}_{l=0}g\(\dfrac{l}{k_1}\)\Delta_{i+l+2}^n \epsilon^{Y}   \\  
 \vdots \\ 
\sum^{k_1-1}_{l=0}g\(\dfrac{l}{k_1}\)\Delta_{i+l+k_2-k_1+1}^n \epsilon^{Y}
\end{pmatrix}
, \quad
\mathcal{E}_i^{X} = 
\begin{pmatrix}
\sum^{k_1-1}_{l=0}g\(\dfrac{l}{k_1}\) \Delta_{i+l+1}^n \(\bepsilon^{X}\)^{\top}  \\ 
\sum^{k_1-1}_{l=0}g\(\dfrac{l}{k_1}\) \Delta_{i+l+2}^n \(\bepsilon^{X}\)^{\top}    \\  
 \vdots \\ 
\sum^{k_1-1}_{l=0}g\(\dfrac{l}{k_1}\) \Delta_{i+l+k_2-k_1+1}^n \(\bepsilon^{X}\)^{\top}
\end{pmatrix}, \cr
&& \mathcal{Y}_i^{'} = 
\begin{pmatrix}
\sum^{k_1-1}_{l=0}g\(\dfrac{l}{k_1}\)\left\{ \Delta_{i+l+1}^n Y^{c} +  \Delta_{i+l+1}^n \epsilon^{Y}  \right\} \, \1 \(| \Delta_i^n\hat{Y}| \leq w_n \) \\ 
\sum^{k_1-1}_{l=0}g\(\dfrac{l}{k_1}\)\left\{ \Delta_{i+l+2}^n Y^{c} + \Delta_{i+l+2}^n \epsilon^{Y} \right\} \, \1 \(| \Delta_{i+1}^n\hat{Y}| \leq w_n \) \\  
 \vdots \\ 
 \sum^{k_1-1}_{l=0}g\(\dfrac{l}{k_1}\)\left\{ \Delta_{i+l+k_2-k_1+1}^n Y^{c} + \Delta_{i+l+k_2-k_1+1}^n \epsilon^{Y} \right\} \, \1 \(| \Delta_{i+k_2-k_1}^n\hat{Y} | \leq w_n \)
\end{pmatrix}, \cr
&& \mathcal{X}_i^{'} = 
\begin{pmatrix}
\left(\Delta_i^n\tilde{\bX}^{\text{trunc}}\right)^{\top} \\ 
\left(\Delta_{i+1}^n\tilde{\bX}^{\text{trunc}}\right)^{\top}  \\  
 \vdots \\ 
\left(\Delta_{i+k_2-k_1}^n\tilde{\bX}^{\text{trunc}}\right)^{\top}
\end{pmatrix},
\end{eqnarray*}
where 
\begin{equation*}
\Delta_i^n\tilde{\bX}^{\text{trunc}}=\left(\sum^{k_1-1}_{l=0}g\(\dfrac{l}{k_1}\)\left\{ \Delta_{i+l+1} ^n  X^{c}_{j} + \Delta_{i+l+1} ^n  \epsilon^{X}_{j} \right\} \, \1 \(| \Delta_i^n\hat{X}_j| \leq v_{j,n} \)\right)_{j=1, \ldots, p}.
\end{equation*}

\begin{proposition}\label{Prop2}
Under the assumptions in Theorem \ref{Thm1}, with the probability at least $1-p^{-3-a}$, we have
\begin{eqnarray}
&& \left\| \dfrac{n}{\phi k_1 k_2} \bF_{i}^{\top} \bF_{i} - \bSigma_{0,f}(i \Delta_n)  \right\|_{\max}  \leq C n^{-1/8}\sqrt{\log p}, \label{Prop2-eq1} \\
&& \left\| \dfrac{n}{\phi k_1 k_2}  \bU_{i}^{\top} \bU_{i} -  \bSigma_{0,u}(i \Delta_n)   \right\|_{\max}  \leq C n^{-1/8}\sqrt{\log p}, \label{Prop2-eq2} \\
&& \left\| \dfrac{n}{\phi k_1 k_2}  \(\mathcal{E}_i^{X}\)^{\top} \mathcal{E}_i^{X} - \dfrac{n \zeta}{\phi k_1^2} \bV^X   \right\|_{\max}  \leq C n^{-1/8}\sqrt{\log p}, \label{Prop2-eq3} \\
&& \left\| \dfrac{n}{\phi k_1 k_2} \bF_{i}^{\top} \bU_{i} \right\|_{\max} \leq C n^{-1/8}\sqrt{\log p}, \label{Prop2-eq4}  \\
&& \left\| \dfrac{n}{\phi k_1 k_2} \bF_{i}^{\top} \mathcal{E}_i^{X} \right\|_{\max} \leq C n^{-1/8}\sqrt{\log p}, \label{Prop2-eq5}  \\
&& \left\| \dfrac{n}{\phi k_1 k_2} \bU_{i}^{\top} \mathcal{E}_i^{X} \right\|_{\max} \leq C n^{-1/8}\sqrt{\log p}, \label{Prop2-eq6}  \\
&& \left\| \dfrac{n}{\phi k_1 k_2} \mathcal{X}_{i}^{\top} \mathcal{X}_{i}  - \dfrac{n}{\phi k_1 k_2} \bB_{0} \bF_{i}^{\top} \bF_{i} \bB_{0}^{\top} -  \bSigma_{0,u}(i \Delta_n) - \dfrac{n \zeta}{\phi k_1^2} \bV^X \right\|_{\max}  \leq C n^{-1/8}\(\log p \)^{3/2}, \label{Prop2-eq7}  \\
&& \left\| \dfrac{n \zeta}{\phi k_1^2} \(\hat{\bV}^X - \bV^X\)  \right\|_{\max} \leq C n^{-1/2}\sqrt{\log p}. \label{Prop2-eq8} 
\end{eqnarray}
\end{proposition}

\textbf{Proof of Proposition \ref{Prop2}.}
Consider \eqref{Prop2-eq1}. 
Without loss of generality, we assume that $k_2 = k_1(L+1) - 1$ for some $L  \in \mathbb{N}$.
Let $\Delta_i^n \bar{f}_{j} = \sum^{k_1-1}_{l=0}g\(\dfrac{l}{k_1}\)\Delta_{i+l+1} ^n f_{j}$ and $\bSigma_{0,f}(t) = (\Sigma_{0,f,jm}(t) ) _{j,m=1,\ldots, p}$.
Define
\begin{equation*}
	  T_{jm}(i\Delta_n) = \sum_{l=0}^{k_1 -1} \left \{ g\left(\frac{l}{k_1}\right)  \right \} ^2 \E \left \{\int_{(i+l)\Delta_n} ^{(i+l+1)\Delta_n} \Sigma_{0,f,jm}(t) dt \middle | i\Delta_n \right \}.
\end{equation*}
Then,  we have
\begin{equation*}
	\E \Big[\Delta_i^n \bar{f}_{j} \Delta_i^n \bar{f}_{m} \Big | \FF_{i\Delta_n} \Big] =  T_{jm}(i\Delta_n)  \text{ a.s.}
\end{equation*}
Also, we have
\begin{eqnarray*}
&& \left| \dfrac{n}{\phi k_1 k_2} \sum_{k=0}^{k_2-k_1}\Delta_{i+k}^n \bar{f}_{j} \Delta_{i+k}^n \bar{f}_{m} -  \int_{i \Delta_n}^{(i+k_2) \Delta_n} \Sigma_{0,f,jm}(t)dt /(k_2 \Delta_n)   \right|  \cr
&& \leq \left| \dfrac{n}{\phi k_1 k_2} \sum_{k=0}^{k_2-k_1} \left[ \Delta_{i+k}^n \bar{f}_{j} \Delta_{i+k}^n \bar{f}_{m} -T_{jm}((i+k)\Delta_n) \right]  \right| \cr
&& \quad + \left| \dfrac{n}{\phi k_1 k_2} \sum_{k=0}^{k_2-k_1} T_{jm}((i+k)\Delta_n) -  \int_{i \Delta_n}^{(i+k_2) \Delta_n} \Sigma_{0,f,jm}(t)dt /(k_2 \Delta_n)   \right|.
\end{eqnarray*}
Consider the first term. We have, for $x=0, \ldots, k_1 -1$,
\begin{eqnarray*}
&& \left| \dfrac{n}{\phi k_1 k_2} \sum_{k=0}^{k_2-k_1} \left[ \Delta_{i+k}^n \bar{f}_{j} \Delta_{i+k}^n \bar{f}_{m} -T_{jm}((i+k)\Delta_n) \right]  \right| \cr
&& \leq \sum_{x=0}^{k_1 -1} \left| \dfrac{n}{\phi k_1 k_2} \sum_{k=0}^{L} \left[ \Delta_{i+k_1 k +x}^n \bar{f}_{j} \Delta_{i+k_1 k +x}^n \bar{f}_{m} -T_{jm}((i+k_1 k +x)\Delta_n) \right]  \right|.
\end{eqnarray*}
Note that $\Delta_i^n \bar{f}_{j}$ has the sub-Gaussian tail with the order of $n^{-1/4}$. 
Thus, by Bernstein’s inequality for martingales, we have
\begin{eqnarray*}
&&	\Pr \left\{ \left| \dfrac{n}{\phi k_1 k_2} \sum_{k=0}^{L} \left[ \Delta_{i+k_1 k +x}^n \bar{f}_{j} \Delta_{i+k_1 k +x}^n \bar{f}_{m} -T_{jm}((i+k_1 k +x)\Delta_n) \right]  \right| \leq Cn^{-5/8}\sqrt{\log p} \right\} \cr
&& \geq 1-p^{-6-a-1/c_1}.
\end{eqnarray*}
Then, by Assumption \ref{assumption1}(h), we have, for large $n$,
\begin{equation*}
	\Pr \left\{ \left| \dfrac{n}{\phi k_1 k_2} \sum_{k=0}^{k_2-k_1} \left[ \Delta_{i+k}^n \bar{f}_{j} \Delta_{i+k}^n \bar{f}_{m} -T_{jm}((i+k)\Delta_n) \right]  \right|  \leq Cn^{-1/8}\sqrt{\log p} \right\} \geq 1-p^{-6-a}.
\end{equation*}
Consider the second term. By the boundedness of $\Sigma_{0,f,jm}(t)$, we have
\begin{eqnarray*}
&& \left| \dfrac{n}{\phi k_1 k_2} \sum_{k=0}^{k_2-k_1} T_{jm}((i+k)\Delta_n) -  \int_{i \Delta_n}^{(i+k_2) \Delta_n} \Sigma_{0,f,jm}(t)dt /(k_2 \Delta_n)   \right| \cr
&& = \left| \dfrac{n}{\phi k_1 k_2} \sum_{l=0}^{k_1 -1} \left \{ g\left(\frac{l}{k_1}\right)  \right \} ^2 \left(\sum_{k=0}^{k_2-k_1} \left[   \E \left \{\int_{(i+k+l)\Delta_n} ^{(i+k+l+1)\Delta_n} \Sigma_{0,f,jm}(t) dt \middle | \(i+k\)\Delta_n \right \} \right]  \right. \right. \\
&& \left. \left. \qquad \qquad \qquad \qquad \qquad \qquad \qquad \qquad \qquad \qquad \qquad \quad- \int_{i \Delta_n}^{(i+k_2) \Delta_n} \Sigma_{0,f,jm}(t)dt \right)  \right| \cr
&& \leq \left| \dfrac{n}{\phi k_1 k_2} \sum_{l=0}^{k_1 -1} \left \{ g\left(\frac{l}{k_1}\right)  \right \} ^2 \sum_{k=0}^{k_2-k_1} \left[   \E \left \{\int_{(i+k+l)\Delta_n} ^{(i+k+l+1)\Delta_n} \Sigma_{0,f,jm}(t) dt \middle | \(i+k\)\Delta_n \right \}  \right. \right. \\
&& \left. \left. \qquad \qquad \qquad \qquad \qquad \qquad \qquad \qquad \qquad \qquad  - \int_{(i+k+l)\Delta_n} ^{(i+k+l+1)\Delta_n} \Sigma_{0,f,jm}(t)dt \right]  \right| +Cn^{-1/4}.
\end{eqnarray*}
Note that $\sum_{k=0}^{k_2-k_1} \left[   \E \left \{\int_{(i+k+l)\Delta_n} ^{(i+k+l+1)\Delta_n} \Sigma_{0,f,jm}(t) dt \middle | \(i+k\)\Delta_n \right \} - \int_{(i+k+l)\Delta_n} ^{(i+k+l+1)\Delta_n} \Sigma_{0,f,jm}(t)dt \right] $ is the sum of $l+1$ martingales.
Hence, using the Azuma-Hoeffding inequality for each martingale, we can show for all $0\leq l \leq k_1-1$,
\begin{eqnarray*}
&&	\Pr \Bigg\{ \left | \sum_{k=0}^{k_2-k_1} \left[   \E \left \{\int_{(i+k+l)\Delta_n} ^{(i+k+l+1)\Delta_n} \Sigma_{0,f,jm}(t) dt \middle | \(i+k\)\Delta_n \right \} - \int_{(i+k+l)\Delta_n} ^{(i+k+l+1)\Delta_n} \Sigma_{0,f,jm}(t)dt \right] \right | \cr
&& \qquad \qquad \qquad \qquad \qquad \qquad \qquad \qquad \qquad \qquad \qquad \qquad  \qquad \qquad \quad \leq C n^{-3/8}\sqrt{\log p}  \Bigg\}  \cr
&& \geq 1-p^{-6-a-1/c_1},
\end{eqnarray*}
which implies
\begin{equation*}
	\Pr \Bigg\{ \left| \dfrac{n}{\phi k_1 k_2} \sum_{k=0}^{k_2-k_1} T_{jm}((i+k)\Delta_n) -  \int_{i \Delta_n}^{(i+k_2) \Delta_n} \Sigma_{0,f,jm}(t)dt /(k_2 \Delta_n)   \right| \leq Cn^{-1/8}\sqrt{\log p} \Bigg\} \geq 1-p^{-6-a}
\end{equation*}
for large $n$.
Thus, we have
\begin{equation*}
\Pr \left\{ \left\| \dfrac{n}{\phi k_1 k_2} \bF_{i}^{\top} \bF_{i} - \int_{i \Delta_n}^{(i+k_2) \Delta_n} \bSigma_{0,f}(t)dt /(k_2 \Delta_n)  \right\|_{\max}  \leq C n^{-1/8}\sqrt{\log p} \right\} \geq 1-2p^{-4-a}.
\end{equation*}
Then, by Assumption \ref{assumption1}(g), we have
\begin{equation}\label{Prop2-eq9}
\Pr \left\{ \left\| \dfrac{n}{\phi k_1 k_2} \bF_{i}^{\top} \bF_{i} - \bSigma_{0,f}(i \Delta_n)  \right\|_{\max}  \leq C n^{-1/8}\sqrt{\log p} \right\} \geq 1-2p^{-4-a}.
\end{equation}
{Consider \eqref{Prop2-eq3}. For $j=1,\ldots,p$, let
\begin{equation*}
\Delta_i^n \bar{\epsilon}_{j}
=
\sum_{l=0}^{k_1-1} g\left(\frac{l}{k_1}\right)\Delta_{i+l+1}^n \epsilon_j^X .
\end{equation*}
Then, we have
\begin{equation*}
\E\Big[\Delta_i^n \bar{\epsilon}_{j}\Delta_i^n \bar{\epsilon}_{m}\Big]
=
\sum_{l=0}^{k_1-1}
\left\{
g\left(\frac{l+1}{k_1}\right)-g\left(\frac{l}{k_1}\right)
\right\}^2
V_{jm}^X
=
\frac{\zeta}{k_1}V_{jm}^X .
\end{equation*}
Therefore,
\begin{eqnarray*}
&& \left|
\frac{n}{\phi k_1 k_2}\sum_{k=0}^{k_2-k_1}
\Delta_{i+k}^n \bar{\epsilon}_{j}\Delta_{i+k}^n \bar{\epsilon}_{m}
-
\frac{n\zeta}{\phi k_1^2}V_{jm}^X
\right| \\
&& \leq
\left|
\frac{n}{\phi k_1 k_2}\sum_{k=0}^{k_2-k_1}
\left[
\Delta_{i+k}^n \bar{\epsilon}_{j}\Delta_{i+k}^n \bar{\epsilon}_{m}
-
\frac{\zeta}{k_1}V_{jm}^X
\right]
\right|
+
V_{jm}^X \left|
\frac{n\zeta}{\phi k_1^2}
\left(
\frac{k_2-k_1+1}{k_2}-1
\right)
\right|.
\end{eqnarray*}
By Assumption \ref{assumption1}(e), the second term is bounded by $Cn^{-1/4}$.
For the first term, note that for each $x=0,\ldots,k_1-1$, the variables
\begin{equation*}
\Delta_{i+k_1k+x}^n \bar{\epsilon}_{j}\Delta_{i+k_1k+x}^n \bar{\epsilon}_{m}
-
\frac{\zeta}{k_1}V_{jm}^X,
\quad k=0,\ldots,L,
\end{equation*}
are independent.
Also, $\Delta_i^n \bar{\epsilon}_{j}$ has a sub-Gaussian tail with the order of $n^{-1/4}$. 
Thus, by Bernstein’s inequality, we have for each $x=0,\ldots,k_1-1$,
\begin{equation*}
\Pr\left\{
\left|
\frac{n}{\phi k_1 k_2}\sum_{k=0}^{L}
\left[
\Delta_{i+k_1k+x}^n \bar{\epsilon}_{j}\Delta_{i+k_1k+x}^n \bar{\epsilon}_{m}
-
\frac{\zeta}{k_1}V_{jm}^X
\right]
\right|
\leq
Cn^{-5/8}\sqrt{\log p}
\right\}
\geq 1-p^{-6-a-1/c_1},
\end{equation*}
which implies
\begin{equation*}
\Pr\left\{
\left|
\frac{n}{\phi k_1 k_2}\sum_{k=0}^{k_2-k_1}
\left[
\Delta_{i+k}^n \bar{\epsilon}_{j}\Delta_{i+k}^n \bar{\epsilon}_{m}
-
\frac{\zeta}{k_1}V_{jm}^X
\right]
\right|
\leq
Cn^{-1/8}\sqrt{\log p}
\right\}
\geq 1-p^{-6-a}
\end{equation*}
and
\begin{equation*}
\Pr\left\{
 \left\| \dfrac{n}{\phi k_1 k_2}  \(\mathcal{E}_i^{X}\)^{\top} \mathcal{E}_i^{X} - \dfrac{n \zeta}{\phi k_1^2} \bV^X   \right\|_{\max}  \leq C n^{-1/8}\sqrt{\log p} 
\right\} \geq 1-p^{-4-a}.
\end{equation*}
Similarly, we can show that \eqref{Prop2-eq2} and \eqref{Prop2-eq4}--\eqref{Prop2-eq6} hold with the probability at least $1-p^{-4-a}$.}

Now, it is enough to show that \eqref{Prop2-eq7} and \eqref{Prop2-eq8} hold with the probability at least $1-Cp^{-4-a}$.
Consider \eqref{Prop2-eq7}. 
We have
\begin{eqnarray}\label{Prop2-eq10}
&& \left\| \dfrac{n}{\phi k_1 k_2} \mathcal{X}_{i}^{\top} \mathcal{X}_{i}  - \dfrac{n}{\phi k_1 k_2} \bB_{0} \bF_{i}^{\top} \bF_{i} \bB_{0}^{\top} -  \bSigma_{0,u}(i \Delta_n) - \dfrac{n \zeta}{\phi k_1^2} \bV^X \right\|_{\max}  \cr
&& \leq  \dfrac{n}{\phi k_1 k_2} \left\| \mathcal{X}_{i}^{\top} \mathcal{X}_{i} -\(\mathcal{X}_{i}^{c} + \mathcal{E}_i^{X} \)^{\top} \(\mathcal{X}_{i}^{c} + \mathcal{E}_i^{X}\) \right\|_{\max} \cr
&& \quad + \dfrac{n}{\phi k_1 k_2} \left\| \(\mathcal{X}_{i}^{c} + \mathcal{E}_i^{X} \)^{\top} \(\mathcal{X}_{i}^{c} + \mathcal{E}_i^{X}\)  - \(\bF_i \bB_{0}^{\top} + \bU_{i} + \mathcal{E}_i^{X} \)^{\top} \(\bF_i \bB_{0}^{\top} + \bU_{i} + \mathcal{E}_i^{X} \) \right\|_{\max} \cr
&& \quad + \dfrac{n}{\phi k_1 k_2} \left\| \(\bF_i \bB_{0}^{\top} \)^{\top} \(\bU_{i} + \mathcal{E}_i^{X} \) +   \bU_{i}^{\top} \(\bF_i \bB_{0}^{\top} + \mathcal{E}_i^{X} \)  +  \( \mathcal{E}_i^{X} \)^{\top} \(\bF_i \bB_{0}^{\top} + \bU_{i}\) \right\|_{\max} \cr
&& \quad + \left\| \dfrac{n}{\phi k_1 k_2}  \bU_{i}^{\top} \bU_{i} -  \bSigma_{0,u}(i \Delta_n)   \right\|_{\max} + \left\| \dfrac{n}{\phi k_1 k_2}  \(\mathcal{E}_i^{X}\)^{\top} \mathcal{E}_i^{X} - \dfrac{n \zeta}{\phi k_1^2} \bV^X   \right\|_{\max} \cr
&& = (\uppercase\expandafter{\romannumeral1})+(\uppercase\expandafter{\romannumeral2})+(\uppercase\expandafter{\romannumeral3})+(\uppercase\expandafter{\romannumeral4})+(\uppercase\expandafter{\romannumeral5}).
\end{eqnarray}
For some large constant $C>0$, define 
\begin{eqnarray*}
Q_1&=&\left\{ \max_{i}  \left\| \mathcal{Y}_i^{c} \right\|_{\infty} \leq   C s_p \sqrt{\dfrac{k_1 \log p}{n}} \right\} \cap \left\{  \max_{i}  \left\| \mathcal{E}_i^{Y} \right\|_{\infty} \leq   C \sqrt{\dfrac{\log p}{k_1}} \right\}, \cr
Q_2&=& \left\{ \max_{i}  \left\| \mathcal{X}_i^c \right\|_{\max} \leq   C \sqrt{\dfrac{k_1 \log p}{n}} \right\} \cap \left\{ \max_{i}  \left\| \bF_i \right\|_{\max} \leq   C \sqrt{\dfrac{k_1 \log p}{n}} \right\} \cr 
&&  \cap \left\{  \max_{i}  \left\| \bU_i \right\|_{\max} \leq   C \sqrt{\dfrac{k_1 \log p}{n}} \right\} \cap \left\{  \max_{i}  \left\| \mathcal{Z}_i \right\|_{\max} \leq   C \sqrt{\dfrac{k_1 \log p}{n}} \right\} \cr
&& \cap \left\{  \max_{i}  \left\| \mathcal{E}_i^{X} \right\|_{\max} \leq   C \sqrt{\dfrac{\log p}{k_1}} \right\}, \cr
Q_3&=&\left\{  \int _{0}^{1} d \Lambda^Y(t)  \leq  C \log p \right\} \cap \left\{  \max_{j}  \int _{0} ^{1} d \Lambda_{j}(t)  \leq  C \log p \right\}, \cr
Q_4&=&\left\{ \sum_{i=0}^ {n-k_1} \1 \(|\Delta_{i}^n \hat{Y} | > w_{n} \)  \leq C  k_1 \log p \right\} \cap \left\{ \max_{j}\sum_{i=0}^ {n-k_1} \1 \(|\Delta_{i}^n \hat{X}_j | > v_{j,n} \)  \leq C  k_1 \log p \right\}, \cr
Q_5&=&\left\{ \max_{i}\left\| \mathcal{\tilde{X}}_i \right\|_{\infty}    \leq C s_p n^{-3/8} \log p  \right\} \cap \left\{ \max_{i}\left\| \tilde{\bF}_i  \right\|_{\max}    \leq C n^{-3/8} \log p  \right\}.
\end{eqnarray*}
We note that  the variables related to the dependent process are also considered to avoid repetition in the proof.
From Assumption \ref{assumption1}(a),(b), we can show
\begin{equation*}
\Pr \left(Q_1 \cap Q_2 \right) \geq 1-p^{-4-a}.
\end{equation*}
By the boundedness of the intensity process, we have
\begin{equation*}
\Pr \left(Q_3 \right) \geq 1-p^{-4-a}.
\end{equation*}
Under $Q_1 \cap Q_2 \cap Q_3$, we have, for large $n$, 
\begin{equation*}
\sum_{i=0}^ {n-k_1} \1 \(|\Delta_{i}^n \hat{Y} | > w_{n} \)  \leq C  k_1 \log p \quad \text{and} \quad \max_{j}\sum_{i=0}^ {n-k_1} \1 \(|\Delta_{i}^n \hat{X}_j | > v_{j,n} \)  \leq C  k_1 \log p.
\end{equation*}
Consider $Q_5$. By Assumption \ref{assumption1}(a),(c), the process  $\sum_{j=1}^{p}\left| \beta_j(t)-\beta_{0,j} \right|$ has the sub-Gaussian tail and $\sum_{j=1}^p \sqrt{\Sigma_{\beta, jj}(t)}  \leq C\sum_{j=1}^p |\Sigma_{\beta, jj}(t)|^{\delta/2} \leq Cs_p  $.
Thus, we have
  \begin{eqnarray*}
 && \Pr \left\{  \sup_{t \in [i \Delta_n,  \(i+k_2\) \Delta_n ] }  \sum_{j=1}^{p}\left| \beta_j(t)-\beta_{0,j} \right| \geq  C s_p n^{-1/8}\sqrt{\log p} \right\} \cr
 &&  \leq  C \Pr \left\{   \sum_{j=1}^{p}\left| \beta_j(\(i+k_2\) \Delta_n )-\beta_{0,j} \right|  \geq  C s_p n^{-1/8}\sqrt{\log p} \right\} \cr
 && \leq p^{-6-a-1/c_1}.   
  \end{eqnarray*}
  Let 
\begin{equation*}
E= \left\{ \max_{i} \sup_{t \in [i \Delta_n,  \(i+k_2\) \Delta_n ] }  \sum_{j=1}^{p}\left| \beta_j(t)-\beta_{0,j} \right| \geq  C s_p n^{-1/8}\sqrt{\log p} \right\}.
\end{equation*}
Since
\begin{equation*}
\Pr \left\{E \right\} \leq p^{-5-a}
\end{equation*}
for large $n$, we have
  \begin{eqnarray*}
  && \Pr \left\{\max_{i}\left\| \mathcal{\tilde{X}}_i \right\|_{\infty}    \geq C s_p n^{-3/8} \log p \right\}   \cr
  &&\leq \Pr \left\{\max_{i}\left\| \mathcal{\tilde{X}}_i \right\|_{\infty}    \geq C s_p n^{-3/8} \log p, E^c \right\} + p^{-5-a} \cr 
  && \leq 2p^{-5-a}.
  \end{eqnarray*}
Similarly, we can show
\begin{equation*}
\Pr \left\{\max_{i}\left\| \tilde{\bF}_i  \right\|_{\max}    \geq C n^{-3/8} \log p \right\} \leq 2p^{-5-a},
\end{equation*}
which implies
\begin{equation*}
\Pr \left(Q_5 \right) \geq 1-p^{-4-a}.
\end{equation*}
Thus, we have
\begin{equation}\label{Prop2-eq11}
\Pr \left(Q_1 \cap Q_2 \cap Q_3 \cap Q_4 \cap Q_5 \right) \geq 1-3p^{-4-a}.
\end{equation}
From \eqref{Prop2-eq11}, we have, with the probability at least $1-3p^{-4-a}$,
\begin{eqnarray}
	(\uppercase\expandafter{\romannumeral1}) &\leq& \dfrac{n}{\phi k_1 k_2} \left\| \mathcal{X}_{i}^{\top} \mathcal{X}_{i} -\mathcal{X}_{i}^{'\top} \mathcal{X}_{i}^{'}  \right\|_{\max} + \dfrac{n}{\phi k_1 k_2} \left\| \mathcal{X}_{i}^{'\top} \mathcal{X}_{i}^{'} -\(\mathcal{X}_{i}^{c} + \mathcal{E}_i^{X} \)^{\top} \(\mathcal{X}_{i}^{c} + \mathcal{E}_i^{X}\) \right\|_{\max}  \cr
&\leq&  C n^{-1/4}\(\log p \)^2, \label{Prop2-eq12} \\
(\uppercase\expandafter{\romannumeral2}) &\leq& \dfrac{2n}{\phi k_1 k_2} \left\|  \(\bF_i \bB_{0}^{\top} + \bU_{i} + \mathcal{E}_i^{X} \)^{\top} \tilde{\bF}_i \right\|_{\max} + \dfrac{n}{\phi k_1 k_2} \left\|  \tilde{\bF}_i^{\top} \tilde{\bF}_i \right\|_{\max} \cr
&\leq&  C n^{-1/8}\(\log p \)^{3/2}. \label{Prop2-eq13}
\end{eqnarray}
For  $(\uppercase\expandafter{\romannumeral3})$, note that the elements of $\bF_i$, $\bU_i$, and $\mathcal{E}_i^X$ have sub-Gaussian tails. 
Thus, from Bernstein’s inequality for martingales, we have
 \begin{equation}\label{Prop2-eq14}
\Pr \left\{   (\uppercase\expandafter{\romannumeral3}) \leq C n^{-1/8} \sqrt{\log p} \right\} \geq 1-p^{-4-a}.
\end{equation}
Consider  $(\uppercase\expandafter{\romannumeral4})$ and $(\uppercase\expandafter{\romannumeral5})$. 
Similar to the proofs of \eqref{Prop2-eq9}, we can show
 \begin{equation}\label{Prop2-eq15}
\Pr \left\{  (\uppercase\expandafter{\romannumeral4}) + (\uppercase\expandafter{\romannumeral5}) \leq C n^{-1/8} \sqrt{\log p} \right\} \geq 1-p^{-4-a}.
\end{equation}
Combining \eqref{Prop2-eq10} and \eqref{Prop2-eq12}--\eqref{Prop2-eq15}, we have, with the probability at least $1-5p^{-4-a}$,
\begin{equation}\label{Prop2-eq16}
\left\| \dfrac{n}{\phi k_1 k_2} \mathcal{X}_{i}^{\top} \mathcal{X}_{i}  - \dfrac{n}{\phi k_1 k_2} \bB_{0} \bF_{i}^{\top} \bF_{i} \bB_{0}^{\top} -  \bSigma_{0,u}(i \Delta_n) - \dfrac{n \zeta}{\phi k_1^2} \bV^X \right\|_{\max}  \leq C n^{-1/8}\(\log p \)^{3/2}.
\end{equation}
Similarly, we can show
\begin{equation}\label{Prop2-eq17}
\Pr \left\{ \left\| \dfrac{n \zeta}{\phi k_1^2} \(\hat{\bV}^X - \bV^X\)  \right\|_{\max} \leq C n^{-1/2}\sqrt{\log p} \right\} \geq 1-p^{-4-a}.
\end{equation}
\endpf

\begin{proposition}\label{Prop3}
Under the assumptions in Theorem \ref{Thm1}, there exists a $r$ by $r$ matrix $\bH_i$ such that with the probability at least $1-p^{-2-a}$, 
\begin{eqnarray}
&& \left\|\hat{\bB}_{i\Delta_n} - \bB_{0} \bH_i \right\|_{\max} \leq C \left\{n^{-1/8} \(\log p \)^{3/2} + p^{-1/2} \right\}, \label{Prop3-eq1} \\
&& \|\bH_i\|_{2} \leq C, \quad \|\bH_i^{-1}\|_{2} \leq C, \label{Prop3-eq2} \\
&& \left\|\hat{\bF}_i - \bF_i \(\bH_i^{\top}\)^{-1} \right\|_{2} \leq C\left\{ \(\log p \)^{3/2} + p^{-1/2} n^{1/8}  \right\}. \label{Prop3-eq3}
\end{eqnarray}
\end{proposition}

\textbf{Proof of Proposition \ref{Prop3}.}
For $j=1, \ldots, r$, let $\lambda_{i,j}$ be the $j$-th largest eigenvalue of $\dfrac{n}{\phi k_1 k_2} \mathcal{X}_i^{\top} \mathcal{X}_i$ and $\xi_{i,j}$ be its corresponding eigenvector. Define
\begin{equation*}
\bLambda_i = \Diag \(\lambda_{i,1} ,\ldots, \lambda_{i,r}\) \quad \text{and} \quad \bH_i =  \dfrac{n}{\phi k_1 k_2} \bF_i^{\top}\bF_i\bB_{0}^{\top}\hat{\bB}_{i\Delta_n}\bLambda_i^{-1}.
\end{equation*}
By \eqref{LSM_factor}, we have 
\begin{equation*}
\hat{\bB}_{i \Delta_n}=\sqrt{p}\(\hat{\xi}_{i,1} ,\ldots, \hat{\xi}_{i,r}\) \quad \text{and} \quad \hat{\bF}_i=p^{-1}\mathcal{X}_i \hat{\bB}_{i \Delta_n}.
\end{equation*}
Then, we have 
\begin{equation*}
\dfrac{n}{\phi k_1 k_2} \mathcal{X}_i^{\top} \mathcal{X}_i \hat{\bB}_{i \Delta_n} = \hat{\bB}_{i \Delta_n} \bLambda_i, \quad \hat{\bB}_{i \Delta_n}^{\top} \hat{\bB}_{i \Delta_n} = p\bI_r, \quad \text{and} \quad  \hat{\bF}_i^{\top} \hat{\bF}_i = \dfrac{\phi k_1 k_2}{np}\bLambda_i.
\end{equation*}
We first investigate $\bLambda_i$.
By the Weyl's theorem and \eqref{Prop2-eq7}, for $1 \leq j \leq r$, we have, with the probability at least $1-p^{-3-a}$,
\begin{eqnarray*}
&& \left| \lambda_{i,j} - \lambda_{j}\(\dfrac{n}{\phi k_1 k_2} \bB_{0} \bF_{i}^{\top} \bF_{i} \bB_{0}^{\top} \) \right| \cr
&& \leq \left\| \dfrac{n}{\phi k_1 k_2} \mathcal{X}_{i}^{\top} \mathcal{X}_{i}  - \dfrac{n}{\phi k_1 k_2} \bB_{0} \bF_{i}^{\top} \bF_{i} \bB_{0}^{\top} \right\|_{2} \cr
&&\leq \left\| \dfrac{n}{\phi k_1 k_2} \mathcal{X}_{i}^{\top} \mathcal{X}_{i}  - \dfrac{n}{\phi k_1 k_2} \bB_{0} \bF_{i}^{\top} \bF_{i} \bB_{0}^{\top} - \bSigma_{0,u}(i \Delta_n) - \dfrac{n \zeta}{\phi k_1^2} \bV^X \right\|_{2} + \left\|  \bSigma_{0,u}(i \Delta_n) + \dfrac{n \zeta}{\phi k_1^2} \bV^X \right\|_{2} \cr
&& \leq Cpn^{-1/8}\(\log p \)^{3/2} +   \left\|  \bSigma_{0,u}(i \Delta_n) \right\|_{1} + \left\| \dfrac{n \zeta}{\phi k_1^2} \bV^X \right\|_{1}   \cr
&& \leq  Cpn^{-1/8}\(\log p \)^{3/2} + C,
\end{eqnarray*}
where the last inequality is due to Assumption \ref{assumption1}(e).
Also, by the Weyl's theorem, for $1 \leq j \leq r$ and large $n$, we have, with the probability at least $1-p^{-3-a}$,
\begin{eqnarray*}
&& \left| p^{-1} \lambda_{j}\(\dfrac{n}{\phi k_1 k_2} \sqrt{\bF_{i}^{\top} \bF_{i}} \bB_{0}^{\top} \bB_{0} \sqrt{\bF_{i}^{\top} \bF_{i}}\) - \lambda_{j}\left( \bSigma_{0,f}(i \Delta_n) \right) \right| \cr
&& \leq \left|p^{-1} \lambda_{j}\(\dfrac{n}{\phi k_1 k_2} \sqrt{\bF_{i}^{\top} \bF_{i}} \bB_{0}^{\top} \bB_{0} \sqrt{\bF_{i}^{\top} \bF_{i}}\) - \lambda_{j}\(\dfrac{n}{\phi k_1 k_2}\bF_{i}^{\top} \bF_{i}\)\right|  \cr
&& \quad + \left| \lambda_{j}\(\dfrac{n}{\phi k_1 k_2}\bF_{i}^{\top} \bF_{i} \) - \lambda_{j}\( \bSigma_{0,f}(i \Delta_n)  \) \right| \cr
&& \leq \left\|\dfrac{n}{\phi k_1 k_2}  \bF_{i}^{\top} \bF_{i} \right\|_{2} \times \left\|p^{-1} \bB_{0}^{\top} \bB_{0} -\bI_{r} \right\|_{2} + \left\|  \dfrac{n}{\phi k_1 k_2}\bF_{i}^{\top} \bF_{i} -  \bSigma_{0,f}(i \Delta_n)   \right\|_{2} \cr
&& \leq \dfrac{1}{2}\lambda_{r}\( \bSigma_{0,f}(i \Delta_n) \),
\end{eqnarray*}
where the last inequality is from \eqref{Prop2-eq1} and Assumption \ref{assumption1}(d).
Note that the non-zero eigenvalues of $\dfrac{n}{\phi k_1 k_2} \bB_{0} \bF_{i}^{\top} \bF_{i} \bB_{0}^{\top}$ and $\dfrac{n}{\phi k_1 k_2} \sqrt{\bF_{i}^{\top} \bF_{i}} \bB_{0}^{\top} \bB_{0} \sqrt{\bF_{i}^{\top} \bF_{i}}$ are the same.
Thus, for large $n$, we have, with the probability at least $1-2p^{-3-a}$,
\begin{equation}\label{Prop3-eq4}
\|\bLambda_{i}^{-1} \|_{\max} \leq Cp^{-1}.
\end{equation}

Consider \eqref{Prop3-eq1}. 
We have, with the probability at least $1-3p^{-3-a}$,
\begin{eqnarray}\label{Prop3-eq5}
&& \left\|\hat{\bB}_{i\Delta_n} - \bB_{0} \bH_i \right\|_{\max} \cr
&& = \left\| \dfrac{n}{\phi k_1 k_2} \mathcal{X}_i^{\top} \mathcal{X}_i \hat{\bB}_{i \Delta_n}\bLambda_i^{-1}- \dfrac{n}{\phi k_1 k_2} \bB_{0} \bF_i^{\top}\bF_i\bB_{0}^{\top}\hat{\bB}_{i\Delta_n}\bLambda_i^{-1}\right\|_{\max} \cr
&& \leq \left\|  \(\dfrac{n}{\phi k_1 k_2} \mathcal{X}_i^{\top} \mathcal{X}_i - \dfrac{n}{\phi k_1 k_2} \bB_{0} \bF_i^{\top}\bF_i\bB_{0}^{\top}-  \bSigma_{0,u}(i \Delta_n) - \dfrac{n \zeta}{\phi k_1^2} \bV^X \)\hat{\bB}_{i\Delta_n}\bLambda_i^{-1}\right\|_{\max} \cr
&& \quad + \left\|  \( \bSigma_{0,u}(i \Delta_n) + \dfrac{n \zeta}{\phi k_1^2} \bV^X \)\hat{\bB}_{i\Delta_n}\bLambda_i^{-1}\right\|_{\max} \cr
&&\leq \left\|  \dfrac{n}{\phi k_1 k_2} \mathcal{X}_i^{\top} \mathcal{X}_i - \dfrac{n}{\phi k_1 k_2} \bB_{0} \bF_i^{\top}\bF_i\bB_{0}^{\top}-  \bSigma_{0,u}(i \Delta_n) - \dfrac{n \zeta}{\phi k_1^2} \bV^X \right\|_{\max} \times \left\|  \hat{\bB}_{i\Delta_n}\right\|_{1} \times \left\|  \bLambda_i^{-1}\right\|_{\max} \cr
&& \quad + \left\|  \bSigma_{0,u}(i \Delta_n) + \dfrac{n \zeta}{\phi k_1^2} \bV^X \right\|_{\infty} \times \left\|  \hat{\bB}_{i\Delta_n}\right\|_{\max} \times \left\|  \bLambda_i^{-1}\right\|_{\max} \cr
&& \leq C \left\{n^{-1/8} \(\log p \)^{3/2} + p^{-1/2} \right\},
\end{eqnarray}
where the last inequality is from Assumption \ref{assumption1}(e), \eqref{Prop2-eq7}, and \eqref{Prop3-eq4}.

For \eqref{Prop3-eq2}, we have, with the probability at least $1-3p^{-3-a}$,
\begin{equation}\label{Prop3-eq6}
\|\bH_i\|_{2} \leq \|\dfrac{n}{\phi k_1 k_2} \bF_i^{\top}\bF_i\|_{2} \times \|\bB_{0} \|_{2} \times \|\hat{\bB}_{i\Delta_n} \|_{2} \times \|\bLambda_{i}^{-1} \|_{2} \leq C,
\end{equation}
where the last inequality is from \eqref{Prop2-eq1} and \eqref{Prop3-eq4}.
Then, by \eqref{Prop3-eq5} and \eqref{Prop3-eq6}, for large $n$, we have, with the probability at least $1-6p^{-3-a}$,
\begin{eqnarray}\label{Prop3-eq7}
\|\bH_i^{\top} \bH_i - \bI_r \|_{2} &\leq&  \|\bH_i^{\top} \bH_i - p^{-1}\bH_i^{\top}\bB_{0}^{\top} \bB_{0}\bH_i \|_{2} + \|p^{-1}\bH_i^{\top}\bB_{0}^{\top} \bB_{0}\bH_i - \bI_r \|_{2} \cr
&\leq& \|\bH_i\|_{2}^{2} \times \left\| p^{-1} \bB_{0}^{\top} \bB_{0} -  \bI_{r} \right\|_{2} + p^{-1} \left\| \bH_i^{\top}\bB_{0}^{\top} \bB_{0}\bH_i -  \hat{\bB}_{i\Delta_n}^{\top} \hat{\bB}_{i\Delta_n}  \right\|_{2} \cr
&\leq& \dfrac{1}{3} + p^{-1}\left\|\hat{\bB}_{i\Delta_n} - \bB_{0} \bH_i \right\|_{2} \times \|\bB_{0} \bH_i\|_{2} +  p^{-1}\left\|\hat{\bB}_{i\Delta_n} - \bB_{0} \bH_i \right\|_{2} \times \|\bB_{0} \|_{2} \cr
&\leq& \dfrac{1}{2}.
\end{eqnarray}
Thus, by the Weyl's theorem, we have, with the probability at least $1-6p^{-3-a}$,
\begin{eqnarray}\label{Prop3-eq8}
   \lambda_r\(\bH_i^{\top} \bH_i\) \geq \dfrac{1}{2}, \quad  \det(\bH_i) \geq C, \quad \text{and} \quad \left\|\bH_i^{-1} \right\|_{2} \leq C.
\end{eqnarray}

Consider \eqref{Prop3-eq3}.
We have 
\begin{eqnarray}\label{Prop3-eq9}
  && \left\|\hat{\bF}_i - \bF_i \(\bH_i^{\top}\)^{-1} \right\|_{2} \cr
  && = \left\|p^{-1}\mathcal{X}_i \hat{\bB}_{i \Delta_n} - \bF_i \(\bH_i^{\top}\)^{-1} \right\|_{2} \cr
  && \leq \left\|p^{-1}\(\mathcal{X}_i - \mathcal{X}^c_i - \mathcal{E}_i^{X} \)\hat{\bB}_{i \Delta_n} \right\|_{2} + \left\|p^{-1}\tilde{\bF}_i \hat{\bB}_{i \Delta_n} \right\|_{2} \cr
  && \quad +  \left\|p^{-1}\(\bF_i \bB_{0}^{\top} + \bU_{i} + \mathcal{E}_i^{X}\) \hat{\bB}_{i \Delta_n} - \bF_i \(\bH_i^{\top}\)^{-1} \right\|_{2} \cr
  && = (\uppercase\expandafter{\romannumeral1}) + (\uppercase\expandafter{\romannumeral2}) + (\uppercase\expandafter{\romannumeral3}).
\end{eqnarray}
For $(\uppercase\expandafter{\romannumeral1})$, let $w_k$ be a $k$th row vector of $\mathcal{X}_{i} - \mathcal{X}_i^c - \mathcal{E}_i^X$.
Note that
\begin{equation*}
\underset{\bbeta \in \mathbb{R}^p, \left\|\bbeta \right\|_2 \leq 1}\sup  \left|  w_{k} \bbeta  \right|  \leq C \max_{j}v_{j,n}  \sqrt{\sum_{j=1}^{p} \1 \(|\Delta_{i+k-1}^n \hat{X}_j | > v_{j,n} \)  }
\end{equation*}
under $Q_2$.
Thus, under $Q_2 \cap Q_4$, we have
\begin{eqnarray*}
(\uppercase\expandafter{\romannumeral1}) &\leq& Cp^{-1/2} \max_{j}v_{j,n}\sqrt{\sum_{i=1}^{k_2 - k_1 +1}\sum_{j=1}^{p} \1 \(|\Delta_{i+k-1}^n \hat{X}_j | > v_{j,n} \)  }  \cr
&\leq& C p^{-1/2} \max_{j}v_{j,n}\sqrt{p  k_1 \log p} \cr
&\leq& C \log p,
\end{eqnarray*}
which implies 
\begin{equation}\label{Prop3-eq10}
\Pr \left\{(\uppercase\expandafter{\romannumeral1}) \leq C \log p \right\} \geq 1-p^{-3-a}.
\end{equation}
For $(\uppercase\expandafter{\romannumeral2})$, under $Q_5$, we have 
\begin{equation*}
(\uppercase\expandafter{\romannumeral2}) \leq p^{-1}\left\|\tilde{\bF}_i\right\|_{2} \times  \left\| \hat{\bB}_{i \Delta_n}\right\|_{2} \leq  C\log p.
\end{equation*}
Thus, we have
\begin{equation}\label{Prop3-eq11}
\Pr \left\{(\uppercase\expandafter{\romannumeral2}) \leq C \log p \right\} \geq 1-p^{-3-a}.
\end{equation}
Consider $(\uppercase\expandafter{\romannumeral3})$. 
We have
\begin{eqnarray*}
&&  \left\|p^{-1}\(\bF_i \bB_{0}^{\top} + \bU_{i} + \mathcal{E}_i^{X}\) \hat{\bB}_{i \Delta_n} - \bF_i \(\bH_i^{\top}\)^{-1} \right\|_{2} \cr
&& = \left\|p^{-1}\(\bF_i \bB_{0}^{\top} + \bU_{i} + \mathcal{E}_i^{X}\) \hat{\bB}_{i \Delta_n} - p^{-1}\bF_i \(\bH_i^{\top}\)^{-1}  \hat{\bB}_{i \Delta_n}^{\top}  \hat{\bB}_{i \Delta_n}\right\|_{2}  \cr
&& = \left\|p^{-1}\bF_i\(\bB_{0}^{\top} \hat{\bB}_{i \Delta_n} -\(\bH_i^{\top}\)^{-1}  \hat{\bB}_{i \Delta_n}^{\top}  \hat{\bB}_{i \Delta_n} \) +p^{-1} \(\bU_{i} + \mathcal{E}_i^{X}\) \hat{\bB}_{i \Delta_n}   \right\|_{2}  \cr
&& \leq \left\|p^{-1}\bF_i \(\bH_i^{\top}\)^{-1} \(\bH_i^{\top} \bB_{0}^{\top}  - \hat{\bB}_{i \Delta_n}^{\top}  \)\hat{\bB}_{i \Delta_n}  \right\|_{2} + \left\| p^{-1} \(\bU_{i} + \mathcal{E}_i^{X}\) \(\hat{\bB}_{i \Delta_n} - \bB_{0} \bH_i \)  \right\|_{2} \cr
&& \quad + \left\| p^{-1} \(\bU_{i} + \mathcal{E}_i^{X}\) \bB_{0} \bH_i  \right\|_{2} \cr
&& = 	(\uppercase\expandafter{\romannumeral3})^{(1)} + (\uppercase\expandafter{\romannumeral3})^{(2)} + (\uppercase\expandafter{\romannumeral3})^{(3)}
\end{eqnarray*}
By \eqref{Prop2-eq1}, \eqref{Prop3-eq5}, and \eqref{Prop3-eq8}, we have, with the probability at least $1-10p^{-3-a}$,
\begin{equation*}
(\uppercase\expandafter{\romannumeral3})^{(1)} \leq p^{-1} \left\| \bF_i \right\|_{2} \times  \left\| \bH_i^{-1} \right\|_{2} \times \left\|\hat{\bB}_{i\Delta_n} - \bB_{0} \bH_i \right\|_{2} \times \left\| \hat{\bB}_{i\Delta_n} \right\|_{2} \leq C \left\{\(\log p \)^{3/2} + p^{-1/2}n^{1/8}  \right\}.
\end{equation*}
Also, by \eqref{Prop2-eq2}, \eqref{Prop2-eq3}, and \eqref{Prop3-eq5}, we have, with the probability at least $1-4p^{-3-a}$,
\begin{eqnarray*}
(\uppercase\expandafter{\romannumeral3})^{(2)} &\leq& p^{-1} \left\| \bU_{i} + \mathcal{E}_i^{X} \right\|_{2} \times \left\| \hat{\bB}_{i \Delta_n} - \bB_{0} \bH_i \right\|_{2} \cr
&\leq& C \left\{ n^{-1/16}\(\log p \)^{7/4}+ p^{-1}n^{1/8} + p^{-1/2}n^{1/16}\(\log p\)^{1/4}       \right\}.
\end{eqnarray*}
Consider $(\uppercase\expandafter{\romannumeral3})^{(3)}$. By Assumption \ref{assumption1}(b), the random variable $\( \sqrt{n}\bv^{\top}\Delta_i^n \(\bu + \bepsilon^{X}\) \Big| \FF_{(i-1)\Delta_n} \)$ is sub-Gaussian with bounded parameter for any unit vector $\bv \in \mathbb{R}^p$.
Thus, each element of $\(\bU_i + \mathcal{E}_i^X  \) \bv$ has the sub-Gaussian tail with the order of $n^{-1/4}$.
Then, using the Bernstein’s inequality for martingales, we can show, for any unit vector $\bv \in \mathbb{R}^p$,
\begin{eqnarray}\label{Prop3-eq12}
&& \Pr\left\{\left| \bv^{\top} \left(  \dfrac{n}{\phi k_1 k_2}\(\bU_i + \mathcal{E}_i^X   \)^{\top} \(\bU_i + \mathcal{E}_i^X  \) - \bSigma_{0,u}(i \Delta_n)   -\dfrac{n\zeta}{\phi k_1^2}\bV^{X} \right) \bv\right|  \leq   Cn^{-1/8}\sqrt{\log p} \right\} \cr
&& \geq 1-p^{-3-a}.
\end{eqnarray}
Thus, we have, with the probability at least $1-p^{-3-a}$,
\begin{eqnarray*}
(\uppercase\expandafter{\romannumeral3})^{(3)} &\leq&  p^{-1} \sqrt{\left\| \bB_{0}^{\top}\(\bU_{i} + \mathcal{E}_i^{X}\)^{\top} \(\bU_{i} + \mathcal{E}_i^{X}\) \bB_{0} \right\|_{\infty}} \cr
&\leq& p^{-1} \sqrt{\left\| \bB_{0}^{\top}\left\{\(\bU_{i} + \mathcal{E}_i^{X}\)^{\top} \(\bU_{i} + \mathcal{E}_i^{X}\)- \dfrac{\phi k_1 k_2}{n}\bSigma_{0,u}(i \Delta_n)  -\dfrac{k_2 \zeta}{k_1} \bV^X \right\} \bB_{0} \right\|_{\infty}} \cr
&& + p^{-1} \sqrt{\left\| \bB_{0}^{\top}\left(\dfrac{\phi k_1 k_2}{n}\bSigma_{0,u}(i \Delta_n) + \dfrac{k_2 \zeta}{k_1} \bV^X \right) \bB_{0} \right\|_{\infty}} \cr
&\leq& Cp^{-1/2}n^{1/16}\(\log p\)^{1/4}+ p^{-1} \sqrt{\left\| \bB_{0}\right\|_{\infty} \times \left\| \bB_{0}\right\|_{1} \times \left\|\dfrac{\phi k_1 k_2}{n}\bSigma_{0,u}(i \Delta_n) + \dfrac{k_2 \zeta}{k_1} \bV^X\right\|_{\infty}} \cr
&\leq& Cp^{-1/2} n^{1/8},
\end{eqnarray*}
which implies
\begin{equation}\label{Prop3-eq13}
\Pr \left((\uppercase\expandafter{\romannumeral3}) \leq C \left\{ \(\log p \)^{3/2} + p^{-1/2} n^{1/8}  \right\}\right) \geq 1-15p^{-3-a}.
\end{equation}
Combining \eqref{Prop3-eq9}--\eqref{Prop3-eq11} and \eqref{Prop3-eq13}, we have
\begin{equation}\label{Prop3-eq14}
\Pr \left(\left\|\hat{\bF}_i - \bF_i \(\bH_i^{\top}\)^{-1} \right\|_{2} \leq 		C\left\{ \(\log p \)^{3/2} + p^{-1/2} n^{1/8}  \right\} \right) \geq 1-17p^{-3-a}.
\end{equation}
Then, the statement can be shown by \eqref{Prop3-eq5}, \eqref{Prop3-eq6}, \eqref{Prop3-eq8}, and \eqref{Prop3-eq14}.
\endpf

\begin{proposition}\label{Prop4}
(Deviation condition) Under the assumptions in Theorem \ref{Thm1}, we have, with the probability at least $1-p^{-1-a}$,
\begin{equation}\label{Prop4-eq1}
 \left\| \nabla {\mathcal{L}}_{i}(\btheta_0) \right\|_{\infty} \leq  \eta/2.
\end{equation}
\end{proposition}

\textbf{Proof of Proposition \ref{Prop4}.}
We have
  \begin{eqnarray}\label{Prop4-eq2}
 	\left\| \nabla {\mathcal{L}}_{i}(\btheta_0) \right\|_{\infty} &\leq& \left\| \left(\dfrac{n}{\phi k_1 k_2} \hat{\bU}_i^{\top} \hat{\bU}_i - \dfrac{n \zeta}{\phi k_1^2} \hat{\bV}^X   - \bSigma_{0,u}(i \Delta_n)  \right)\bbeta_0 \right\|_{\infty}  \cr
 	 	&& +  \left\| \dfrac{n}{\phi k_1 k_2} \hat{\bU}_i^{\top} \mathcal{Y}_i- \bSigma_{0,u}(i \Delta_n)\bbeta_0  \right\|_{\infty} \cr
 	 	&& + \left\| \dfrac{n}{\phi k_1 k_2} \hat{\bF}_i^{\top} \hat{\bF}_i  \bH_i^{\top} \bB_0^{\top} \bbeta_0  - \dfrac{n}{\phi k_1 k_2} \hat{\bF}_i^{\top} \mathcal{Y}_i \right\|_{\infty} \cr
 	&& +  \left\| \dfrac{n}{\phi k_1 k_2} \hat{\bU}_i^{\top} \hat{\bF}_i \bH_i^{\top} \bB_0^{\top} \bbeta_0  \right\|_{\infty}  +  \left\| \dfrac{n}{\phi k_1 k_2} \hat{\bF}_i^{\top} \hat{\bU}_i \bbeta_0  \right\|_{\infty} \cr
 	&=& (\uppercase\expandafter{\romannumeral1})+(\uppercase\expandafter{\romannumeral2})+(\uppercase\expandafter{\romannumeral3})+(\uppercase\expandafter{\romannumeral4})+(\uppercase\expandafter{\romannumeral5}).
 \end{eqnarray} 
For $(\uppercase\expandafter{\romannumeral1})$, we have 
\begin{eqnarray}\label{Prop4-eq3}
&&\left\| \dfrac{n}{\phi k_1 k_2} \hat{\bU}_i^{\top} \hat{\bU}_i - \dfrac{n \zeta}{\phi k_1^2} \hat{\bV}^X   - \bSigma_{0,u}(i \Delta_n)  \right\|_{\max}  \cr
&& \leq \dfrac{n}{\phi k_1 k_2} \left\| \hat{\bU}_i^{\top} \hat{\bU}_i  - \(\mathcal{X}_i -\bF_i \bB_{0}^{\top} \)^{\top}\(\mathcal{X}_i -\bF_i \bB_{0}^{\top}\)  \right\|_{\max}  \cr
&&\quad + \dfrac{n}{\phi k_1 k_2} \left\|\(\mathcal{X}_i -\bF_i \bB_{0}^{\top} \)^{\top}\(\mathcal{X}_i -\bF_i \bB_{0}^{\top}\)  - \(\mathcal{X}_i^c + \mathcal{E}_i^{X} - \bF_i \bB_{0}^{\top} \)^{\top}\(\mathcal{X}_i^c + \mathcal{E}_i^{X} - \bF_i \bB_{0}^{\top}\) \right\|_{\max}  \cr
&&\quad + \dfrac{n}{\phi k_1 k_2} \left\| \(\mathcal{X}_i^c + \mathcal{E}_i^{X} -\bF_i \bB_{0}^{\top} \)^{\top}\(\mathcal{X}_i^c + \mathcal{E}_i^{X} -\bF_i \bB_{0}^{\top}\) \right. \nonumber \\ 
&& \left. \qquad \qquad \quad - \(\mathcal{X}_i^c -\bF_i \bB_{0}^{\top} \)^{\top}\(\mathcal{X}_i^c -\bF_i \bB_{0}^{\top}\) - \dfrac{k_2 \zeta}{k_1} \bV^X \right\|_{\max}  \cr
&& \quad  + \dfrac{n}{\phi k_1 k_2} \left\| \(\mathcal{X}_i^c  -\bF_i \bB_{0}^{\top} \)^{\top}\(\mathcal{X}_i^c -\bF_i \bB_{0}^{\top}\) - \bU_i^{\top} \bU_i  \right\|_{\max} + \left\| \dfrac{n}{\phi k_1 k_2} \bU_i^{\top} \bU_i  - \bSigma_{0,u}(i \Delta_n)  \right\|_{\max}  \cr
&& \quad  + \left\|  \dfrac{n \zeta}{\phi k_1^2} \left( \hat{\bV}^X  - \bV^X   \right) \right\|_{\max} \cr
&&= (\uppercase\expandafter{\romannumeral1})^{(1)}+(\uppercase\expandafter{\romannumeral1})^{(2)}+(\uppercase\expandafter{\romannumeral1})^{(3)}+(\uppercase\expandafter{\romannumeral1})^{(4)}+(\uppercase\expandafter{\romannumeral1})^{(5)}+(\uppercase\expandafter{\romannumeral1})^{(6)}.
\end{eqnarray}
Consider $(\uppercase\expandafter{\romannumeral1})^{(1)}$. 
Let $\mathcal{X}_i -\bF_i \bB_{0}^{\top}=\(a_{jl}\)_{1\leq j \leq k_2-k_1+1, 1 \leq l \leq p}$ and $\mathcal{X}_i -\hat{\bF}_i \hat{\bB}_{i \Delta_n}^{\top}=\(\hat{a}_{jl}\)_{1\leq j \leq k_2-k_1+1, 1 \leq l \leq p}$.
For $1 \leq j \leq k_2-k_1+1$, let $\bff_j$ and $\hat{\bff}_j$ be $j$-th column of $\bF_i^{\top}$ and $\hat{\bF}_i^{\top}$, respectively.
For $1 \leq l \leq p$, let $\bb_l$ and $\hat{\bb}_l$ be $l$-th column of $ \bB_{0}^{\top}$ and $\hat{\bB}_{i \Delta_n}^{\top}$, respectively. 
We have
\begin{eqnarray*}
a_{jl}-\hat{a}_{jl}&=&\hat{\bff}_j^{\top} \hat{\bb}_l - \bff_j^{\top} \bb_l \cr
&=&\(\hat{\bff}_j^{\top} - \bff_j^{\top}\(\bH_i^{\top}\)^{-1} \) \bH_i^{\top} \bb_l  +  \bff_j^{\top} \(\bH_i^{\top}\)^{-1} \(\hat{\bb}_l - \bH_i^{\top}\bb_l \) \cr 
&&  + \(\hat{\bff}_j^{\top} - \bff_j^{\top}\(\bH_i^{\top}\)^{-1} \)\(\hat{\bb}_l - \bH_i^{\top}\bb_l \).
\end{eqnarray*}
Thus, by the Cauchy-Schwarz inequality, we have
\begin{eqnarray*}
&& \sum_{j=1}^{k_2-k_1+1}\(\hat{a}_{jl}-a_{jl}\)^2 \cr
&& \leq 3\sum_{j=1}^{k_2-k_1+1}\left\{\(\hat{\bff}_j^{\top} - \bff_j^{\top}\(\bH_i^{\top}\)^{-1} \) \bH_i^{\top} \bb_l \right\}^{2} +  3\sum_{j=1}^{k_2-k_1+1} \left\{\bff_j^{\top} \(\bH_i^{\top}\)^{-1} \(\hat{\bb}_l - \bH_i^{\top}\bb_l \)\right\}^{2} \cr 
&& \quad  +  3\sum_{j=1}^{k_2-k_1+1} \left\{\(\hat{\bff}_j^{\top} - \bff_j^{\top}\(\bH_i^{\top}\)^{-1} \)\(\hat{\bb}_l - \bH_i^{\top}\bb_l \)\right\}^{2} \cr
&& = (A) + (B) + (C).
\end{eqnarray*}
For $(A)$, we have
\begin{eqnarray*}
(A) &\leq& 3\sum_{j=1}^{k_2-k_1+1} \bb_l^{\top} \bH_i \(\hat{\bff}_j^{\top} - \bff_j^{\top}\(\bH_i^{\top}\)^{-1} \)^{\top}  \(\hat{\bff}_j^{\top} - \bff_j^{\top}\(\bH_i^{\top}\)^{-1} \) \bH_i^{\top} \bb_l \cr
&=& 3   \bb_l^{\top} \bH_i   \(\hat{\bF}_i - \bF_i \(\bH_i^{\top}\)^{-1}\)^{\top}\(\hat{\bF}_i - \bF_i \(\bH_i^{\top}\)^{-1}\) \bH_i^{\top} \bb_l \cr
&\leq&  3 \lambda_{\max}\( \bH_i   \(\hat{\bF}_i - \bF_i \(\bH_i^{\top}\)^{-1}\)^{\top}\(\hat{\bF}_i - \bF_i \(\bH_i^{\top}\)^{-1}\) \bH_i^{\top} \)  \bb_l^{\top} \bb_l \cr
&\leq&  C \left\| \hat{\bF}_i - \bF_i \(\bH_i^{\top}\)^{-1}\right\|_{2}^{2} \times \left\| \bH_i \right\|_{2}^{2}.
\end{eqnarray*}
Similarly, we can show
\begin{eqnarray*}
(B) &\leq&  C \left\| \bF_{i} \right\|_{2}^{2} \times \left\| \bH_i^{-1} \right\|_{2}^{2} \times \left\|\hat{\bB}_{i\Delta_n} - \bB_{0} \bH_i \right\|_{\max}^{2}, \cr
(C) &\leq&  C \left\| \hat{\bF}_i - \bF_i \(\bH_i^{\top}\)^{-1}\right\|_{2}^{2} \times  \left\|\hat{\bB}_{i\Delta_n} - \bB_{0} \bH_i \right\|_{\max}^{2}.
\end{eqnarray*}
Then, by \eqref{Prop2-eq1} and Proposition \ref{Prop3}, we have, with the probability at least $1-2p^{-2-a}$,
\begin{equation*}
\max_{1 \leq l \leq p}\sum_{j=1}^{k_2-k_1+1}\(\hat{a}_{jl}-a_{jl}\)^2 \leq C \left\{ \(\log p\)^3 + p^{-1}n^{1/4}   \right\}.
\end{equation*}
Note that by \eqref{Prop2-eq11}, we have, with the probability at least $1-p^{-2-a}$,
\begin{equation*}
 \max_{1\leq j \leq k_2-k_1+1, 1 \leq l \leq p} \left|a_{jl}\right| \leq   C n^{-1/4}\sqrt{\log p}.
\end{equation*}
Thus, from the Cauchy-Schwarz inequality, we have, with the probability at least $1-3p^{-2-a}$,
\begin{eqnarray*}
&& \max_{1\leq l,m \leq p}\left| \sum_{j=1}^{k_2-k_1+1} \hat{a}_{jl} \hat{a}_{jm} -\sum_{j=1}^{k_2-k_1+1} a_{jl} a_{jm} \right| \cr
&& \leq  \max_{1\leq l,m \leq p}\left| \sum_{j=1}^{k_2-k_1+1} \(\hat{a}_{jl}- a_{jl}\)  \(\hat{a}_{jm}- a_{jm}\) \right| + 2 \max_{1\leq l,m \leq p}\left| \sum_{j=1}^{k_2-k_1+1} a_{jl} \(\hat{a}_{jm}- a_{jm}\)  \right| \cr
&& \leq  \max_{1 \leq l \leq p}\sum_{j=1}^{k_2-k_1+1}\(\hat{a}_{jl}-a_{jl}\)^2  + 2\sqrt{\max_{1 \leq l \leq p}\sum_{j=1}^{k_2-k_1+1}a_{jl}^2  \times \max_{1 \leq l \leq p}\sum_{j=1}^{k_2-k_1+1}\(\hat{a}_{jl}-a_{jl}\)^2 } \cr
&& \leq C \left\{ n^{1/8}  \(\log p\)^{2}  + p^{-1/2}n^{1/4}\sqrt{\log p}    \right\},
\end{eqnarray*}
which implies
\begin{equation}\label{Prop4-eq4}
\Pr\left((\uppercase\expandafter{\romannumeral1})^{(1)} \leq C \left\{ n^{-1/8} \(\log p\)^{2} + p^{-1/2}\sqrt{\log p}  \right\}  \right) \geq  1-3p^{-2-a}.
\end{equation}
Consider $(\uppercase\expandafter{\romannumeral1})^{(2)}$. By \eqref{Prop2-eq11}, we have, with the probability at least $1-p^{-2-a}$,
\begin{eqnarray}\label{Prop4-eq5}
(\uppercase\expandafter{\romannumeral1})^{(2)}  &\leq& \dfrac{n}{\phi k_1 k_2} \left\|\(\mathcal{X}_i -\bF_i \bB_{0}^{\top} \)^{\top}\(\mathcal{X}_i -\bF_i \bB_{0}^{\top}\)  - \(\mathcal{X}_{i}^{'} - \bF_i \bB_{0}^{\top} \)^{\top}\(\mathcal{X}_{i}^{'}  - \bF_i \bB_{0}^{\top}\) \right\|_{\max}    \cr
&& + \dfrac{n}{\phi k_1 k_2} \left\|\(\mathcal{X}_{i}^{'} -\bF_i \bB_{0}^{\top} \)^{\top}\(\mathcal{X}_{i}^{'} -\bF_i \bB_{0}^{\top}\)  - \(\mathcal{X}_i^c + \mathcal{E}_i^{X} - \bF_i \bB_{0}^{\top} \)^{\top}\(\mathcal{X}_i^c + \mathcal{E}_i^{X} - \bF_i \bB_{0}^{\top}\) \right\|_{\max}    \cr
&\leq& C  n^{-1/4} \(\log p\)^{2}.
\end{eqnarray}
For $(\uppercase\expandafter{\romannumeral1})^{(3)}$, note that the elements of $\mathcal{X}_i^c$, $\bF_i$, and $\mathcal{E}_i^X$ have sub-Gaussian tails. 
Hence, by Bernstein’s inequality for martingales, we have
\begin{equation*}
\Pr\left\{ \dfrac{n}{\phi k_1 k_2} \left\| \(\mathcal{X}_i^c -\bF_i \bB_{0}^{\top} \)^{\top} \mathcal{E}_i^{X} \right\|_{\max} \leq C n^{-1/8}\sqrt{\log p} \right\} \geq  1-p^{-2-a}.
\end{equation*}
Then, by \eqref{Prop2-eq3}, we have, with the probability at least $1-2p^{-2-a}$,
\begin{eqnarray}\label{Prop4-eq6}
(\uppercase\expandafter{\romannumeral1})^{(3)} &\leq&  \dfrac{2n}{\phi k_1 k_2} \left\| \(\mathcal{X}_i^c -\bF_i \bB_{0}^{\top} \)^{\top} \mathcal{E}_i^{X} \right\|_{\max} +  \dfrac{n}{\phi k_1 k_2} \left\| \(\mathcal{E}_i^{X}\)^{\top} \mathcal{E}_i^{X} - \dfrac{k_2 \zeta}{k_1} \bV^X  \right\|_{\max} \cr
&\leq&  C n^{-1/8}\sqrt{\log p}.
\end{eqnarray}
Consider $(\uppercase\expandafter{\romannumeral1})^{(4)}$. 
By \eqref{Prop2-eq11}, we have, with the probability at least $1-p^{-2-a}$,
\begin{eqnarray}\label{Prop4-eq7}
(\uppercase\expandafter{\romannumeral1})^{(4)}  &\leq&  \dfrac{2n}{\phi k_1 k_2} \left\| \tilde{\bF}_i^{\top} \bU_i \right\|_{\max} +  \dfrac{n}{\phi k_1 k_2} \left\| \tilde{\bF}_i^{\top} \tilde{\bF}_i  \right\|_{\max} \cr
&\leq&  C n^{-1/8}\(\log p\)^{3/2}.
\end{eqnarray}
For $(\uppercase\expandafter{\romannumeral1})^{(5)}$ and $(\uppercase\expandafter{\romannumeral1})^{(6)}$, by Proposition \ref{Prop2}, we have
\begin{equation}\label{Prop4-eq8}
\Pr\left\{(\uppercase\expandafter{\romannumeral1})^{(5)} + (\uppercase\expandafter{\romannumeral1})^{(6)}  \leq C  n^{-1/8} \sqrt{\log p} \right\} \geq  1-p^{-2-a}.
\end{equation}
Combining \eqref{Prop4-eq3}--\eqref{Prop4-eq8}, we have, with the probability at least $1-8p^{-2-a}$,
\begin{equation*}
 \left\|\dfrac{n}{\phi k_1 k_2} \hat{\bU}_i^{\top} \hat{\bU}_i - \dfrac{n \zeta}{\phi k_1^2} \hat{\bV}^X   - \bSigma_{0,u}(i \Delta_n)  \right\|_{\max}   \leq C \left\{ n^{-1/8} \(\log p\)^{2} + p^{-1/2}\sqrt{\log p}  \right\},
\end{equation*}
which implies
\begin{equation}\label{Prop4-eq9}
\Pr\left((\uppercase\expandafter{\romannumeral1})  \leq C \left\{s_p n^{-1/8} \(\log p\)^{2} + p^{-1/2}s_p\sqrt{\log p}  \right\} \right) \geq  1-8p^{-2-a}.
\end{equation}
Consider $(\uppercase\expandafter{\romannumeral2})$. 
We have
\begin{eqnarray}\label{Prop4-eq10}
&& (\uppercase\expandafter{\romannumeral2}) \cr
&& \leq \dfrac{n}{\phi k_1 k_2} \left\|  \(\hat{\bF}_i \hat{\bB}_{i \Delta_n}^{\top} - \bF_i \bB_{0}^{\top}\)^{\top} \mathcal{Y}_i  \right\|_{\infty}  \cr
&& \quad + \dfrac{n}{\phi k_1 k_2} \left\| \( \mathcal{X}_i - \bF_i \bB_{0}^{\top}\)^{\top} \mathcal{Y}_i - \( \mathcal{X}_i^c + \mathcal{E}_i^X - \bF_i \bB_{0}^{\top}\)^{\top}\(\mathcal{Y}_i^c  + \mathcal{E}_i^{Y}\)  \right\|_{\infty}  \cr
&& \quad + \dfrac{n}{\phi k_1 k_2} \left\|  \( \mathcal{X}_i^c + \mathcal{E}_i^X - \bF_i \bB_{0}^{\top}\)^{\top}\(\mathcal{Y}_i^c  + \mathcal{E}_i^{Y}\)  -\(\bU_i+ \mathcal{E}_i^X\)^{\top}\(\bU_i \bbeta_0  + \bF_i \bB_{0}^{\top} \bbeta_0 +\mathcal{Z}_i + \mathcal{E}_i^{Y}\) \right\|_{\infty}  \cr
&& \quad + \dfrac{n}{\phi k_1 k_2} \left\|  \(\bU_i+ \mathcal{E}_i^X\)^{\top}\(\bU_i \bbeta_0  + \bF_i \bB_{0}^{\top} \bbeta_0 +\mathcal{Z}_i + \mathcal{E}_i^{Y}\) -  \bU_i^{\top} \bU_i \bbeta_0 \right\|_{\infty}  \cr
&& \quad +  \left\|  \dfrac{n}{\phi k_1 k_2} \bU_i^{\top} \bU_i \bbeta_0 - \bSigma_{0,u}(i \Delta_n)  \bbeta_0 \right\|_{\infty}  \cr
&=& (\uppercase\expandafter{\romannumeral2})^{(1)}+(\uppercase\expandafter{\romannumeral2})^{(2)}+(\uppercase\expandafter{\romannumeral2})^{(3)}+(\uppercase\expandafter{\romannumeral2})^{(4)}+(\uppercase\expandafter{\romannumeral2})^{(5)}.
\end{eqnarray}
For $(\uppercase\expandafter{\romannumeral2})^{(1)}$, by the Cauchy-Schwarz inequality,  we have, with the probability at least $1-2p^{-2-a}$,
\begin{eqnarray}\label{Prop4-eq11}
(\uppercase\expandafter{\romannumeral2})^{(1)} &\leq&  Cn^{-1/4}\sqrt{\max_{1 \leq l \leq p}\sum_{j=1}^{k_2-k_1+1}\(\hat{a}_{jl}-a_{jl}\)^2 \times \(k_2-k_1+1\)\left\|\mathcal{Y}_i  \right\|_{\infty}^2} \cr
&\leq& C \left\{s_p n^{-1/8} \(\log p\)^{2} + p^{-1/2}s_p\sqrt{\log p}  \right\}.
\end{eqnarray}
For $(\uppercase\expandafter{\romannumeral2})^{(2)}$, by \eqref{Prop2-eq11}, we have, with the probability at least $1-p^{-2-a}$,
\begin{eqnarray}\label{Prop4-eq12}
(\uppercase\expandafter{\romannumeral2})^{(2)} &\leq& \dfrac{n}{\phi k_1 k_2} \left\| \( \mathcal{X}_i - \bF_i \bB_{0}^{\top}\)^{\top} \(\mathcal{Y}_i - \mathcal{Y}_i^{'}\)\right\|_{\infty} + \dfrac{n}{\phi k_1 k_2} \left\| \( \mathcal{X}_i - \bF_i \bB_{0}^{\top}\)^{\top} \(\mathcal{Y}_i^{'}-\mathcal{Y}_i^{c}-\mathcal{E}_i^{Y}\)\right\|_{\infty} \cr
&& + \dfrac{n}{\phi k_1 k_2} \left\| \( \mathcal{X}_i - \mathcal{X}_i^{'}\)^{\top} \(\mathcal{Y}_i^c  + \mathcal{E}_i^{Y}\)  \right\|_{\infty} + \dfrac{n}{\phi k_1 k_2} \left\| \( \mathcal{X}_i^{'} - \mathcal{X}_i^{c}-\mathcal{E}_i^{X}\)^{\top} \(\mathcal{Y}_i^c  + \mathcal{E}_i^{Y}\)  \right\|_{\infty} \cr
&\leq& Cs_p n^{-1/4}\(\log p\)^2.
\end{eqnarray}
Also, by  \eqref{Prop2-eq11}, we have, with the probability at least $1-p^{-2-a}$,
\begin{eqnarray}\label{Prop4-eq13}
 (\uppercase\expandafter{\romannumeral2})^{(3)}    &\leq&  \dfrac{n}{\phi k_1 k_2} \left\|  \(\bU_i+ \mathcal{E}_i^X\)^{\top}\(\tilde{\bF}_i\bbeta_0 +\mathcal{\tilde{X}}_i\)   \right\|_{\infty}  + \dfrac{n}{\phi k_1 k_2} \left\|  \tilde{\bF}_i^{\top}\(\bU_i \bbeta_0  + \bF_i \bB_{0}^{\top} \bbeta_0 +\mathcal{Z}_i + \mathcal{E}_i^{Y}\)   \right\|_{\infty} \cr
&& + \dfrac{n}{\phi k_1 k_2} \left\|  \tilde{\bF}_i^{\top}\(\tilde{\bF}_i\bbeta_0 +\mathcal{\tilde{X}}_i\)   \right\|_{\infty}  \cr
&\leq& C s_p n^{-1/8} \(\log p\)^{3/2}.
\end{eqnarray}
Consider $(\uppercase\expandafter{\romannumeral2})^{(4)}$.
Note that the elements of $\bU_i$, $\bF_i$, $\mathcal{Z}_i$,  $\mathcal{E}_i^Y$, and $\mathcal{E}_i^X$ have sub-Gaussian tails. Hence, by Bernstein’s inequality for martingales, we have
 \begin{equation}\label{Prop4-eq14}
\Pr \left\{   (\uppercase\expandafter{\romannumeral2})^{(4)} \leq C s_p n^{-1/8} \sqrt{\log p} \right\} \geq 1-p^{-2-a}.
\end{equation}
Consider $(\uppercase\expandafter{\romannumeral2})^{(5)}$.
By  \eqref{Prop2-eq2}, we have, with the probability at least $1-p^{-2-a}$,
\begin{eqnarray}\label{Prop4-eq15}
(\uppercase\expandafter{\romannumeral2})^{(5)}    &\leq& \left\|  \dfrac{n}{\phi k_1 k_2} \bU_i^{\top} \bU_i  - \bSigma_{0,u}(i \Delta_n)  \right\|_{\max}  \times \left\| \bbeta_0  \right\|_{1}  \cr
&\leq& Cs_p n^{-1/8}\sqrt{\log p}.
\end{eqnarray}
Combining \eqref{Prop4-eq10}--\eqref{Prop4-eq15}, we have 
 \begin{equation}\label{Prop4-eq16}
\Pr \left( (\uppercase\expandafter{\romannumeral2}) \leq  C \left\{s_p n^{-1/8} \(\log p\)^{2} + p^{-1/2}s_p\sqrt{\log p}  \right\}  \right) \geq 1-6p^{-2-a}.
\end{equation}
Also, similar to the proofs of \eqref{Prop4-eq9} and \eqref{Prop4-eq16}, we can show
\begin{equation}\label{Prop4-eq17}
\Pr \left( (\uppercase\expandafter{\romannumeral3}) \leq  C \left\{s_p n^{-1/8} \(\log p\)^{2} + p^{-1/2}s_p\sqrt{\log p}  \right\}  \right) \geq 1-2p^{-2-a}.
\end{equation}
Consider $(\uppercase\expandafter{\romannumeral4})$.
We have
\begin{eqnarray*}
\dfrac{n}{\phi k_1 k_2} \left\|\hat{\bU}_i^{\top} \hat{\bF}_i \bH_i^{\top} \right\|_{\max} &\leq&  \dfrac{n}{\phi k_1 k_2} \left\| \bU_i^{\top} \bF_i \right\|_{\max} + \dfrac{n}{\phi k_1 k_2} \left\| \hat{\bU}_i^{\top} \(\hat{\bF}_i \bH_i^{\top} - \bF_i \) \right\|_{\max} \cr
&& + \dfrac{n}{\phi k_1 k_2} \left\| \(\hat{\bU}_i-\bU_i \)^{\top} \bF_i \right\|_{\max} \cr
&=& (\uppercase\expandafter{\romannumeral4})^{(1)} + (\uppercase\expandafter{\romannumeral4})^{(2)} + (\uppercase\expandafter{\romannumeral4})^{(3)}.
\end{eqnarray*}
For $(\uppercase\expandafter{\romannumeral4})^{(1)}$, by Proposition \ref{Prop2}, we have
\begin{equation*}
\Pr \left\{ (\uppercase\expandafter{\romannumeral4})^{(1)}  \leq  C n^{-1/8} \sqrt{\log p}  \right\} \geq 1-p^{-2-a}.
\end{equation*}
For $(\uppercase\expandafter{\romannumeral4})^{(2)}$, by Proposition \ref{Prop3}, we have, with the probability at least $1-p^{-2-a}$,
\begin{equation*}
\left\|\hat{\bU}_i\right\|_{\max} \leq Cn^{-1/4}\sqrt{\log p} \quad \text{and} \quad \left\| \hat{\bF}_i \bH_i^{\top} - \bF_i \right\|_{F} \leq C \left\{ \(\log p\)^{3/2} +p^{-1/2} n^{1/8}  \right\}.
\end{equation*}
Thus, by the Cauchy-Schwarz inequality, we have
\begin{equation*}
\Pr \left( (\uppercase\expandafter{\romannumeral4})^{(2)}  \leq  C \left\{n^{-1/8} \(\log p\)^{2} + p^{-1/2}\sqrt{\log p} \right\} \right) \geq 1-p^{-2-a}.
\end{equation*}
Consider $(\uppercase\expandafter{\romannumeral4})^{(3)}$. 
We have 
\begin{equation*}
(\uppercase\expandafter{\romannumeral4})^{(3)} \leq \dfrac{n}{\phi k_1 k_2}\left\| \(\bF_i \bB_{0}^{\top} - \hat{\bF}_i \hat{\bB}_{i \Delta_n}^{\top} + \tilde{\bF}_i + \mathcal{X}_i - \mathcal{X}_i^{c} - \mathcal{E}_i^X \)^{\top} \bF_i \right\|_{\max} + \dfrac{n}{\phi k_1 k_2}\left\| \(\mathcal{E}_i^X\)^{\top} \bF_i \right\|_{\max}.
\end{equation*}
For the first term, recall that 
\begin{equation*}
\Pr \left( \max_{1 \leq l \leq p}\sum_{j=1}^{k_2-k_1+1}\(\hat{a}_{jl}-a_{jl}\)^2 \leq C \left\{ \(\log p\)^3 +p^{-1}n^{1/4}  \right\}\right) \geq 1-2p^{-2-a}.
\end{equation*}
Thus, by  \eqref{Prop2-eq11} and Cauchy-Schwarz inequality, we have, with the probability at least $1-3p^{-2-a}$,
\begin{equation*}
\dfrac{n}{\phi k_1 k_2}\left\| \(\bF_i \bB_{0}^{\top} - \hat{\bF}_i \hat{\bB}_{i \Delta_n}^{\top} + \tilde{\bF}_i + \mathcal{X}_i - \mathcal{X}_i^{c} - \mathcal{E}_i^X \)^{\top} \bF_i \right\|_{\max} \leq C \left\{n^{-1/8} \(\log p\)^{2} + p^{-1/2}\sqrt{\log p} \right\}.
\end{equation*}
For the second term, by Proposition \ref{Prop2}, we have
\begin{equation*}
\Pr \left\{ \dfrac{n}{\phi k_1 k_2}\left\| \(\mathcal{E}_i^X\)^{\top} \bF_i \right\|_{\max}  \leq  C n^{-1/8} \sqrt{\log p}  \right\} \geq 1-p^{-2-a}.
\end{equation*}
Thus, we have 
\begin{equation*}
\Pr \left( (\uppercase\expandafter{\romannumeral4})^{(3)}  \leq  C \left\{n^{-1/8} \(\log p\)^{2} + p^{-1/2}\sqrt{\log p} \right\} \right) \geq 1-4p^{-2-a},
\end{equation*}
which implies
\begin{equation*}
\Pr \left( \dfrac{n}{\phi k_1 k_2} \left\|\hat{\bU}_i^{\top} \hat{\bF}_i \bH_i^{\top} \right\|_{\max} \leq  C \left\{n^{-1/8} \(\log p\)^{2} + p^{-1/2}\sqrt{\log p} \right\} \right) \geq 1-6p^{-2-a}
\end{equation*}
and
\begin{equation}\label{Prop4-eq18}
\Pr \left( (\uppercase\expandafter{\romannumeral4})  \leq  C \left\{s_p n^{-1/8} \(\log p\)^{2} + p^{-1/2} s_p \sqrt{\log p} \right\} \right) \geq 1-6p^{-2-a}.
\end{equation}
Similarly, we can show
\begin{equation}\label{Prop4-eq19}
\Pr \left( (\uppercase\expandafter{\romannumeral5})  \leq  C \left\{s_p n^{-1/8} \(\log p\)^{2} + p^{-1/2} s_p \sqrt{\log p} \right\} \right) \geq 1-6p^{-2-a}.
\end{equation}
Then, \eqref{Prop4-eq1} is obtained by \eqref{Prop4-eq2}, \eqref{Prop4-eq9}, and \eqref{Prop4-eq16}--\eqref{Prop4-eq19}.
\endpf

\begin{proposition}\label{Prop5}
(RE condition) Under the assumptions in Theorem \ref{Thm1}, there exist positive constants $\alpha_1$, $\alpha_2$, and $\kappa$ such that, with the probability at least $1-p^{-1-a}$,
\begin{eqnarray}
 \btheta^{\top} \nabla^2 {\mathcal{L}}_{i}(\btheta) \btheta \geq \alpha_1 \left\|\btheta \right\|_2^2 - \kappa \(n^{-1/4} \(\log p\)^{4} + p^{-1}\)  \left\|\btheta \right\|_1^2 \quad \text{ for all } \, \btheta \in \mathbb{R}^{p+r}, \label{Prop5-eq1} \\
 \btheta^{\top} \nabla^2 {\mathcal{L}}_{i}(\btheta) \btheta \leq \alpha_2 \left\|\btheta \right\|_2^2 + \kappa \(n^{-1/4} \(\log p\)^{4} + p^{-1}\)  \left\|\btheta \right\|_1^2 \quad \text{ for all } \, \btheta \in \mathbb{R}^{p+r}. \label{Prop5-eq2}
\end{eqnarray}
\end{proposition}
\textbf{Proof of Proposition \ref{Prop5}.}
The drift term $\bmu(t)$ has a negligible order comparing with the Brownian motion term. 
Thus, for simplicity, we assume that $\bmu(t)=0$ for $0 \leq t \leq 1$ without loss of generality.
For any parameter $s \geq 1$, define
\begin{equation*}
\mathbb{K}_1^{s} =\left\{\bx \in \mathbb{R}^{p+r} \,\,\, | \,\,\,  \|\bx\|_2 \leq 1, \, \|\bx\|_0 \leq s   \right\} \quad \text{and} \quad \mathbb{K}_2^{s} =\left\{\bx \in \mathbb{R}^p \,\,\, | \,\,\,  \|\bx\|_2 \leq 1, \, \|\bx\|_0 \leq s   \right\}.
\end{equation*}
Also, define
\begin{equation*}
\mathbb{B}^r =\left\{\bx \in \mathbb{R}^r \,\,\, | \,\,\,  \|\bx\|_2 \leq 1  \right\} \quad \text{and} \quad \bSigma_{0,g}(i \Delta_n) = 
\begin{pmatrix}
    \begin{array}{c|c}
        \bSigma_{0,u}(i \Delta_n) & \mathbf{0}_{p \times r} \\ \hline
        \mathbf{0}_{r \times p} & \bLambda_{0,f}(i \Delta_n)
    \end{array}
\end{pmatrix},
\end{equation*}
where $\bLambda_{0,f}(i \Delta_n) = \Diag \(\lambda_{0,f,1}(i \Delta_n) ,\ldots, \lambda_{0,f,r}(i \Delta_n)\)$ and $\lambda_{0,f,j}(i \Delta_n)$ is the $j$-th largest eigenvalue of $\bSigma_{0,f}(i \Delta_n)$ for $j=1, \ldots, r$.
We note that by Assumption \ref{assumption1}(a),(c),(d), the eigenvalues of $\bSigma_{0,g}(i \Delta_n)$ are bounded above and below.
Since
\begin{equation*}
	\nabla^2 {\mathcal{L}}_{i}(\btheta) = \begin{pmatrix}
    \begin{array}{c|c}
        \dfrac{n}{\phi k_1 k_2} \hat{\bU}_i^{\top} \hat{\bU}_i -  \dfrac{n \zeta}{\phi k_1^2} \hat{\bV}^X  & \dfrac{n}{\phi k_1 k_2} \hat{\bU}_i^{\top} \hat{\bF}_i \\ \hline
       \dfrac{n}{\phi k_1 k_2} \hat{\bF}_i^{\top} \hat{\bU}_i & \dfrac{n}{\phi k_1 k_2} \hat{\bF}_i^{\top} \hat{\bF}_i
    \end{array}
\end{pmatrix},
\end{equation*}
we have
\begin{eqnarray}\label{Prop5-eq3}
&&  \underset{\btheta \in \mathbb{K}_1^{s+r}}\sup \left|  \btheta^{\top} \left( \nabla^2 {\mathcal{L}}_{i}(\btheta) - \bSigma_{0,g}(i \Delta_n) \right) \btheta \right| \cr
&& \leq \underset{\bbeta \in \mathbb{K}_2^{s},\bgamma \in \mathbb{B}^r}\sup \left|  \(\bbeta^{\top}, \bgamma^{\top}\) \left( \nabla^2 {\mathcal{L}}_{i}(\btheta) - \bSigma_{0,g}(i \Delta_n) \right) \(\bbeta^{\top}, \bgamma^{\top}\)^{\top} \right| \cr
&& \leq \underset{\bbeta \in \mathbb{K}_2^{s}}\sup \left| \bbeta^{\top} \left(  \dfrac{n}{\phi k_1 k_2} \hat{\bU}_i^{\top} \hat{\bU}_i - \bSigma_{0,u}(i \Delta_n) -\dfrac{n\zeta}{\phi k_1^2}\bV^X  \right) \bbeta \right| \cr
&& \quad + \underset{\bbeta \in \mathbb{K}_2^{s}}\sup \left| \bbeta^{\top} \left( \dfrac{n\zeta}{\phi k_1^2}\hat{\bV}^X  - \dfrac{n\zeta}{\phi k_1^2}\bV^{X}  \right) \bbeta \right| + \underset{\bbeta \in \mathbb{K}_2^{s}, \bgamma \in \mathbb{B}^r}\sup \left| \dfrac{2n}{\phi k_1 k_2}  \bbeta^{\top}  \hat{\bU}_i^{\top} \hat{\bF}_i  \bgamma \right|  \cr
&& \quad + \underset{\bgamma \in \mathbb{B}^r}\sup \left| \bgamma^{\top} \left(  \dfrac{n}{\phi k_1 k_2} \hat{\bF}_i^{\top} \hat{\bF}_i - \bSigma_{0,f}(i \Delta_n)  \right) \bgamma \right| \cr
&&= (\uppercase\expandafter{\romannumeral1})  + (\uppercase\expandafter{\romannumeral2})+ (\uppercase\expandafter{\romannumeral3})+(\uppercase\expandafter{\romannumeral4}).
\end{eqnarray}
For $(\uppercase\expandafter{\romannumeral1})$, we have
\begin{eqnarray}\label{Prop5-eq4}
(\uppercase\expandafter{\romannumeral1}) &\leq& \underset{\bbeta \in \mathbb{K}_2^{s}}\sup \left|  \bbeta^{\top} \left(  \dfrac{n}{\phi k_1 k_2} \(\bU_i + \mathcal{E}_i^X  \)^{\top} \(\bU_i + \mathcal{E}_i^X  \) - \bSigma_{0,u}(i \Delta_n) -\dfrac{n\zeta}{\phi k_1^2}\bV^{X}  \right) \bbeta \right| \cr
&& + \underset{\bbeta \in \mathbb{K}_2^{s}}\sup \, \dfrac{n}{\phi k_1 k_2} \left| \bbeta^{\top}  \left( \hat{\bU}_i^{\top} \hat{\bU}_i   -   \(\bU_i + \mathcal{E}_i^X  \)^{\top} \(\bU_i + \mathcal{E}_i^X  \)  \right) \bbeta \right| \cr
&=& (\uppercase\expandafter{\romannumeral1})^{(1)}  + (\uppercase\expandafter{\romannumeral1})^{(2)}.
\end{eqnarray}
Consider $(\uppercase\expandafter{\romannumeral1})^{(1)}$.
By Assumption \ref{assumption1}(e), the random variable $\( \sqrt{n}\bv^{\top}\Delta_i^n \bu \Big| \FF_{(i-1)\Delta_n} \)$ is sub-Gaussian with bounded parameter for any unit vector $\bv \in \mathbb{R}^p$.
Thus, each element of $\(\bU_i + \mathcal{E}_i^X  \) \bbeta$ has the sub-Gaussian tail with the order of $n^{-1/4}\left\|\bbeta \right\|_2$.
Then, using the Bernstein’s inequality for martingales, we can show, for any fixed unit vector $\bbeta \in \mathbb{R}^p$,
\begin{eqnarray}\label{Prop5-eq5}
&& \Pr\left\{\left| \bbeta^{\top} \left(  \dfrac{n}{\phi k_1 k_2}\(\bU_i + \mathcal{E}_i^X  \)^{\top} \(\bU_i + \mathcal{E}_i^X  \) - \bSigma_{0,u}(i \Delta_n)   -\dfrac{n\zeta}{\phi k_1^2}\bV^{X} \right) \bbeta\right|  \leq   \dfrac{\lambda_{\min}\left\{\bSigma_{0,g}\left(i \Delta_n\right)\right\}}{216} \right\} \cr
&& \geq 1-\exp\(-c_3 n^{1/4}\)
\end{eqnarray}
for some constant $c_3>0$.
For any subset $U \subset \left\{1, \ldots, p\right\}$, define
\begin{equation*}
A_U =\left\{\bbeta \in \mathbb{R}^p \,\,\, | \,\,\,  \|\bbeta\|_2 \leq 1, \, \supp \(\bbeta\) \subset U  \right\}.
\end{equation*}
By \eqref{Prop5-eq5} and discretization argument in Lemma 15 \citep{loh2012high}, for any $A_U$ with $|U| \leq s$, we have, with the probability at least $1-9^{s}\exp\(-c_3 n^{1/4}\)$,
\begin{eqnarray*}
\underset{\bbeta \in A_U}\sup  \left| \bbeta^{\top} \left(  \dfrac{n}{\phi k_1 k_2} \(\bU_i+ \mathcal{E}_i^X  \)^{\top} \(\bU_i + \mathcal{E}_i^X  \) - \bSigma_{0,u}(i \Delta_n)-\dfrac{n\zeta}{\phi k_1^2}\bV^{X}   \right) \bbeta \right|  \leq    \dfrac{\lambda_{\min}\left\{\bSigma_{0,g}\left(i \Delta_n\right)\right\}}{216}.
\end{eqnarray*}
Note that  $\mathbb{K}_2^{s} = \cup_{|U| \leq s} A_U$ and $\binom{p}{s} \leq p^s$.
Hence, we have
\begin{equation}\label{Prop5-eq6}
\Pr\left\{ (\uppercase\expandafter{\romannumeral1})^{(1)} \leq    \dfrac{\lambda_{\min}\left\{\bSigma_{0,g}\left(i \Delta_n\right)\right\}}{216} \right\} \geq 1-\exp\(-c_3 n^{1/4} + s \log 9p\).
\end{equation}
Consider $(\uppercase\expandafter{\romannumeral1})^{(2)}$.
We have
\begin{equation*}
\hat{\bU}_i - \bU_i - \mathcal{E}_i^X   =  \mathcal{X}_i - \mathcal{X}_i^{c} - \mathcal{E}_i^X  + \tilde{\bF}_i +  \bF_i \bB_{0}^{\top} - \hat{\bF}_i \hat{\bB}_{i \Delta_n}^{\top}.
\end{equation*}
Thus, by the Cauchy-Schwarz inequality, we have
\begin{eqnarray}\label{Prop5-eq7}
(\uppercase\expandafter{\romannumeral1})^{(2)} \leq \(2\sqrt{D_1}+\sqrt{D_2}+\sqrt{D_3}+\sqrt{D_4}\)\(\sqrt{D_2}+\sqrt{D_3}+\sqrt{D_4}\),
\end{eqnarray}
where
\begin{eqnarray*}
&& D_1= \underset{\bbeta \in \mathbb{K}_2^{s}}\sup \, \dfrac{n}{\phi k_1 k_2} \left| \bbeta^{\top}   \(\bU_i + \mathcal{E}_i^X  \)^{\top} \(\bU_i + \mathcal{E}_i^X  \) \bbeta \right|, \cr
&& D_2= \underset{\bbeta \in \mathbb{K}_2^{s}}\sup \, \dfrac{n}{\phi k_1 k_2} \left| \bbeta^{\top}   \(\mathcal{X}_i - \mathcal{X}_i^{c} - \mathcal{E}_i^X\)^{\top} \(\mathcal{X}_i - \mathcal{X}_i^{c} - \mathcal{E}_i^X \) \bbeta \right|, \cr
&&D_3= \underset{\bbeta \in \mathbb{K}_2^{s}}\sup \, \dfrac{n}{\phi k_1 k_2} \left| \bbeta^{\top}   \tilde{\bF}_i^{\top} \tilde{\bF}_i \bbeta \right|, \cr
&& D_4= \underset{\bbeta \in \mathbb{K}_2^{s}}\sup \, \dfrac{n}{\phi k_1 k_2} \left| \bbeta^{\top}   \(\bF_i \bB_{0}^{\top} - \hat{\bF}_i \hat{\bB}_{i \Delta_n}^{\top}  \)^{\top} \(\bF_i \bB_{0}^{\top} - \hat{\bF}_i \hat{\bB}_{i \Delta_n}^{\top}  \) \bbeta \right|.
\end{eqnarray*}
For $D_1$, by Assumption \ref{assumption1}(e), each element of $\bU_i + \mathcal{E}_i^X$ has the sub-Gaussian tail with the order of $n^{-1/4}$.
Thus, we can show
\begin{equation*}
\Pr \left\{  \underset{\left\|\bbeta\right\|_{2} \leq 1}\sup \, \left\|\(\bU_i + \mathcal{E}_i^X  \) \bbeta\right\|_{\infty}  \leq Cn^{-1/4}\sqrt{\log p}\right\} \geq 1-p^{-2-a},
\end{equation*}
which implies
\begin{equation}\label{Prop5-eq8}
\Pr \left\{ D_1 \leq C \log p \right\} \geq 1-p^{-2-a}.
\end{equation}
Consider $D_2$. For some large constant $C>0$, let
\begin{eqnarray*}
&& A_1=\left\{  \sum_{j=1}^{p} \int _{i \Delta_n} ^{\(i+ k_2\)\Delta_n} d \Lambda_j(t)  \leq  C \log p \right\}, \cr
&& A_2=\left\{ \sum_{j=1}^{p} \sum_{ k=1}^ {k_2-k_1+1} \1 \(|\Delta_{i+k-1}^n \hat{X}_j | > v_{j,n} \)  \leq C  k_1 \log p \right\}.
\end{eqnarray*}
By Assumption \ref{assumption1}(f), similar to the proofs of \eqref{Prop2-eq11}, we can show
\begin{equation*}
\Pr \left(Q_2 \cap A_1 \cap A_2 \right) \geq 1-p^{-2-a}.
\end{equation*}
Also, let $w_k$ be a $k$th row vector of $\mathcal{X}_{i} - \mathcal{X}_i^c - \mathcal{E}_i^X$.
Under $Q_2$, we have
\begin{equation*}
\underset{\left\|\bbeta \right\|_2 \leq 1}\sup  \left|  w_{k} \bbeta  \right|  \leq C \max_{j}v_{j,n}  \sqrt{\sum_{j=1}^{p} \1 \(|\Delta_{i+k-1}^n \hat{X}_j | > v_{j,n} \)  }.
\end{equation*}
Thus, we have, with the probability at least $1-p^{-2-a}$,
\begin{equation}\label{Prop5-eq9}
D_2 \leq  C \(\max_{j}v_{j,n}\)^2 n^{1/4} \log p \leq Cn^{-1/4}\(\log p\)^2.
\end{equation}
For $D_3$, by \eqref{Prop2-eq11}, we have
\begin{equation}\label{Prop5-eq10}
\Pr \left\{ D_3 \leq C n^{-1/4}\(\log p \)^2 s \right\} \geq 1-p^{-2-a}.
\end{equation}
Consider $D_4$. By the proofs of \eqref{Prop4-eq4}, we have
\begin{equation*}
\Pr \left(  \left\|\(\bF_i \bB_{0}^{\top} - \hat{\bF}_i \hat{\bB}_{i \Delta_n}^{\top}  \)^{\top} \(\bF_i \bB_{0}^{\top} - \hat{\bF}_i \hat{\bB}_{i \Delta_n}^{\top}  \)\right\|_{\max} \leq C \left\{\(\log p \)^3 +p^{-1}n^{1/4} \right\} \right) \geq 1-2p^{-2-a},
\end{equation*}
which implies
\begin{equation}\label{Prop5-eq11}
\Pr \left( D_4 \leq C \left\{ n^{-1/4}\(\log p \)^3 s +p^{-1}s  \right\} \right) \geq 1-2p^{-2-a}.
\end{equation}
Combining \eqref{Prop5-eq7}--\eqref{Prop5-eq11}, we have
\begin{equation}\label{Prop5-eq12}
\Pr \left\{ (\uppercase\expandafter{\romannumeral1})^{(2)} \leq C \(n^{-1/8}\(\log p\)^{2}\sqrt{s}  + n^{-1/4}\(\log p\)^3 s + p^{-1}s \)  \right\} \geq 1-5p^{-2-a}.
\end{equation}
Then, by \eqref{Prop5-eq4}, \eqref{Prop5-eq6}, and \eqref{Prop5-eq12}, we have
\begin{align*}
&\Pr \left\{(\uppercase\expandafter{\romannumeral1}) \leq \dfrac{\lambda_{\min}\left\{\bSigma_{0,g}\left(i \Delta_n\right)\right\}}{216} + c_4 \(n^{-1/8}\(\log p\)^{2}\sqrt{s}  + n^{-1/4}\(\log p\)^3 s + p^{-1}s \)  \right\}  \\
& \geq 1-\exp\(-c_3 n^{1/4} + s \log 9p\)-5p^{-2-a}
\end{align*}
for some constant $c_4>0$.
Choose
\begin{equation}\label{Prop5-eq13}
s=\dfrac{\min\left\{\lambda_{\min}\left\{\bSigma_{0,g}\left(i \Delta_n\right)\right\} / 432 , \(\lambda_{\min}\left\{\bSigma_{0,g}\left(i \Delta_n\right)\right\} / 432\)^2   \right\}}{c_4\left\{n^{-1/4} \(\log p\)^{4} + p^{-1}\right\}}.
\end{equation}
We have, for large $n$,
\begin{equation*}
 c_4 \(n^{-1/8}\(\log p\)^{2}\sqrt{s}  + n^{-1/4}\(\log p\)^3 s + p^{-1}s \) \leq \dfrac{\lambda_{\min}\left\{\bSigma_{0,g}\left(i \Delta_n\right)\right\}}{216}.
\end{equation*}
Also, using the fact that  $c_3 n^{1/4} \geq (2a+4) \log p$ for large $n$, we have
\begin{equation*}
\exp\(-c_3 n^{1/4} + s \log 9p\) + 5p^{-2-a} \leq \exp\(-c_3 n^{1/4} + c_3 n^{1/4}/2  \) + 5p^{-2-a} \leq 6p^{-2-a}.
\end{equation*}
Thus, we have
\begin{equation}\label{Prop5-eq14}
\Pr\left\{(\uppercase\expandafter{\romannumeral1}) \leq    \dfrac{\lambda_{\min}\left\{\bSigma_{0,g}\left(i \Delta_n\right)\right\}}{108}\right\} \geq 1-6p^{-2-a}.
\end{equation}
Similarly, for the same $s$, we can show
\begin{equation}\label{Prop5-eq15}
\Pr\left\{(\uppercase\expandafter{\romannumeral2}) + (\uppercase\expandafter{\romannumeral3}) + (\uppercase\expandafter{\romannumeral4}) \leq    \dfrac{\lambda_{\min}\left\{\bSigma_{0,g}\left(i \Delta_n\right)\right\}}{108}\right\} \geq 1-6p^{-2-a}.
\end{equation}
From \eqref{Prop5-eq3}, \eqref{Prop5-eq14}, and \eqref{Prop5-eq15}, we have
\begin{equation}\label{Prop5-eq16}
\Pr\left\{ \underset{\btheta \in \mathbb{K}_1^{s+r}}\sup \left|  \btheta^{\top} \left( \nabla^2 {\mathcal{L}}_{i}(\btheta) - \bSigma_{0,g}(i \Delta_n) \right) \btheta \right| \leq \dfrac{\lambda_{\min}\left\{\bSigma_{0,g}\left(i \Delta_n\right)\right\}}{54}\right\} \geq 1-p^{-1-a}
\end{equation}
for large $n$.
Then, by Lemma 13 \citep{loh2012high}, we have, with the probability at least $1-p^{-1-a}$,
\begin{eqnarray}
 \btheta^{\top} \nabla^2 {\mathcal{L}}_{i}(\btheta) \btheta \geq \dfrac{\lambda_{\min}\left\{\bSigma_{0,g}\left(i \Delta_n\right)\right\}}{2} \left\|\btheta \right\|_2^2 - \dfrac{\lambda_{\min}\left\{\bSigma_{0,g}\left(i \Delta_n\right)\right\}}{s}\left\|\btheta \right\|_1^2 \quad \text{ for all } \, \btheta \in \mathbb{R}^{p+r}, \\ \label{Prop5-eq17}
 \btheta^{\top} \nabla^2 {\mathcal{L}}_{i}(\btheta) \btheta \leq \dfrac{3\lambda_{\max}\left\{\bSigma_{0,g}\left(i \Delta_n\right)\right\}}{2} \left\|\btheta \right\|_2^2 + \dfrac{\lambda_{\min}\left\{\bSigma_{0,g}\left(i \Delta_n\right)\right\}}{s} \left\|\btheta \right\|_1^2 \quad \text{ for all } \, \btheta \in \mathbb{R}^{p+r},  \label{Prop5-eq18}
\end{eqnarray}
which completes the proof.
\endpf

\textbf{Proof of Theorem \ref{Thm1}.} 
By Propositions \ref{Prop3}--\ref{Prop5}, it is enough to show the statement under \eqref{Prop3-eq2}, \eqref{Prop4-eq1}, and \eqref{Prop5-eq1}.
From  \eqref{sparsity_beta} and \eqref{Prop3-eq2}, we have
\begin{equation*}
\sum_{j=1}^{p+r}    |\theta_{0,j} | ^{\delta}  \leq c_5 s_p
\end{equation*}
for some constant $c_5>0$.
Define $S_{\theta} = \{j: \text{ jth element}$ of $| \btheta_0 | > \eta \}$.
We have
\begin{equation}\label{Thm1-eq1}
\eta^{\delta}|S_{\theta}| \leq c_5 s_p \quad \text{ and } \quad \|(\btheta_{0})_{S_{\theta}^c} \|_{1} = \Sigma_{j \in S_{\theta}^c}|(\btheta_{0})_j|^{\delta} |(\btheta_{0})_j|^{1-\delta} \leq c_5 s_p \eta^{1-\delta}. 
\end{equation}
From the optimality of $\hat{\btheta}_{i \Delta_n}$, we have
\begin{eqnarray*}
0 &\geq&  {\mathcal{L}}_{i}(\hat{\btheta}_{i \Delta_n}) - {\mathcal{L}}_{i}(\btheta_0) + \eta \left(\| \hat{\btheta}_{i \Delta_n}\|_{1} - \left\| \btheta_0 \right\|_{1} \right) \cr
&=& \eta \left(\| \hat{\btheta}_{i \Delta_n}\|_{1} - \left\| \btheta_0 \right\|_{1} \right) + \langle \hat{\btheta}_{i \Delta_n}-\btheta_0, \nabla {\mathcal{L}}_{i}(\btheta_0) \rangle  + \frac{1}{2}  \( \hat{\btheta}_{i \Delta_n}-\btheta_0\right)^{\top }\nabla^2 {\mathcal{L}}_{i}(\btheta)  \left(\hat{\btheta}_{i \Delta_n}-\btheta_0\).
\end{eqnarray*}
Note that 
\begin{eqnarray*}
\left\| \hat{\btheta}_{i \Delta_n}\right\|_{1} - \left\| \btheta_0 \right\|_{1} &=&  \left\| \(\hat{\btheta}_{i \Delta_n}\)_{S_{\theta}^c}\right\|_{1} - \left\| \(\btheta_0\)_{S_{\theta}^c} \right\|_{1} +  \left\| \(\hat{\btheta}_{i \Delta_n}\)_{S_{\theta}}\right\|_{1}  -  \left\| \(\btheta_0\)_{S_{\theta}} \right\|_{1}  \cr
&\geq& \left\| \left( \hat{\btheta}_{i \Delta_n}-\btheta_0\right)_{S_{\theta}^c} \right\|_{1} - \left\|  \left(\hat{\btheta}_{i \Delta_n}-\btheta_0\right)_{S_{\theta}} \right\|_{1}-  2\left\| \(\btheta_0\)_{S_{\theta}^c} \right\|_{1} \cr
&\geq& \left\| \left( \hat{\btheta}_{i \Delta_n}-\btheta_0\right)_{S_{\theta}^c} \right\|_{1} - \left\|  \left(\hat{\btheta}_{i \Delta_n}-\btheta_0\right)_{S_{\theta}} \right\|_{1}-  2 c_5 s_p \eta^{1-\delta}
\end{eqnarray*}
and
\begin{equation*}
\left| \langle \hat{\btheta}_{i \Delta_n}-\btheta_0, \nabla {\mathcal{L}}_{i}(\btheta_0) \rangle \right|  \leq \left( \left\| \(\hat{\btheta}_{i \Delta_n}-\btheta_0\)_{S_{\theta}} \right\|_{1} +  \left\| \(\hat{\btheta}_{i \Delta_n}-\btheta_0\)_{S_{\theta}^c} \right\|_{1} \right) \eta /2.
\end{equation*}
Hence, we have 
\begin{eqnarray}\label{Thm1-eq2}
&& \dfrac{3\eta}{2}\left\|  \left(\hat{\btheta}_{i \Delta_n}-\btheta_0\right)_{S_{\theta}} \right\|_1 -\dfrac{\eta}{2} \left\|  \left(\hat{\btheta}_{i \Delta_n}-\btheta_0\right)_{S_{\theta}^c} \right\|_1 + 2 c_5 s_p \eta^{2-\delta} \cr 
&& \geq  \frac{1}{2}  \( \hat{\btheta}_{i \Delta_n}-\btheta_0\right)^{\top }\nabla^2 {\mathcal{L}}_{i}(\btheta)  \left(\hat{\btheta}_{i \Delta_n}-\btheta_0\) \cr
&& \geq  \dfrac{\alpha_1}{2}\left\|\hat{\btheta}_{i \Delta_n}-\btheta_0 \right\|_2^2  -\dfrac{\kappa \(n^{-1/4} \(\log p\)^{4} + p^{-1}\) }{2}\left\|\hat{\btheta}_{i \Delta_n}-\btheta_0 \right\|_1^2,
\end{eqnarray}
where the last inequality is from \eqref{Prop5-eq1}.
Also, using the fact that $\left\|\hat{\btheta}_{i \Delta_n}-\btheta_0 \right\|_1 \leq C s_p$, we have, for large $n$,
\begin{equation*}
\eta \geq 2 \kappa \(n^{-1/4} \(\log p\)^{4} + p^{-1}\)  \left\|\hat{\btheta}_{i \Delta_n}-\btheta_0 \right\|_1,
\end{equation*}
which implies
\begin{equation*}
\left\|  \left(\hat{\btheta}_{i \Delta_n}-\btheta_0\right)_{S_{\theta}^c} \right\|_1 \leq 7 \left\|  \left(\hat{\btheta}_{i \Delta_n}-\btheta_0\right)_{S_{\theta}} \right\|_1 + 8 c_5 s_p \eta^{1-\delta}.
\end{equation*}
Thus, by \eqref{Thm1-eq1} and Cauchy–Schwarz inequality, we have
\begin{equation}\label{Thm1-eq3}
\left\| \hat{\btheta}_{i \Delta_n}-\btheta_0\right\|_1 \leq 8 \sqrt{c_5 s_p}\eta^{-\delta/2} \left\|  \hat{\btheta}_{i \Delta_n}-\btheta_0 \right\|_2 + 8 c_5 s_p \eta^{1-\delta}.
\end{equation}
Combining \eqref{Thm1-eq2} and \eqref{Thm1-eq3}, we have, for large $n$,
\begin{eqnarray}\label{Thm1-eq4}
&& 12\sqrt{c_5 s_p}\eta^{1-\delta/2} \left\|  \hat{\btheta}_{i \Delta_n}-\btheta_0 \right\|_2 + 14 c_5 s_p \eta^{2-\delta} \cr
&&\geq \dfrac{3\eta}{2}\left\|  \hat{\btheta}_{i \Delta_n}-\btheta_0 \right\|_1 + 2 c_5 s_p \eta^{2-\delta}  \cr
&&\geq \dfrac{3\eta}{2}\left\|  \left(\hat{\btheta}_{i \Delta_n}-\btheta_0\right)_{S_{\theta}} \right\|_1 -\dfrac{\eta}{2} \left\|  \left(\hat{\btheta}_{i \Delta_n}-\btheta_0\right)_{S_{\theta}^c} \right\|_1 + 2 c_5 s_p \eta^{2-\delta}  \cr
&&\geq  \dfrac{\alpha_1}{2}\left\|\hat{\btheta}_{i \Delta_n}-\btheta_0 \right\|_2^2 - 64\kappa \(n^{-1/4} \(\log p\)^{4} + p^{-1}\)  \left(c_5 s_p \eta^{-\delta} \left\|  \hat{\btheta}_{i \Delta_n}-\btheta_0 \right\|_2^2 +  c_5^2 s_p^2 \eta^{2-2\delta}  \right) \cr
&&\geq \dfrac{\alpha_1}{4} \left\|\hat{\btheta}_{i \Delta_n}-\btheta_0 \right\|_2^2 - c_5 s_p \eta^{2-\delta}.
\end{eqnarray}
Then, using the fact that
\begin{equation*}
\dfrac{\alpha_1}{8} \left\|\hat{\btheta}_{i \Delta_n}-\btheta_0 \right\|_2^2  + \dfrac{288}{\alpha_1} c_5 s_p \eta^{2-\delta} \geq 12\sqrt{c_5 s_p}\eta^{1-\delta/2} \left\|  \hat{\btheta}_{i \Delta_n}-\btheta_0 \right\|_2,
\end{equation*}
we have 
\begin{equation}\label{Thm1-eq5}
\left\|\hat{\btheta}_{i \Delta_n}-\btheta_0 \right\|_2 \leq C\sqrt{s_p}\eta^{1-\delta/2}.
\end{equation}
Also, by \eqref{Thm1-eq3} and \eqref{Thm1-eq5}, we have
\begin{equation}\label{Thm1-eq6}
\left\|\hat{\btheta}_{i \Delta_n}-\btheta_0 \right\|_1 \leq C s_p \eta^{1-\delta},
\end{equation}
which completes the proof.
\endpf

\subsection{Proof of Theorem \ref{Thm2}}\label{Proof-Thm2}
\textbf{Proof of Theorem \ref{Thm2}.} 
To obtain the upper bound for $\|  \hat{I\beta}  - I\beta_0 \| _{\max} $, we first investigate  $\hat{\bOmega}_{i \Delta_n}$.
We have
\begin{eqnarray*}
&& \sup_{0 \leq i \leq n-k_2}\left\| \left(\dfrac{n}{\phi k_1 k_2} \hat{\bU}_i^{\top} \hat{\bU}_i - \dfrac{n \zeta}{\phi k_1^2}\hat{\bV}^X \right)\bOmega_0\(i \Delta_n\)  -\bI  \right\|_{\max} \cr
&& \leq   \sup_{0 \leq i \leq n-k_2} \left\| \dfrac{n}{\phi k_1 k_2} \hat{\bU}_i^{\top} \hat{\bU}_i - \dfrac{n \zeta}{\phi k_1^2}\hat{\bV}^X  -\bSigma_{0,u}\(i \Delta_n\)  \right\|_{\max} \times \sup_{0 \leq i \leq n-k_2}\left\| \bOmega_0\(i \Delta_n\) \right\|_{1}. 
\end{eqnarray*}
By the proofs of \eqref{Prop4-eq9}, we can show, with the probability at least $1-p^{-2-a}$,
\begin{equation*}
\sup_{0 \leq i \leq n-k_2} \left\| \dfrac{n}{\phi k_1 k_2} \hat{\bU}_i^{\top} \hat{\bU}_i - \dfrac{n \zeta}{\phi k_1^2}\hat{\bV}^X  -\bSigma_{0,u}\(i \Delta_n\)  \right\|_{\max}  \leq C \left\{ n^{-1/8}\(\log p\)^{2}  + p^{-1/2}\sqrt{\log p} \right\}.
\end{equation*}
Thus,  we have
\begin{equation}\label{Thm2-eq1}
\Pr\left\{\sup_{0 \leq i \leq n-k_2} \| \hat{\bOmega}_{i \Delta_n} \|_{1} \leq C \right\} \geq 1-p^{-2-a}.
\end{equation}
Also, we have, with the probability at least $1-p^{-2-a}$,
\begin{eqnarray}\label{Thm2-eq2}
\sup_{0 \leq i \leq n-k_2} \left\| \bSigma_{0,u}\(i \Delta_n\) \hat{\bOmega}_{i \Delta_n}  -\bI \right\|_{\max} &\leq& \sup_{0 \leq i \leq n-k_2} \left\| \(\bSigma_{0,u}\(i \Delta_n\)  - \dfrac{n}{\phi k_1 k_2} \hat{\bU}_i^{\top} \hat{\bU}_i + \dfrac{n \zeta}{\phi k_1^2}\hat{\bV}^X  \)\hat{\bOmega}_{i \Delta_n} \right\|_{\max} \cr
&& +  \sup_{0 \leq i \leq n-k_2} \left\|\(\dfrac{n}{\phi k_1 k_2} \hat{\bU}_i^{\top} \hat{\bU}_i - \dfrac{n \zeta}{\phi k_1^2}\hat{\bV}^X  \) \hat{\bOmega}_{i \Delta_n} -\bI \right\|_{\max} \cr
&\leq& C \left\{ n^{-1/8}\(\log p\)^{2}  + p^{-1/2}\sqrt{\log p} \right\},
 \end{eqnarray}
which implies
\begin{eqnarray*}
&& \sup_{0 \leq i \leq n-k_2} \| \hat{\bOmega}_{i \Delta_n} - \bOmega_0\(i \Delta_n\) \|_{\max} \cr
&& \leq \sup_{0 \leq i \leq n-k_2} \| \bOmega_0\(i \Delta_n\) \|_{\infty} \times \sup_{0 \leq i \leq n-k_2} \|\bSigma_{0,u}\(i \Delta_n\)  \hat{\bOmega}_{i \Delta_n} - \bI \|_{\max}   \cr
&& \leq C \left\{ n^{-1/8}\(\log p\)^{2}  + p^{-1/2}\sqrt{\log p} \right\}.
\end{eqnarray*}
Then, similar to the proofs of Theorem 1 \citep{kim2026high}, we can show
\begin{equation}\label{Thm2-eq3}
\Pr\left\{ \sup_{0 \leq i \leq n-k_2} \| \hat{\bOmega}_{i \Delta_n} - \bOmega_0\(i \Delta_n\) \|_{1} \leq  Cs_{\omega, p } \tau^{1-q}\right\} \geq 1-p^{-2-a}.
\end{equation}
Let  
\begin{eqnarray*}
&&\tilde{ \bbeta}_{i \Delta_n}^{(2)} = \hat{\bbeta}_{i \Delta_n}  +   \dfrac{n}{\phi k_1 k_2}  \hat{\bOmega}_{i \Delta_n}^{\top} \Big[ \(\mathcal{X}^c_i + \mathcal{E}_i^X - \hat{\bF}_i \hat{\bB}_{i \Delta_n}^{\top}\)^{\top} \left(\mathcal{Y}_{i}^c + \mathcal{E}_{i}^{Y}  \right) \cr
&& \qquad \qquad \qquad \qquad \qquad \qquad - \left( \(\mathcal{X}^c_i + \mathcal{E}_i^X - \hat{\bF}_i \hat{\bB}_{i \Delta_n}^{\top} \)^{\top} \(\mathcal{X}^c_i + \mathcal{E}_i^X \) - \dfrac{k_2 \zeta}{k_1} \hat{\bV}^X \right) \hat{\bbeta}_{i \Delta_n} \Big], \cr
&&\tilde{ \bbeta}_{i \Delta_n}^{(3)} = \bbeta_0\(i\Delta_n\) + \dfrac{n}{\phi k_1 k_2}\hat{\bOmega}_{i \Delta_n}^{\top}  \Big[ \(\bU_{i} + \mathcal{E}_i^X + \tilde{\bF}_i + \bF_i \bB_{0}^{\top}\(i\Delta_n\) - \hat{\bF}_i \hat{\bB}_{i \Delta_n}^{\top}\)^{\top} \(\mathcal{Z}_{i}  +  \mathcal{\tilde{X}}_{i} + \mathcal{E}_{i}^{Y}  \) \cr
&& \qquad \qquad \qquad \qquad \qquad  \quad - \left(\(\bU_{i} + \mathcal{E}_i^X + \tilde{\bF}_i + \bF_i \bB_{0}^{\top}\(i\Delta_n\) - \hat{\bF}_i \hat{\bB}_{i \Delta_n}^{\top}\)^{\top} \mathcal{E}_{i}^{X} - \dfrac{k_2\zeta}{k_1} \hat{\bV}^X \right) \hat{\bbeta}_{i \Delta_n} \Big],  \cr
&&\tilde{ \bbeta}_{i \Delta_n}^{(4)} = \bbeta_0\(i\Delta_n\)  + \dfrac{n}{\phi k_1 k_2}\bOmega_0\(i \Delta_n\) \left[ \(\bU_{i} + \mathcal{E}_i^X + \tilde{\bF}_i + \bF_i \bB_{0}^{\top}\(i\Delta_n\) - \hat{\bF}_i \hat{\bB}_{i \Delta_n}^{\top}\)^{\top}  \right. \\
&& \left.  \quad \times \(\mathcal{Z}_{i}  +  \mathcal{\tilde{X}}_{i} + \mathcal{E}_{i}^{Y}  \) - \left(\(\bU_{i} + \mathcal{E}_i^X + \tilde{\bF}_i + \bF_i \bB_{0}^{\top}\(i\Delta_n\) - \hat{\bF}_i \hat{\bB}_{i \Delta_n}^{\top}\)^{\top} \mathcal{E}_{i}^{X} - \dfrac{k_2\zeta}{k_1} \bV^X \right) \bbeta_0\(i\Delta_n\) \right].
\end{eqnarray*}
Then, we have
\begin{eqnarray}\label{Thm2-eq4}
\|  \hat{I\beta}  - I\beta_0 \| _{\max} &\leq&   \left\|  \sum_{i=0}^{[1/(k_2 \Delta_n) ]-1}\left(\tilde{\bbeta}_{ik_2 \Delta_n} - \tilde{\bbeta}_{ik_2 \Delta_n}^{(2)} \right) k_2 \Delta_n \right\|_{\infty}    \cr
&&+  \left\|  \sum_{i=0}^{[1/(k_2 \Delta_n) ]-1}\left(\tilde{\bbeta}_{ik_2 \Delta_n}^{(2)} - \tilde{\bbeta}_{i k_2 \Delta_n}^{(3)} \right) k_2 \Delta_n \right\|_{\infty}     \cr
&&+  \left\|  \sum_{i=0}^{[1/( k_2 \Delta_n) ]-1}\left(\tilde{\bbeta}_{i  k_2 \Delta_n}^{(3)} - \tilde{\bbeta}_{i  k_2 \Delta_n}^{(4)} \right) k_2 \Delta_n \right\|_{\infty}     \cr
&&+ \left\|  \sum_{i=0}^{[1/( k_2 \Delta_n) ]-1}\left(\tilde{\bbeta}_{i k_2 \Delta_n}^{(4)} - \bbeta_0(i  k_2 \Delta_n) \right)k_2 \Delta_n  \right\|_{\infty}     \cr
&&+   \left\|  \sum_{i=0}^{[1/( k_2 \Delta_n) ]-1} \int_{i  k_2 \Delta_n}^{(i+1) k_2 \Delta_n}\left(\bbeta_0(i  k_2\Delta_n) -\bbeta_0(t) \right)  dt \right\|_{\infty}    \cr
&&+   \left\|  \int_{ [1/(k_2 \Delta_n)] k_2\Delta_n}^{1} \bbeta_0(t) dt \right\|_{\infty}    \cr
& = & (\uppercase\expandafter{\romannumeral1})+(\uppercase\expandafter{\romannumeral2})+(\uppercase\expandafter{\romannumeral3})+(\uppercase\expandafter{\romannumeral4})+(\uppercase\expandafter{\romannumeral5})+(\uppercase\expandafter{\romannumeral6}).
\end{eqnarray}
Consider $(\uppercase\expandafter{\romannumeral1})$.
By \eqref{Thm1-result1}, \eqref{Prop2-eq11}, Proposition \ref{Prop3}, and \eqref{Thm2-eq1}, we have
\begin{equation}\label{Thm2-eq5}
\Pr \left\{(\uppercase\expandafter{\romannumeral1}) \leq C s_p n^{-1/4} (\log p)^2 \right\} \geq 1-p^{-1-a}.
\end{equation}
Consider $(\uppercase\expandafter{\romannumeral2})$.
We have
\begin{eqnarray*}
&&\sup_{0 \leq i \leq n- k_2} \left\| \dfrac{n}{\phi k_1 k_2} \(\bU_{i} + \mathcal{E}_i^X + \tilde{\bF}_i + \bF_i \bB_{0}^{\top}\(i\Delta_n\) - \hat{\bF}_i \hat{\bB}_{i \Delta_n}^{\top}\)^{\top} \right. \\
&& \left. \qquad \qquad \qquad \quad \,\, \times \(\bU_{i} +  \bF_i \bB_{0}^{\top}\(i\Delta_n\) + \tilde{\bF}_i \)  - \bSigma_{0,u}\(i\Delta_n \) \right\|_{\max} \cr 
&&\leq \sup_{0 \leq i \leq n- k_2} \left\| \dfrac{n}{\phi k_1 k_2} \bU_{i}^{\top}\bU_{i} - \bSigma_{0,u}\(i\Delta_n \) \right\|_{\max} \cr
&& \quad + \sup_{0 \leq i \leq n- k_2} \left\| \dfrac{n}{\phi k_1 k_2}\(\bU_{i}^{\top}\bF_i \bB_{0}^{\top}\(i\Delta_n\) + \(\mathcal{E}_i^X\)^{\top}\bU_{i} +  \(\mathcal{E}_i^X\)^{\top} \bF_i \bB_{0}^{\top}\(i\Delta_n\) \)\right\|_{\max} \cr
&& \quad + \sup_{0 \leq i \leq n- k_2} \left\| \dfrac{n}{\phi k_1 k_2}\(2 \bU_{i}^{\top}\tilde{\bF}_i + \(\mathcal{E}_i^X\)^{\top}\tilde{\bF}_i +\tilde{\bF}_i^{\top}\bF_i \bB_{0}^{\top}\(i\Delta_n\) +  \tilde{\bF}_i^{\top}\tilde{\bF}_i  \) \right\|_{\max} \cr
&& \quad + \sup_{0 \leq i \leq n- k_2} \left\| \dfrac{n}{\phi k_1 k_2}\(\bF_i \bB_{0}^{\top}\(i\Delta_n\) - \hat{\bF}_i \hat{\bB}_{i \Delta_n}^{\top} \)^{\top} \(\bU_{i} +  \bF_i \bB_{0}^{\top}\(i\Delta_n\) + \tilde{\bF}_i  \) \right\|_{\max} \cr
&& = (\uppercase\expandafter{\romannumeral2})^{(1)} + (\uppercase\expandafter{\romannumeral2})^{(2)} + (\uppercase\expandafter{\romannumeral2})^{(3)} + (\uppercase\expandafter{\romannumeral2})^{(4)}.
\end{eqnarray*}
By Proposition \ref{Prop2}, we have
\begin{equation*}
\Pr \left\{(\uppercase\expandafter{\romannumeral2})^{(1)} \leq C n^{-1/8} \sqrt{\log p} \right\} \geq 1-p^{-2-a}.
\end{equation*}
For $(\uppercase\expandafter{\romannumeral2})^{(2)}$, note that the elements of $\bU_i$, $\bF_i$, and $\mathcal{E}_i^X$ have sub-Gaussian tails. 
Thus, by the Bernstein’s inequality for martingales, we have
\begin{equation*}
\Pr \left\{(\uppercase\expandafter{\romannumeral2})^{(2)} \leq C n^{-1/8} \sqrt{\log p} \right\} \geq 1-p^{-2-a}.
\end{equation*}
Also, by \eqref{Prop2-eq11}, we have
\begin{equation*}
\Pr \left\{(\uppercase\expandafter{\romannumeral2})^{(3)} \leq C n^{-1/8} \(\log p\)^{3/2} \right\} \geq 1-p^{-2-a}.
\end{equation*}
For $(\uppercase\expandafter{\romannumeral2})^{(4)}$, let $\bF_i \bB_{0}^{\top}\(i\Delta_n\) - \hat{\bF}_i \hat{\bB}_{i \Delta_n}^{\top}=\(T_{jl}^{(i)}\)_{1\leq j \leq k_2-k_1+1, 1 \leq l \leq p}$. 
By the proofs of \eqref{Prop4-eq4}, we have
\begin{equation*}
\Pr \left( \sup_{0 \leq i \leq n- k_2} \max_{1 \leq l \leq p}\sum_{j=1}^{k_2-k_1+1}\(T_{jl}^{(i)}\)^2 \leq C \left\{ \(\log p\)^3 + p^{-1}n^{1/4} \right\}\right) \geq 1-p^{-2-a}.
\end{equation*}
Then, from the Cauchy-Schwarz inequality and \eqref{Prop2-eq11},  we have, with the probability at least $1-2p^{-2-a}$,
\begin{eqnarray*}
(\uppercase\expandafter{\romannumeral2})^{(4)} &\leq&  \dfrac{Cn}{\phi k_1 k_2} \sqrt{\sup_{0 \leq i \leq n- k_2} \max_{1 \leq l \leq p}\sum_{j=1}^{k_2-k_1+1} \(T_{jl}^{(i)}\)^2 \times \(k_2-k_1+1\)n^{-1/2} \log p} \cr
&\leq& C \left\{n^{-1/8}\(\log p\)^{2}  + p^{-1/2}\sqrt{\log p}   \right\}.
\end{eqnarray*}
Thus, we have, with the probability at least $1-5p^{-2-a}$,
\begin{eqnarray}\label{Thm2-eq6}
&&\sup_{0 \leq i \leq n- k_2} \left\| \dfrac{n}{\phi k_1 k_2} \(\bU_{i} + \mathcal{E}_i^X + \tilde{\bF}_i + \bF_i \bB_{0}^{\top}\(i\Delta_n\) - \hat{\bF}_i \hat{\bB}_{i \Delta_n}^{\top}\)^{\top} \right. \nonumber \\
&& \left. \qquad \qquad \qquad \quad \,\,\, \times  \(\bU_{i} +  \bF_i \bB_{0}^{\top}\(i\Delta_n\) + \tilde{\bF}_i \) - \bSigma_{0,u}\(i\Delta_n \) \right\|_{\max} \cr 
&&\leq C\left\{n^{-1/8}\(\log p\)^{2}  + p^{-1/2}\sqrt{\log p}   \right\}.
\end{eqnarray}
Combining \eqref{Thm2-eq1}, \eqref{Thm2-eq2}, and \eqref{Thm2-eq6}, we have, with the probability at least $1-6p^{-2-a}$,
\begin{eqnarray*}
&&\sup_{0 \leq i \leq n-k_2} \left\|\dfrac{n}{\phi k_1 k_2} \hat{\bOmega}_{i \Delta_n}^{\top}  \(\bU_{i} + \mathcal{E}_i^X + \tilde{\bF}_i + \bF_i \bB_{0}^{\top}\(i\Delta_n\) - \hat{\bF}_i \hat{\bB}_{i \Delta_n}^{\top}\)^{\top} \right.  \\
&& \left.  \qquad \qquad \qquad \qquad \qquad \qquad \quad \, \,\, \times \(\bU_{i} +  \bF_i \bB_{0}^{\top}\(i\Delta_n\) + \tilde{\bF}_i \) - \bI  \right\|_{\max} \cr 
&&\leq C\left\{n^{-1/8}\(\log p\)^{2}  + p^{-1/2}\sqrt{\log p}   \right\}.
\end{eqnarray*}
Then, from \eqref{Thm1-result1}, we have, with the probability at least $1-p^{-1-a}$,
\begin{eqnarray}\label{Thm2-eq7}
(\uppercase\expandafter{\romannumeral2}) &\leq& \sup_{0 \leq i \leq n-k_2} \left\|\dfrac{n}{\phi k_1 k_2} \hat{\bOmega}_{i \Delta_n}^{\top}  \(\bU_{i} + \mathcal{E}_i^X + \tilde{\bF}_i + \bF_i \bB_{0}^{\top}\(i\Delta_n\) - \hat{\bF}_i \hat{\bB}_{i \Delta_n}^{\top}\)^{\top} \right. \nonumber \\
&& \left. \qquad \qquad \qquad \qquad \qquad \qquad \quad \, \, \, \times \(\bU_{i} +  \bF_i \bB_{0}^{\top}\(i\Delta_n\) + \tilde{\bF}_i \) - \bI  \right\|_{\max} \cr
&& \times \sup_{0 \leq i \leq n- k_2} \left\| \hat{\bbeta}_{i \Delta_n} - \bbeta_0\(i\Delta_n\)  \right\|_{1} \cr
&\leq& C \left\{s_p^{2-\delta} n^{(-2+\delta)/8}\(\log p \)^{4-2\delta} + s_p^{2-\delta}p^{(-2+\delta)/2} \(\log p\)^{(2-\delta)/2}  \right\}.
\end{eqnarray}
Consider $(\uppercase\expandafter{\romannumeral3})$. 
We have
\begin{eqnarray*}
&&(\uppercase\expandafter{\romannumeral3}) \cr
&&\leq \sup_{0 \leq i \leq n- k_2} \left\| \hat{\bOmega}_{i \Delta_n}^{\top} \right\|_{\infty} \times \sup_{0 \leq i \leq n- k_2} \left\| \dfrac{n \zeta}{\phi k_1^2}\(\hat{\bV}^X - \bV^X \) \right\|_{\max} \times \sup_{0 \leq i \leq n- k_2} \left\| \hat{\bbeta}_{i \Delta_n}  \right\|_{1} \cr
&& \quad + \sup_{0 \leq i \leq n-k_2} \left\|  \hat{\bOmega}_{i \Delta_n}^{\top} - \bOmega_0\(i \Delta_n\)  \right\|_{\infty} \cr
&& \quad \quad \times \sup_{0 \leq i \leq n- k_2} \left\| \dfrac{n}{\phi k_1 k_2} \(\bU_{i} + \mathcal{E}_i^X + \tilde{\bF}_i + \bF_i \bB_{0}^{\top}\(i\Delta_n\) - \hat{\bF}_i \hat{\bB}_{i \Delta_n}^{\top}\)^{\top} \(\mathcal{Z}_{i}  +  \mathcal{\tilde{X}}_{i} + \mathcal{E}_{i}^{Y}  \) \right\|_{\max} \cr
&& \quad + \sup_{0 \leq i \leq n-k_2} \left\|  \hat{\bOmega}_{i \Delta_n}^{\top} - \bOmega_0\(i \Delta_n\)  \right\|_{\infty} \cr
&& \quad \quad \times  \sup_{0 \leq i \leq n- k_2}  \left\| \dfrac{n}{\phi k_1 k_2} \(\bU_{i} + \mathcal{E}_i^X + \tilde{\bF}_i + \bF_i \bB_{0}^{\top}\(i\Delta_n\) - \hat{\bF}_i \hat{\bB}_{i \Delta_n}^{\top}\)^{\top} \mathcal{E}_{i}^{X}  - \dfrac{n \zeta}{\phi k_1^2} {\bV}^X  \right\|_{\max} \cr
&& \quad \quad \times \sup_{0 \leq i \leq n- k_2} \left\| \hat{\bbeta}_{i \Delta_n} \right\|_{1} \cr
&& \quad + \sup_{0 \leq i \leq n- k_2} \left\| \bOmega_0\(i \Delta_n\) \right\|_{\infty}  \cr
&& \quad \quad \times \sup_{0 \leq i \leq n- k_2} \left\| \dfrac{n}{\phi k_1 k_2}\(\bU_{i} + \mathcal{E}_i^X + \tilde{\bF}_i + \bF_i \bB_{0}^{\top}\(i\Delta_n\) - \hat{\bF}_i \hat{\bB}_{i \Delta_n}^{\top}\)^{\top} \mathcal{E}_{i}^{X}  -  \dfrac{n \zeta}{\phi k_1^2 }\bV^X  \right\|_{\max} \cr
&& \quad \quad \times \sup_{0 \leq i \leq n- k_2} \left\| \hat{\bbeta}_{i \Delta_n} - \bbeta_0\(i\Delta_n\) \right\|_{1}.
\end{eqnarray*}
By Proposition \ref{Prop2}, we have
\begin{equation*}
\Pr \left\{ \sup_{0 \leq i \leq n- k_2} \left\| \dfrac{n \zeta}{\phi k_1^2}\(\hat{\bV}^X - \bV^X \) \right\|_{\max} \leq C n^{-1/2} \sqrt{\log p} \right\} \geq 1-p^{-2-a}.
\end{equation*}
Similar to the proofs of \eqref{Thm2-eq6}, we can show, with the probability at least $1-p^{-2-a}$,
\begin{eqnarray*}
&& \sup_{0 \leq i \leq n- k_2} \left\| \dfrac{n}{\phi k_1 k_2} \(\bU_{i} + \mathcal{E}_i^X + \tilde{\bF}_i + \bF_i \bB_{0}^{\top}\(i\Delta_n\) - \hat{\bF}_i \hat{\bB}_{i \Delta_n}^{\top}\)^{\top} \(\mathcal{Z}_{i}  +  \mathcal{\tilde{X}}_{i} + \mathcal{E}_{i}^{Y}  \) \right\|_{\max} \cr
&& \leq C \left\{s_p n^{-1/8}\(\log p\)^{3/2}  +  n^{-1/8} \(\log p\)^2 + p^{-1/2}\sqrt{\log p} \right\}\quad \text{and} \cr
&& \sup_{0 \leq i \leq n- k_2} \left\| \dfrac{n}{\phi k_1 k_2}\(\bU_{i} + \mathcal{E}_i^X + \tilde{\bF}_i + \bF_i \bB_{0}^{\top}\(i\Delta_n\) - \hat{\bF}_i \hat{\bB}_{i \Delta_n}^{\top}\)^{\top} \mathcal{E}_{i}^{X} -  \dfrac{n \zeta}{\phi k_1^2}\bV^X  \right\|_{\max} \cr
&& \leq C \left\{n^{-1/8}\(\log p\)^{2}  + p^{-1/2}\sqrt{\log p}   \right\}. 
\end{eqnarray*}
Then, from \eqref{Thm1-result1}, \eqref{Thm2-eq1}, and \eqref{Thm2-eq3}, we have
\begin{eqnarray}\label{Thm2-eq8}
&& \Pr \left( (\uppercase\expandafter{\romannumeral3}) \leq C \left\{s_p^{2-\delta} n^{(-2+\delta)/8}\(\log p\)^{4-2\delta}   + s_p^{2-\delta} p^{(-2+\delta)/2}\(\log p\)^{(2-\delta)/2}    \right. \right. \nonumber \\
&& \qquad \qquad \qquad   \quad \left. \left. + s_p s_{\omega, p} n^{(-2+q)/8}\(\log p\)^{4-2q} + s_p s_{\omega, p} p^{(-2+q)/2}\(\log p\)^{(2-q)/2}  \right\} \right) \cr
&& \geq 1-p^{-1-a}.
\end{eqnarray}
Consider $(\uppercase\expandafter{\romannumeral4})$. 
We have
\begin{eqnarray*}
(\uppercase\expandafter{\romannumeral4}) &\leq & C n^{-1/2} \Bigg\| \sum_{i=0}^{[1/(k_2 \Delta_n) ]-1}   \bOmega_0\(i k_2 \Delta_n\)  \Big[ \(\bU_{i k_2} + \mathcal{E}_{i k_2}^X \)^{\top} \left(\mathcal{Z}_{i  k_2}  +  \mathcal{E}_{i  k_2}^{Y}  \right) \cr
&& \qquad \qquad  \qquad \qquad \qquad \qquad \qquad \, -\left(\(\bU_{i  k_2} + \mathcal{E}_{i  k_2}^X \)^{\top} \mathcal{E}_{i  k_2}^{X} -\dfrac{k_2\zeta}{k_1} \bV^X \right) \bbeta_0\(i  k_2\Delta_n\) \Big]\Bigg\|_{\infty} \cr
&&  + C n^{-1/2} \Bigg\| \sum_{i=0}^{[1/( k_2 \Delta_n) ]-1}   \bOmega_0\(i  k_2 \Delta_n\)  \Big[ \(\bU_{i  k_2} + \mathcal{E}_{i k_2}^X \)^{\top} \mathcal{\tilde{X}}_{i  k_2} \cr
&& \qquad \qquad  \qquad \qquad \qquad  \qquad  \qquad \quad + \tilde{\bF}_{i k_2}^{\top}\(  \mathcal{Z}_{i  k_2}  +  \mathcal{E}_{i  k_2}^{Y} +  \mathcal{E}_{i  k_2}^{X} \bbeta_0\(i  k_2\Delta_n\) \) \Big] \Bigg\|_{\infty} \cr
&&  + C n^{-1/2} \Bigg\| \sum_{i=0}^{[1/( k_2 \Delta_n) ]-1}   \bOmega_0\(i  k_2 \Delta_n\) \Big[ \(\tilde{\bF}_{i k_2} + \bF_{i k_2} \bB_0^{\top}\(i k_2 \Delta_n \) - \hat{\bF}_{i k_2} \hat{\bB}_{i k_2 \Delta_n}^{\top}\)^{\top} \mathcal{\tilde{X}}_{i  k_2} \Big] \Bigg\|_{\infty} \cr
&& +  C n^{-1/2} \Bigg\| \sum_{i=0}^{[1/(k_2 \Delta_n) ]-1}   \bOmega_0\(i k_2 \Delta_n\)  \Big[ \(\bF_{i k_2} \bB_0^{\top}\(i k_2 \Delta_n \) - \hat{\bF}_{i k_2} \hat{\bB}_{i k_2 \Delta_n}^{\top} \)^{\top} \left(\mathcal{Z}_{i  k_2}  +  \mathcal{E}_{i  k_2}^{Y}  \right) \cr
&& \qquad \qquad  \qquad \qquad \qquad \qquad \qquad   -  \(\bF_{i k_2} \bB_0^{\top}\(i k_2 \Delta_n \) - \hat{\bF}_{i k_2} \hat{\bB}_{i k_2 \Delta_n}^{\top} \)^{\top} \mathcal{E}_{i  k_2}^{X} \bbeta_0\(i  k_2\Delta_n\) \Big]\Bigg\|_{\infty} \cr
&=& (\uppercase\expandafter{\romannumeral4})^{(1)} + (\uppercase\expandafter{\romannumeral4})^{(2)} + (\uppercase\expandafter{\romannumeral4})^{(3)} + (\uppercase\expandafter{\romannumeral4})^{(4)}.
\end{eqnarray*} 
For $(\uppercase\expandafter{\romannumeral4})^{(1)}$, note that each element of $\bU_i$, $\mathcal{E}_i^X$, $\mathcal{Z}_i$, and $\mathcal{E}_i^Y$ has sub-Gaussian tail. 
Thus, by  Bernstein’s inequality for martingales, we have, with the probability at least $1-p^{-2-a}$,
\begin{equation*}
(\uppercase\expandafter{\romannumeral4})^{(1)} \leq C s_p n^{-1/4} \sqrt{\log p}.
\end{equation*}
Consider $(\uppercase\expandafter{\romannumeral4})^{(2)}$. 
Let 
\begin{equation*}
\chi_1 = \left\{\sup_{0 \leq i \leq n-k_2} \sup_{t \in [i\Delta_n,\(i+ k_2 \) \Delta_n]} \sum_{j=1}^{p}\left|\beta_{0,j}(t) - \beta_{0,j}(i\Delta_n) \right| \leq Cs_p n^{-1/8}\sqrt{\log p}\right\}.
\end{equation*}
By the proofs of \eqref{Prop2-eq11}, we have
\begin{equation*}
\Pr \(  \chi_1 \) \geq 1-p^{-2-a}.
\end{equation*}
Under the event $\chi_1$, each element of $\bU_i + \mathcal{E}_i^X$ and  $\mathcal{\tilde{X}}_{i }$ has sub-Gaussian tail with the order of $n^{-1/4}$ and $s_p n^{-3/8} \sqrt{\log p}$, respectively.
Thus, by Bernstein’s inequality for martingales, we can show, with the probability at least $1-2p^{-2-a}$,
\begin{equation*}
\left\|  \sum_{i=0}^{[1/(k_2 \Delta_n) ]-1} \bOmega_0\(i k_2 \Delta_n\)  \(\bU_{i k_2} + \mathcal{E}_{i k_2}^X \)^{\top}  \mathcal{\tilde{X}}_{i k_2}  \right\|_{\infty} \leq C s_p n^{1/4} \(\log p\)^{3/2}.
\end{equation*}
Similarly, let 
\begin{equation*}
\chi_2 = \left\{\sup_{0 \leq i \leq n-k_2} \sup_{t \in [i\Delta_n,\(i+ k_2 \) \Delta_n]} \left\|\bB_{0}(t) - \bB_{0}(i\Delta_n) \right\|_{\infty} \leq C n^{-1/8}\sqrt{\log p}\right\}.
\end{equation*}
We have
\begin{equation*}
\Pr \(  \chi_2 \) \geq 1-p^{-2-a}.
\end{equation*}
Then, we can show, with the probability at least $1-2p^{-2-a}$,
\begin{equation*}
\left\|  \sum_{i=0}^{[1/(k_2 \Delta_n) ]-1} \bOmega_0\(i k_2 \Delta_n\)  \tilde{\bF}_{i k_2}^{\top}\(  \mathcal{Z}_{i  k_2}  +  \mathcal{E}_{i  k_2}^{Y} +  \mathcal{E}_{i  k_2}^{X} \bbeta_0\(i  k_2\Delta_n\) \)  \right\|_{\infty} \leq C n^{1/4} \(\log p\)^{3/2},
\end{equation*}
which implies
\begin{equation*}
\Pr\left\{ (\uppercase\expandafter{\romannumeral4})^{(2)} \leq C s_p n^{-1/4} \(\log p\)^{3/2} \right\} \geq 1-4p^{-2-a}.
\end{equation*}
Consider $(\uppercase\expandafter{\romannumeral4})^{(3)}$.
Recall that 
\begin{equation*}
\Pr \left( \sup_{0 \leq i \leq n- k_2} \max_{1 \leq l \leq p}\sum_{j=1}^{k_2-k_1+1}\(T_{jl}^{(i)}\)^2 \leq C \left\{ \(\log p\)^3+ p^{-1}n^{1/4}  \right\}\right) \geq 1-p^{-2-a},
\end{equation*}
where $\bF_i \bB_{0}^{\top}\(i\Delta_n\) - \hat{\bF}_i \hat{\bB}_{i \Delta_n}^{\top}=\(T_{jl}^{(i)}\)_{1\leq j \leq k_2-k_1+1, 1 \leq l \leq p}$.
Thus, from the Cauchy-Schwarz inequality and \eqref{Prop2-eq11},  we have, with the probability at least $1-2p^{-2-a}$,
\begin{eqnarray*}
(\uppercase\expandafter{\romannumeral4})^{(3)} &\leq&   C n^{-1/4} \sqrt{\left\{\(\log p\)^3  +p^{-1}n^{1/4} \right\} \times \(k_2-k_1+1\)\(s_p n^{-3/8} \log p\)^2 } \cr
&\leq& C \left\{s_p n^{-1/4}\(\log p\)^{5/2} + p^{-1/2}s_p n^{-1/8} \log p   \right\}.
\end{eqnarray*}
For $(\uppercase\expandafter{\romannumeral4})^{(4)}$, we first consider 
\begin{equation*}
\Bigg\| \sum_{i=0}^{[1/(k_2 \Delta_n) ]-1}   \bOmega_0\(i k_2 \Delta_n\)  \(\bF_{i k_2} \bB_0^{\top}\(i k_2 \Delta_n \) - \hat{\bF}_{i k_2} \hat{\bB}_{i k_2 \Delta_n}^{\top} \)^{\top} \mathcal{E}_{i  k_2}^{X} \bbeta_0\(i  k_2\Delta_n\) \Bigg\|_{\infty}.
\end{equation*}
By Proposition \ref{Prop3}, we have, with the probability at least $1-p^{-2-a}$,
\begin{eqnarray*}
&&\bF_i\(\bH_i^{\top}\)^{-1} - \hat{\bF}_i \cr
&& = \bF_i\(\bH_i^{\top}\)^{-1} - p^{-1}\(\bF_i \bB_{0}^{\top}\(i\Delta_n\) + \bU_{i} + \tilde{\bF}_i + \mathcal{E}_{i}^X  \)\hat{\bB}_{i \Delta_n} - p^{-1}\(\mathcal{X}_i - \mathcal{X}^c_i - \mathcal{E}_{i}^X\)\hat{\bB}_{i \Delta_n} \cr
&& = \bF_i\(\bH_i^{\top}\)^{-1} \(\bI_r - p^{-1}\bH_i^{\top} \bB_{0}^{\top}\(i\Delta_n\) \hat{\bB}_{i \Delta_n} \) - p^{-1}\(\bU_{i} + \tilde{\bF}_i + \mathcal{E}_{i}^X  \)\hat{\bB}_{i \Delta_n} \cr 
&& \quad - p^{-1}\(\mathcal{X}_i - \mathcal{X}^c_i - \mathcal{E}_{i}^X\)\hat{\bB}_{i \Delta_n} \cr
&& = p^{-1} \bF_i\(\bH_i^{\top}\)^{-1} \(\hat{\bB}_{i \Delta_n} - \bB_{0}\(i\Delta_n\)\bH_i \)^{\top}\hat{\bB}_{i \Delta_n}- p^{-1}\(\bU_{i} + \tilde{\bF}_i + \mathcal{E}_{i}^X  \)\hat{\bB}_{i \Delta_n} \cr
&& \quad - p^{-1}\(\mathcal{X}_i - \mathcal{X}^c_i - \mathcal{E}_{i}^X\)\hat{\bB}_{i \Delta_n}
\end{eqnarray*}
for all $0 \leq i \leq n-k_2$.
Thus, we have, with the probability at least $1-p^{-2-a}$,
\begin{eqnarray*}
&& \Bigg\| \sum_{i=0}^{[1/(k_2 \Delta_n) ]-1}   \bOmega_0\(i k_2 \Delta_n\)  \(\bF_{i k_2} \bB_0^{\top}\(i k_2 \Delta_n \) - \hat{\bF}_{i k_2} \hat{\bB}_{i k_2 \Delta_n}^{\top} \)^{\top} \mathcal{E}_{i  k_2}^{X} \bbeta_0\(i  k_2\Delta_n\) \Bigg\|_{\infty} \cr
&& = \Bigg\| \sum_{i=0}^{[1/(k_2 \Delta_n) ]-1}   \bOmega_0\(i k_2 \Delta_n\)  \(\bF_{ik_2}\(\bH_{ik_2}^{\top}\)^{-1} \left[\(\bB_{0}\(ik_2\Delta_n\) \bH_{ik_2} \)^{\top} - \hat{\bB}_{ik_2 \Delta_n}^{\top} \right] \right. \\
&&  \qquad  \qquad \qquad \qquad \quad  \,\, \, \left.  + \left[ \bF_{ik_2}\(\bH_{ik_2}^{\top}\)^{-1} - \hat{\bF}_{ik_2}  \right] \hat{\bB}_{ik_2 \Delta_n}^{\top} \)^{\top} \mathcal{E}_{i  k_2}^{X} \bbeta_0\(i  k_2\Delta_n\) \Bigg\|_{\infty} \cr
&& \leq \Bigg\| \sum_{i=0}^{[1/(k_2 \Delta_n) ]-1}   \bOmega_0\(i k_2 \Delta_n\) \left( \hat{\bB}_{ik_2 \Delta_n} - \bB_{0}\(ik_2\Delta_n\) \bH_{ik_2}  \right)  \bH_{ik_2}^{-1} \bF_{ik_2}^{\top} \mathcal{E}_{i  k_2}^{X} \bbeta_0\(i  k_2\Delta_n\) \Bigg\|_{\infty} \cr
&& \quad + p^{-1} \Bigg\| \sum_{i=0}^{[1/(k_2 \Delta_n) ]-1}   \bOmega_0\(i k_2 \Delta_n\) \hat{\bB}_{ik_2 \Delta_n} \hat{\bB}_{ik_2 \Delta_n}^{\top} \(\hat{\bB}_{ik_2 \Delta_n} - \bB_{0}\(ik_2\Delta_n\)\bH_{ik_2} \)  \cr
&&  \qquad  \qquad \qquad \qquad \qquad \qquad \qquad \qquad \qquad  \qquad  \quad \times \bH_{ik_2}^{-1} \bF_{i k_2}^{\top } \mathcal{E}_{i  k_2}^{X} \bbeta_0\(i  k_2\Delta_n\)\Bigg\|_{\infty} \cr
&& \quad + p^{-1} \Bigg\| \sum_{i=0}^{[1/(k_2 \Delta_n) ]-1}   \bOmega_0\(i k_2 \Delta_n\) \(\hat{\bB}_{ik_2 \Delta_n} \hat{\bB}_{ik_2 \Delta_n}^{\top} - \bB_{0}\(ik_2\Delta_n\)  \bB_{0}^{\top} \(ik_2\Delta_n\)  \) \cr
&&  \qquad  \qquad \qquad \qquad \qquad \qquad \qquad  \,\,\,  \times \(\bU_{ik_2} + \tilde{\bF}_{ik_2} + \mathcal{E}_{ik_2}^X  \)^{\top} \mathcal{E}_{i  k_2}^{X} \bbeta_0\(i  k_2\Delta_n\) \Bigg\|_{\infty} \cr
&& \quad + p^{-1} \Bigg\| \sum_{i=0}^{[1/(k_2 \Delta_n) ]-1}   \bOmega_0\(i k_2 \Delta_n\) \bB_{0}\(ik_2\Delta_n\)  \bB_{0}^{\top} \(ik_2\Delta_n\)  \cr
&&  \qquad  \qquad \qquad \qquad  \,\,\,  \times \left[\(\bU_{ik_2} + \tilde{\bF}_{ik_2} + \mathcal{E}_{ik_2}^X  \)^{\top} \mathcal{E}_{i  k_2}^{X} - \dfrac{k_2 \zeta}{k_1} \bV^X\right] \bbeta_0\(i  k_2\Delta_n\) \Bigg\|_{\infty} \cr
&& \quad + \dfrac{k_2 \zeta}{p k_1} \Bigg\| \sum_{i=0}^{[1/(k_2 \Delta_n) ]-1}   \bOmega_0\(i k_2 \Delta_n\) \bB_{0}\(ik_2\Delta_n\)  \bB_{0}^{\top}\(ik_2\Delta_n\)  \bV^X  \bbeta_0\(i  k_2\Delta_n\)  \Bigg\|_{\infty} \cr
&& \quad + p^{-1} \Bigg\| \sum_{i=0}^{[1/(k_2 \Delta_n) ]-1}   \bOmega_0\(i k_2 \Delta_n\) \hat{\bB}_{ik_2 \Delta_n} \hat{\bB}_{ik_2 \Delta_n}^{\top} \(\mathcal{X}_{ik_2} - \mathcal{X}^c_{ik_2} - \mathcal{E}_{ik_2}^X  \)^{\top} \mathcal{E}_{i  k_2}^{X} \bbeta_0\(i  k_2\Delta_n\) \Bigg\|_{\infty} \cr
&& = (A) + (B) + (C) + (D) + (E) + (F).
\end{eqnarray*}
Consider $(A)$--$(C)$.
By the proofs of Proposition \ref{Prop2}, we have, with the probability at least $1-p^{-2-a}$, 
\begin{eqnarray*}
&& \max_{i}\left\| \bF_{i}^{\top} \mathcal{E}_{i}^X \bbeta_0\(i  k_2\Delta_n\)\right\|_{\infty}    \leq C s_p n^{1/8} \sqrt{\log p}  \quad \text{and} \cr
&& \max_{i}\left\| \(\bU_{i} + \tilde{\bF}_{i} + \mathcal{E}_{i}^X  \)^{\top} \mathcal{E}_{i}^X \bbeta_0\(i  k_2\Delta_n\) \right\|_{1}    \leq C \left\{ p s_p n^{1/8} \sqrt{\log p} + s_p n^{1/4}  \right\}.
\end{eqnarray*}
Then, by Proposition \ref{Prop3}, we can show, with the probability at least $1-2p^{-2-a}$, 
\begin{equation*}
(A) + (B) + (C)   \leq C \left\{s_p n^{1/4} \(\log p\)^2  + p^{-1/2} s_p n^{3/8} \sqrt{\log p} + p^{-3/2}s_p n^{1/2} \right\}.
\end{equation*}
Consider $(D)$.
Under the event $\chi_2$, each element of  $\mathcal{\tilde{F}}_{i }$ has sub-Gaussian tail with the order of $n^{-3/8} \sqrt{\log p}$. 
Also, each element of  $\bU_i$ and $\mathcal{E}_{i}^X$ has sub-Gaussian tail with the order of $n^{-1/4}$. 
Thus, from Bernstein’s inequality for martingales, we can show, with the probability at least $1-2p^{-2-a}$,  
\begin{equation*}
(D)   \leq Cs_p n^{1/4} \sqrt{\log p}.
\end{equation*}
For $(E)$, by Assumption \ref{assumption1}(a),(e), we have
\begin{equation*}
(E)   \leq C p^{-1 }s_p n^{1/2} \text{ a.s.}
\end{equation*}
For $(F)$, by \eqref{Prop2-eq11} and Proposition \ref{Prop3}, we have, with the probability at least $1-2p^{-2-a}$,
\begin{equation*}
(F)   \leq C n^{1/4} \(\log p\)^2.
\end{equation*}
Thus, we have, with the probability at least $1-7p^{-2-a}$,
\begin{eqnarray*}
&& \Bigg\| \sum_{i=0}^{[1/(k_2 \Delta_n) ]-1}   \bOmega_0\(i k_2 \Delta_n\)  \(\bF_{i k_2} \bB_0^{\top}\(i k_2 \Delta_n \) - \hat{\bF}_{i k_2} \hat{\bB}_{i k_2 \Delta_n}^{\top} \)^{\top} \mathcal{E}_{i  k_2}^{X} \bbeta_0\(i  k_2\Delta_n\) \Bigg\|_{\infty}\cr
&&\leq C \left\{s_p n^{1/4} \(\log p\)^2  + p^{-1/2} s_p n^{3/8} \sqrt{\log p} + p^{-1}s_p n^{1/2} \right\}.
\end{eqnarray*} 
Similarly, we can show, with the probability at least $1-7p^{-2-a}$,
\begin{eqnarray*}
&& \Bigg\| \sum_{i=0}^{[1/(k_2 \Delta_n) ]-1}   \bOmega_0\(i k_2 \Delta_n\)  \(\bF_{i k_2} \bB_0^{\top}\(i k_2 \Delta_n \) - \hat{\bF}_{i k_2} \hat{\bB}_{i k_2 \Delta_n}^{\top} \)^{\top} \(\mathcal{Z}_{i  k_2} + \mathcal{E}_{i  k_2}^{Y}  \)\Bigg\|_{\infty}  \cr
&& \leq C \left\{ n^{1/4} \(\log p\)^2  + p^{-1/2} n^{3/8} \sqrt{\log p}   \right\},
\end{eqnarray*} 
which implies
\begin{equation*}
\Pr \left((\uppercase\expandafter{\romannumeral4})^{(4)} \leq C \left\{s_p n^{-1/4} \(\log p\)^2  + p^{-1/2} s_p n^{-1/8} \sqrt{\log p} + p^{-1}s_p \right\} \right) \geq 1-14p^{-2-a}
\end{equation*} 
and
\begin{equation}\label{Thm2-eq9}
\Pr \left((\uppercase\expandafter{\romannumeral4}) \leq C\left\{s_p n^{-1/4}\(\log p \)^{5/2} + p^{-1/2} s_p n^{-1/8} \log p + p^{-1}s_p  \right\}\right) \geq 1-p^{-1-a}.
\end{equation} 
Consider $(\uppercase\expandafter{\romannumeral5})$. Since the process $\bbeta_0(t)$ has the sub-Gaussian tail, we can show
\begin{equation}\label{Thm2-eq10}
\Pr \left\{ (\uppercase\expandafter{\romannumeral5}) \leq C n^{-1/4}\sqrt{\log p} \right\} \geq 1-p^{-1-a}.
\end{equation}
For $(\uppercase\expandafter{\romannumeral6})$, by Assumption \ref{assumption1}(a), we have
\begin{equation}\label{Thm2-eq11}
(\uppercase\expandafter{\romannumeral6}) \leq C n^{-1/4} \text{ a.s.}
\end{equation}
Combining \eqref{Thm2-eq4}--\eqref{Thm2-eq5} and \eqref{Thm2-eq7}--\eqref{Thm2-eq11}, we have, with the probability at least $1-p^{-a}$,
\begin{eqnarray}\label{Thm2-eq12}
&& \|  \hat{I\beta}  - I\beta_0 \| _{\max} \cr
&&  \leq C \left\{s_p^{2-\delta} n^{(-2+\delta)/8}\(\log p\)^{4-2\delta}   + s_p^{2-\delta} p^{(-2+\delta)/2}\(\log p\)^{(2-\delta)/2}   \right. \nonumber \\
&& \qquad \,\,\,  \left.  + s_p s_{\omega, p} n^{(-2+q)/8}\(\log p\)^{4-2q} + s_p s_{\omega, p} p^{(-2+q)/2}\(\log p\)^{(2-q)/2}  \right\}.
\end{eqnarray}
\endpf

\subsection{Proof of Theorem \ref{Thm3}}\label{Proof-Thm3}
\textbf{Proof of Theorem \ref{Thm3}.}
By \eqref{Thm2-result1}, there exists a constant $C_{h,a}$ such that
\begin{equation*}
 \Pr \left\{ \|  \hat{I\beta}  - I\beta_0 \| _{\max}  \leq h_n/2 \right\} \geq 1-p^{-a}.
\end{equation*}
Thus, it suffices to show the statement under  $\{\|  \hat{I\beta}  - I\beta_0 \| _{\max}  \leq h_n/2 \}$.
Similar to the proofs of Theorem 1 \citep{kim2026high}, we can obtain
\begin{eqnarray*} 
	\|  \tilde{I\beta}  - I\beta_0 \|_1 \leq  C s_p h_n^{1-\delta}.  
\end{eqnarray*}
\endpf

\subsection{Proof of Proposition \ref{Prop1}}

\textbf{Proof of Proposition \ref{Prop1}.}
By Proposition \ref{Prop5} and \eqref{Thm1-eq1}, we can show Proposition \ref{Prop1} similar to the proofs of Theorem 2 \citep{agarwal2012fast}.
\endpf

\section{Discussion on the jump truncation levels}\label{jump-discuss}
{
In this section, we provide additional details on the choice of the truncation parameters $w_n$, $v_{j,n}$, and $v_{j,n}^{(2)}$ in \eqref{jump_trunc1} and \eqref{jump_trunc2}.
We first explain the choice of $w_n$ and $v_{j,n}$ in \eqref{jump_trunc1}. 
By the proofs of Proposition \ref{Prop2}, for any positive constant $a$, we can show
\begin{equation*}
\Pr \Bigg[
\left\{ \max_{i}  \left\| \mathcal{Y}_i^{c} \right\|_{\infty} \leq   C s_p \sqrt{\dfrac{k_1 \log p}{n}} \right\}
\cap
\left\{  \max_{i}  \left\| \mathcal{X}_i^c \right\|_{\max} \leq   C \sqrt{\dfrac{k_1 \log p}{n}} \right\}
\Bigg] \geq 1-p^{-a}.
\end{equation*}
Hence, if the truncation levels are chosen larger than these bounds, then the continuous parts are retained with high probability, while sufficiently large jumps can be detected and truncated.
Since $k_1 = Cn^{1/2}$, this leads to the orders
$
w_n = C s_p \sqrt{\log p}\, n^{-1/4}
$ 
and
$
v_{j,n} =C \sqrt{\log p}\, n^{-1/4}.
$
The $\log p$ term is needed to control the continuous parts with high probability, and the additional $s_p$ term in $w_n$ is the cost of handling the contribution of factors to the dependent process.
On the other hand, the truncation levels must also be sufficiently sharp in order to establish the restricted eigenvalue condition.
In particular, as in \eqref{Prop5-eq9} in the proof of Proposition \ref{Prop5}, we need
\begin{equation*}
\left(\max_j v_{j,n}\right)^2 n^{1/4}\log p \to 0
\quad \text{ as } n,p \to \infty.
\end{equation*}
Thus, the choice in \eqref{jump_trunc1} balances two requirements: it is large enough to detect jumps with high probability, but still sharp enough for the restricted eigenvalue argument.
We next explain the choice of $v^{(2)}_{j,n}$ in \eqref{jump_trunc2}. 
The goal is to estimate the noise covariance matrix from the observed log-returns, so the required bound is simpler.
Using similar arguments, we can show that the noise term is bounded by $C\sqrt{\log p}$ with high probability.
Hence, to detect jumps with high probability, it is required to choose
$
v^{(2)}_{j,n} \ge C\sqrt{\log p}.
$
At the same time, the truncation level must remain sharp to establish the restricted eigenvalue condition.
Specifically, we require
\begin{equation*}
\left(\max_j v_{j,n}^{(2)}\right)^2 \log p / n \to 0
\quad \text{as } n,p \to \infty.
\end{equation*}
Therefore, the sharp truncation level in \eqref{jump_trunc2} is used.
Finally, when the jump sizes are finite, truncation for the observed log-returns is not required.
However, since we do not impose any restriction on the jump size process, the proposed truncation step is used to handle the heavy-tailedness of the jump sizes.
}

\section{Miscellaneous materials}\label{SEC-materials}

\begingroup
\scriptsize
\setlength{\tabcolsep}{4pt}
\renewcommand{\arraystretch}{0.65}
\setlength\LTleft{\fill}
\setlength\LTright{\fill}

\begin{longtable}{@{}p{2.1cm}p{3.2cm}p{8.2cm}@{}}
\caption{{Groups, symbols, and descriptions of the 138 characteristic-sorted and industry portfolios used in Section \ref{SEC-5}. We adopt the same portfolio symbols as in \citet{fama1997industry, jensen2023there, y2022open}}.}
\label{Table4} \\

\toprule
Group & Symbol & Description \\
\midrule
\endfirsthead

\multicolumn{3}{@{}l}{\tablename~\thetable\ (continued)}\\
\toprule
Group & Symbol & Description \\
\midrule
\endhead

\midrule
\multicolumn{3}{r@{}}{Continued on the next page}
\endfoot

\bottomrule
\endlastfoot


Value & abnormalaccruals & Abnormal Accruals \\
& adexp & Advertising Expense \\
& ageipo & IPO and age \\
& at\_me & Assets-to-market \\
& be\_me & Book-to-market equity \\
& bev\_mev & Book-to-market enterprise value \\
& bmdec & Book to market using December ME \\
& compequiss & Composite equity issuance \\
& debt\_me & Debt-to-market \\
& debtissuance & Debt Issuance \\
& div12m\_me & Dividend yield \\
& ebitda\_mev & Ebitda-to-market enterprise value \\
& eq\_dur & Equity duration \\
& eqnpo\_12m & Equity net payout \\
& eqnpo\_me & Net payout yield \\
& eqpo\_me & Payout yield \\
& frontier & Efficient frontier index \\
& grsaleotgrovrehead & Change sales minus change SGA \\
& meanrankrevgrowth & Revenue Growth Rank \\
& ni\_me & Earnings-to-price \\
& ocf\_me & Operating cash flow-to-market \\
& orderbacklog & Order backlog \\
& rdability & R\&D ability \\
& sale\_me & Sales-to-market \\
& sfe & Earnings Forecast to price \\
\addlinespace[9pt]

Size & ami\_126d & Amihud Measure \\
& dolvol\_126d & Dollar trading volume \\
& exchswitch & Exchange Switch \\
& market\_equity & Market Equity \\
& prc & Price per share \\
& rd\_me & R\&D-to-market \\
& rdipo & IPO and no R\&D spending \\
& rio\_mb & Inst Own and Market to Book \\
& rio\_turnover & Inst Own and Turnover \\
& rio\_volatility & Inst Own and Idio Vol \\
& shortinterest & Short Interest \\
& varcf & Cash-flow to price variance \\

\addlinespace[9pt]

Profitability & brandinvest & Brand capital investment \\
& dolvol\_var\_126d & Coefficient of variation for dollar trading volume \\
& earnsupbig & Earnings surprise of big firms \\
& ebit\_bev & Return on net operating assets \\
& ebit\_sale & Profit margin \\
& f\_score & Piotroski F-score \\
& fr & Pension Funding Status \\
& intrinsic\_value & Intrinsic value-to-market \\
& ni\_be & Return on equity \\
& niq\_be & Quarterly return on equity \\
& o\_score & Ohlson O-score \\
& ocf\_at & Operating cash flow to assets \\
& ope\_be & Operating profits-to-book equity \\
& ope\_bel1 & Operating profits-to-lagged book equity \\
& turnover\_var\_126d & Coefficient of variation for share turnover \\

\addlinespace[9pt]

Investment & aliq\_at & Liquidity of book assets \\
& at\_gr1 & Asset Growth \\
& be\_gr1a & Change in common equity \\
& capx\_gr1 & CAPEX growth (1 year) \\
& capx\_gr2 & CAPEX growth (2 years) \\
& capx\_gr3 & CAPEX growth (3 years) \\
& coa\_gr1a & Change in current operating assets \\
& col\_gr1a & Change in current operating liabilities \\
& emp\_gr1 & Hiring rate \\
& inv\_gr1 & Inventory growth \\
& inv\_gr1a & Inventory change \\
& lnoa\_gr1a & Change in long-term net operating assets \\
& mispricing\_mgmt & Mispricing factor: Management \\
& ncoa\_gr1a & Change in noncurrent operating assets \\
& nncoa\_gr1a & Change in net noncurrent operating assets \\
& noa\_gr1a & Change in net operating assets \\
& ppeinv\_gr1a & Change PPE and Inventory \\
& rds & Real dirty surplus \\
& ret\_60\_12 & Long-term reversal \\
& sale\_gr1 & Sales Growth (1 year) \\
& sale\_gr3 & Sales Growth (3 years) \\
& saleq\_gr1 & Sales growth (1 quarter) \\
& seas\_2\_5na & Years 2--5 lagged returns, nonannual \\

\addlinespace[9pt]

Momentum & announcementreturn & Earnings announcement return \\
& customermomentum & Customer momentum \\
& firmagemom & Firm Age - Momentum \\
& indmom & Industry momentum \\
& mom12m & Momentum over 12 months \\
& mom6m & Momentum over 6 months \\
& prc\_highprc\_252d & Current price to high price over last year \\
& ret\_12\_1 & Price momentum t-12 to t-1 \\
& ret\_3\_1 & Price momentum t-3 to t-1 \\
& ret\_6\_1 & Price momentum t-6 to t-1 \\
& ret\_9\_1 & Price momentum t-9 to t-1 \\
& resff3\_12\_1 & Residual momentum t-12 to t-1 \\
& resff3\_6\_1 & Residual momentum t-6 to t-1 \\
& seas\_1\_1na & Year 1-lagged return, nonannual \\

\addlinespace[9pt]

Industry & Aero  & Aircraft \\
& Agric  & Agriculture \\
& Autos  & Automobiles and Trucks \\
& Banks  & Banking \\
& Beer   & Alcoholic Beverages \\
& BldMt  & Construction Materials \\
& Books  & Printing and Publishing \\
& Boxes  & Shipping Containers \\
& BusSv  & Business Services \\
& Chems  & Chemicals \\
& Chips  & Electronic Equipment \\
& Clths  & Apparel \\
& Cnstr  & Construction \\
& Coal   & Coal \\
& Comps  & Computers \\
& Drugs  & Pharmaceutical Products \\
& ElcEq  & Electrical Equipment \\
& Enrgy  & Petroleum and Natural Gas \\
& FabPr  & Fabricated Products \\
& Fin    & Trading \\
& Food   & Food Products \\
& Fun    & Entertainment \\
& Gold   & Precious Metals \\
& Guns   & Defense \\
& Hlth   & Healthcare \\
& Hshld  & Consumer Goods \\
& Insur  & Insurance \\
& LabEq  & Measuring and Control Equipment \\
& Mach   & Machinery \\
& Meals  & Restaurants, Hotels, and Motels \\
& MedEq  & Medical Equipment \\
& Mines  & Nonmetallic Mining \\
& Misc   & Miscellaneous \\
& Paper  & Business Supplies \\
& PerSv  & Personal Services \\
& RlEst  & Real Estate \\
& Rtail  & Retail \\
& Rubbr  & Rubber and Plastic Products \\
& Ships  & Shipbuilding and Railroad Equipment \\
& Smoke  & Tobacco Products \\
& Soda   & Candy and Soda \\
& Steel  & Steel Works, Etc. \\
& Telcm  & Telecommunications \\
& Toys   & Recreational Products \\
& Trans  & Transportation \\
& Txtls  & Textiles \\
& Util   & Utilities \\
& Whlsl  & Wholesale \\

\end{longtable}
\endgroup

\begin{algorithm}[H]
         \caption{FATEN-LASSO estimation procedure}   \label{FATEN-LASSO}
         \footnotesize
         \begin{algorithmic}

  \State \textbf{1.} Estimate the factor loading matrix and smoothed latent factor variable:
\begin{equation*}
(\hat{\bB}_{i \Delta_n}, \hat{\bF}_i)=\arg \min_{\bB \in \mathbb{R}^{p \times r}, \bF \in \mathbb{R}^{(k_2 - k_1 +1) \times r}} \|\mathcal{X}_i-\bF \bB^{\top} \|_{F}^{2},
\end{equation*}
subject to 
\begin{equation*}
p^{-1} \bB^{\top} \bB = \bI_r \quad \text{and} \quad \bF^{\top} \bF \text{ is an $r \times r$ diagonal matrix}.
\end{equation*}

   \State \textbf{2.}  Estimate the smoothed idiosyncratic variable:
\begin{equation*}
\hat{\bU}_i = \mathcal{X}_i-\hat{\bF}_i \hat{\bB}^{\top}_{i \Delta_n}.
\end{equation*}

 \State \textbf{3.} Obtain the noise covariance matrix estimator:
 \begin{equation*}  
 \hat{\bV}^X = \dfrac{1}{2n} \sum_{i=1}^{n} \Delta_i^n \bX^{\text{trunc}}   \(\Delta_i^n \bX^{\text{trunc}}\)^{\top}.
\end{equation*}

 \State \textbf{4.} Estimate the instantaneous coefficient:
  \begin{equation*}  
 \hat{\bbeta}_{i\Delta_n}=\(\hat{\theta}_{i\Delta_n,j}\)_{j=1,\ldots,p},
\end{equation*}
where
 \begin{equation*}  
 \hat{\btheta}_{i\Delta_n} = \underset{\| \btheta \|_{1} \leq \rho}{\operatorname{arg \, min}}  \, \dfrac{n}{2\phi k_1 k_2}\left\| \mathcal{Y}_{i} - \hat{\bG}_i \btheta \right\|_2^2 - \dfrac{n \zeta}{2\phi k_1^2} \btheta^{\top} \hat{\bV} \btheta + \eta \left\| \btheta \right\| _{1},  \quad \hat{\bV} = 
\begin{pmatrix}
    \begin{array}{c|c}
        \hat{\bV}^X & \mathbf{0}_{p \times r} \\ \hline
        \mathbf{0}_{r \times p} & \mathbf{0}_{r \times r}
    \end{array}
\end{pmatrix},
\end{equation*}
$\rho = C_{\rho} s_p$,  $k_1 = c_{k_1} n^{1/2}$, $k_2 = c_{k_2} n^{3/4}$, and $\eta=C_{\eta} \left\{s_p n^{-1/8} \(\log p\)^{2} + p^{-1/2} s_p \sqrt{\log p} \right\}$  for some  constants $C_{\rho}$, $c_{k_1}$,  $c_{k_2}$, and $C_\eta$.
 
 \State  \textbf{5.}  Estimate the inverse instantaneous idiosyncratic volatility matrix: 
 \begin{equation*}  
	\hat{\bOmega}_{i \Delta_n} = \arg \min \| \bOmega\|_{1} \quad \text{s.t.} \quad  \left\|    \left(\dfrac{n}{\phi k_1 k_2} \hat{\bU}_i^{\top} \hat{\bU}_i - \dfrac{n \zeta}{\phi k_1^2}\hat{\bV}^X \right)\bOmega  - \bI  \right\|_{\max} \leq \tau, 
\end{equation*}
where $\tau=C_\tau  \left\{ n^{-1/8}\(\log p\)^{2}  + p^{-1/2}\sqrt{\log p} \right\} $    for some large constant $C_\tau$.

 \State  \textbf{6.} Obtain the debiased instantaneous coefficient estimator:
 \begin{equation*}
	\tilde{ \bbeta}_{i \Delta_n} = \hat{\bbeta}_{i \Delta_n}  +  \dfrac{n}{\phi k_1 k_2}  \hat{\bOmega}_{i \Delta_n}^{\top}  \left\{  \hat{\bU}_i^{\top} \mathcal{Y}_{i} - \left(  \hat{\bU}_i^{\top} \mathcal{X}_i - \dfrac{k_2 \zeta}{k_1} \hat{\bV}^X \right) \hat{\bbeta}_{i \Delta_n} \right\}.
\end{equation*}

\State  \textbf{7.}  Obtain the debiased integrated coefficient estimator:
\begin{equation*}
	\hat{I \beta}  = \sum_{i=0}^{[1/(k_2 \Delta_n) ]-1}\tilde{\bbeta}_{i k_2 \Delta_n} k_2 \Delta_n.
	\end{equation*}

\State   \textbf{8.} Threshold the debiased integrated coefficient estimator:
	\begin{equation*}
	\tilde{I\beta}_j= s (\hat{I\beta}_j)  \1 \(  |\hat{I\beta}_j  | \geq h_n  \) \quad \text{and} \quad  \tilde{I \beta} = \( \tilde{I\beta}_j \)_{j=1,\ldots,p},
\end{equation*}
where  $s (x)$ satisfies $|s (x)-x| \leq h_n$ and  {$h_n= C_h \Big[s_p^{2-\delta} n^{(-2+\delta)/8}\(\log p\)^{4-2\delta}   + s_p^{2-\delta} p^{(-2+\delta)/2}\(\log p\)^{(2-\delta)/2} + s_p s_{\omega, p} n^{(-2+q)/8}\(\log p\)^{4-2q} + s_p s_{\omega, p} p^{(-2+q)/2}\(\log p\)^{(2-q)/2} \Big]$} for some large constant $C_h$.

\end{algorithmic}
\end{algorithm}


\end{spacing}
\end{document}